%% file: Main-Suppl_ArXiv_final.tex
\documentclass[utf8]{FrontiersinHarvard} % for articles in journals using the Harvard Referencing Style (Author-Date), for Frontiers Reference Styles by Journal: https://zendesk.frontiersin.org/hc/en-us/articles/360017860337-Frontiers-Reference-Styles-by-Journal
\usepackage{url,lineno,microtype,subcaption}
\usepackage[hidelinks]{hyperref}
\usepackage[onehalfspacing]{setspace}
\usepackage{booktabs}
\usepackage{siunitx}
\usepackage{amssymb}
\usepackage{graphicx}
\usepackage{float}
\usepackage{longtable}

\usepackage{caption}
\usepackage{lscape}

\def\keyFont{\fontsize{8}{11}\helveticabold }
\def\firstAuthorLast{van 't Hoff {et~al.}} %use et al only if is more than 1 author
\def\Authors{Merel L.R. van 't Hoff\,$^{1,*}$, {\L}ukasz Tychoniec\,$^{2}$, John J. Tobin\,$^{3}$, and Daniel Harsono,$^{4}$ }
\def\Address{$^{1}$Purdue University, Department of Physics and Astronomy, West-Lafayette, IN, USA  \\
$^{2}$Leiden Observatory, Leiden University, Leiden, the Netherlands\\
$^{3}$National Radio Astronomy Observatory, Charlottesville, VA, USA \\
$^{4}$National Tsing Hua University, Institute of Astronomy, Hsinchu, Taiwan}

\def\corrAuthor{Purdue University, Department of Physics and Astronomy, 525 Northwestern Ave, West Lafayette, IN 47907, USA}

\def\corrEmail{vanthoff@purdue.edu}

\begin{document}
\onecolumn
\firstpage{1}

\title[L1527 Chemical Inventory]{A Chemical Inventory of the Disk around the Class 0 Protostar L1527 IRS with ALMA} 

\author[\firstAuthorLast ]{\Authors} %This field will be automatically populated
\address{} %This field will be automatically populated
\correspondance{} %This field will be automatically populated

\extraAuth{}% If there are more than 1 corresponding author, comment this line and uncomment the next one.
%\extraAuth{corresponding Author2 \\ Laboratory X2, Institute X2, Department X2, Organization X2, Street X2, City X2 , State XX2 (only USA, Canada and Australia), Zip Code2, X2 Country X2, email2@uni2.edu}

\maketitle

\begin{abstract}

%%% Leave the Abstract empty if your article does not require one, please see the Summary Table for full details.
\section{Planet formation starts in disks that are still embedded within their natal envelopes. Here, we compile an extensive inventory of the chemical composition of the disk and envelope ($<$ 3500 au) around the Class 0 protostar L1527 IRS. Using all publicly available ALMA (Atacama Large Millimeter/submillimeter Array) data, we report the detection of 40 molecules, including isotopologues. Of these, 23 are different molecular species and 29 are reported here for the first time toward L1527 in ALMA observations. CH$_3$OH is the only complex organic molecule detected, while the hydrocarbon CH$_3$CCH is the largest molecule detected. Overall, only a few programs are sensitive enough to detect emission unambiguously originating from the disk based on the kinematics. Nitrogen-bearing molecules are predominantly detected on more extended scales, while hydrocarbons show a distinct tail roughly along the southeastern outflow cavity wall, probably due to a stronger UV field in the eastern outflow lobe. The L1527 IRS protostellar system is not rich in sulfur-bearing molecules, with only strong emission observed for CS and SO. Overall, the envelope appears dominated by a carbon-rich chemistry, which seems to transition into an oxygen-rich chemistry in the disk. We calculate column densities of all detected species, providing a starting point to quantify the chemical diversity among young disks and the chemical evolution of the planet-forming material. }

\tiny
 \keyFont{ \section{Keywords:} astrochemistry, chemical composition, planet formation, protostellar disks, submillimeter (sub-mm)} %All article types: you may provide up to 8 keywords; at least 5 are mandatory.
\end{abstract}

% ==================================================================
% INTRODUCTION
% ==================================================================

\section{Introduction}

The formation of a planetary system starts with the collapse of a dense molecular cloud core, where conservation of angular momentum dictates the formation of an accretion disk around the protostar. As the disk evolves, the initially (sub-)micrometer-sized dust grains grow into pebbles, boulders, and eventually planetary bodies. Rocky cores massive enough to gravitationally attract gas will become gas giants, while leftover smaller bodies may become comets and asteroids. Alternatively, giant planets may form through gravitational instabilities in the disk (see e.g., \citealt{Armitage2024,Ikoma2025,Youdin2025} for recent reviews). Understanding the outcome of planet formation and planetary diversity, therefore, requires detailed knowledge of the composition of the dust, gas, and ice in planet-forming disks and how that evolves on planet-formation timescales. 

The advent of the Atacama Large Millimeter/submillimeter Array (ALMA), has opened the door to detailed chemical studies of disks on scales larger than typically $\sim$10~au. In tandem with chemical modeling efforts, this has resulted in a global understanding of the chemical structure of mature Class II protoplanetary disks. For example, the most detailed study to date, although for only five disks, suggests the presence of large reservoirs of organics in the inner disk regions and elevated C/O ratios across most disks (MAPS Large Program; \citealt{Bosman2021,Guzman2021,Ilee2021,LeGal2021,Oberg2021}). Quantitatively, the abundance of the most abundant molecule after H$_2$, CO, is typically found to be depleted by 1--2 orders of magnitude \citep{Favre2013,McClure2016,Zhang2019,Zhang2021}. Complementary to ALMA, JWST is now allowing for an in-depth view of the inner few au, immediately exposing stark differences in composition, with some inner disks rich in organics and others dominated by water \citep[e.g.,][]{Perotti2023,Tabone2023,Arulanatham2025,Grant2025}.

While our knowledge of the composition and distribution of material in Class II disks has been improving, so has our understanding of the planet-formation process, and it has become evident that the first steps of grain growth most likely already take place when the disk itself is still forming \citep[e.g.,][]{Miotello2014,Harsono2018,Tychoniec2020,Cacciapuoti2023}. Young protostellar disks, or Class 0 and I disks, would thus represent the initial conditions for planet formation. Due to these young disks still being embedded within an infalling envelope and the presence of strong molecular outflows, the composition of protostellar disks is more challenging to unravel. A first systematic attempt to study young disks at high resolution was made by the eDisk Large Program \citep{Ohashi2023}. In terms of chemistry, hints of changes in the planet-forming material were observed as CH$_3$OH was only detected toward Class 0 sources \citep{Sharma2025}. Other evidence of chemical evolution has been inferred from smaller surveys, all indicating CO being present at canonical abundances in young disks, but severely depleted in mature disks \citep{Bergner2019,Zhang2019,vantHoff2020}. 

Despite the enormous amount of progress over the past 15 years, full chemical inventories of disks at diverse stages are still missing. We therefore present here an overview of all molecular species, and their column densities, detected with ALMA toward the Class 0 protostellar system L1527 IRS (also known as IRAS 04368+2557 and hereafter L1527). L1527 is the first protostar around which a rotationally supported disk was established \citep{Tobin2012}, and overall is one of the most studied embedded systems (see e.g., \citealt{vantHoff_eDisk} for an overview). Briefly, located in the Taurus star-forming region ($\sim$140 pc; \citealt{Kenyon1994,Zucker2019}), the $\sim$0.3--0.5 $M_{\odot}$ protostar \citep{Aso2017,vantHoff_eDisk} has a bolometric luminosity of 1.3 $L_{\odot}$ \citep{Ohashi2023}, and is surrounded by a $\sim$100 au disk \citep{vantHoff_eDisk} that is viewed almost completely edge-on \citep{Oya2015}. 

In terms of composition, single-dish observations have found a rich hydrocarbon chemistry \citep{Sakai2008,Yoshida2019}, and ALMA observations revealed different morphologies and kinematics for different molecular species. In particular, SO emission seems to be enhanced in a ring at the interface between the disk and envelope \citep{Ohashi2014,Sakai2014,Sakai2014_Nature}. Molecular line observations have also shown that, in contrast to protoplanetary disks, this young disk is warm, with the CO snowline located in the inner envelope \citep{vantHoff2018,vantHoff_eDisk}. The location of the water snowline is proposed to be around $\sim$2--4 au \citep{vantHoff2022}, and the weak detection of CH$_3$OH \citep{Sakai2014} may confirm a much smaller hot area compared to protostellar systems without a (large) disk, or indicate that most of the region inside the water snowline is hidden by optically thick dust. 

Overall, L1527 is an excellent system for probing the initial chemical conditions for planet formation, including how the composition may change from the envelope into the disk. The edge-on view of the disk is favorable for disentangling the disk and envelope, as the broadening of the emission lines is maximized and emission in the line wings originates only from the disk \citep{vantHoff2018}. In addition, the system displays some of the strongest molecular emission among young disk systems \citep{Sharma2025}. Moreover, the deeply embedded nature in combination with signs of grain growth (e.g., \citealt{Sheehan2022,Cacciapuoti2023}) suggests an early stage of planet formation. Finally, L1527 has been targeted by 40 ALMA programs, of which only a small fraction has been published. Therefore, we present a molecular inventory of L1527 on disk and inner envelope scales from all available ALMA observations. Details of the ALMA programs are described in Sect.~\ref{sec:Observations}. The results are presented in Sect.~\ref{sec:Results}, with an overview of detected molecules in Sect.~\ref{sec:Detections} and a discussion of the dominant molecular species in each component of the protostellar system in Sect.~\ref{sec:Structure}. Then, in Sect.~\ref{sec:Columndensities}, we present column densities and isotope ratios. A comparison with previous work on L1527, including recent JWST observations, is made in Sect.~\ref{sec:Discussion}, as well as comparisons with other young and mature disks. Finally, the main findings are summarized in Sect.~\ref{sec:Conclusion}.

% For Original Research articles, please note that the Material and Methods section can be placed in any of the following ways: before Results, before Discussion or after Discussion.

% ==================================================================
% OBSERVATIONS
% ==================================================================

\section{Observations} \label{sec:Observations}

\begin{figure}
    \centering
    \vspace{-0.3cm}
    \includegraphics[width=0.95\linewidth]{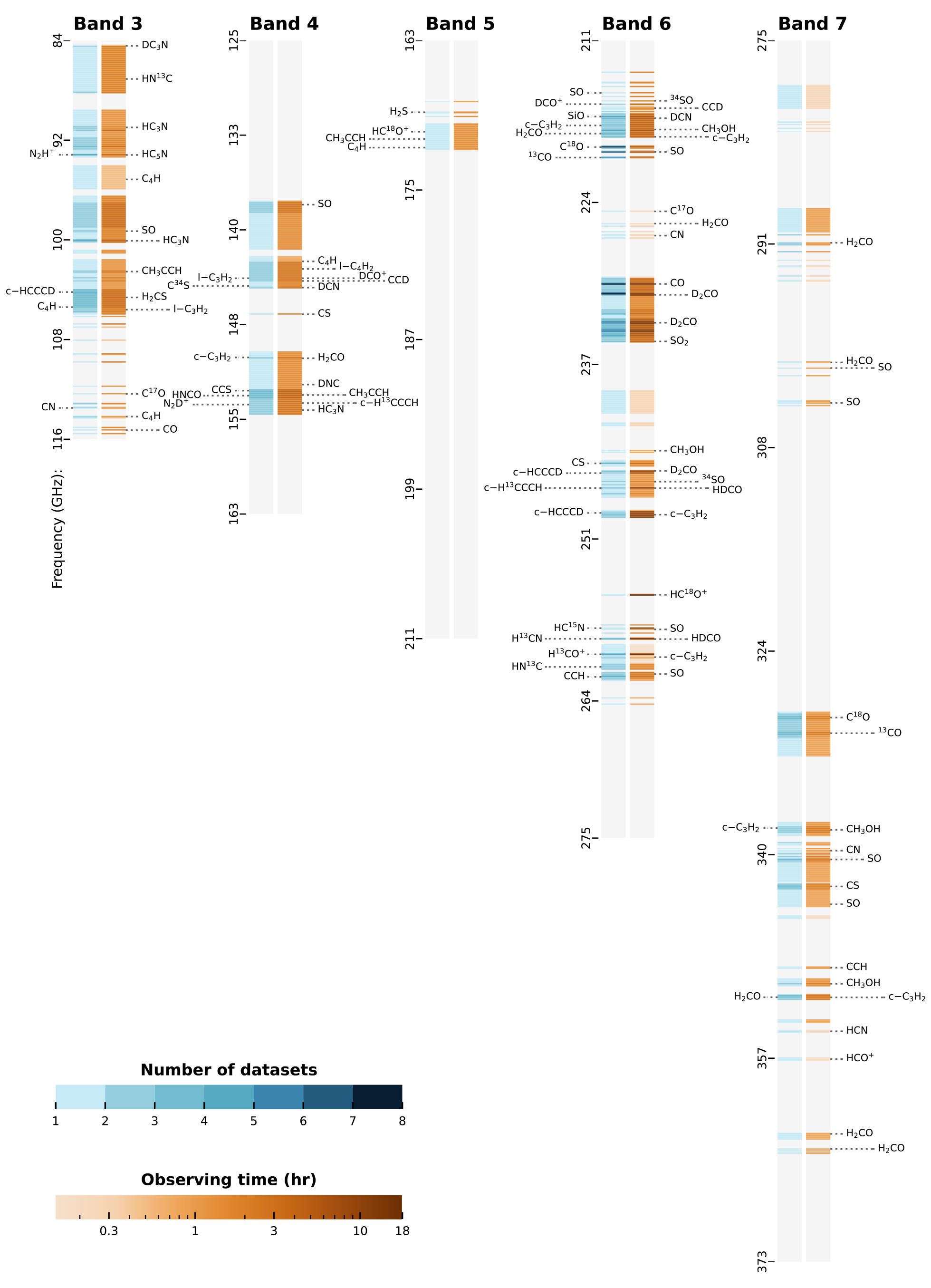}
    \vspace{-0.4cm}
    \caption{Overview of the ALMA observations toward L1527. The different shades of blue indicate the number of datasets that cover a given frequency, the shades of orange indicate the total observing time. Regions in gray have not been observed. Molecular lines that are detected are annotated.}
    \label{fig:Observations}
\end{figure}

L1527 has been observed with ALMA by 40 projects (unique project code). Of those, 36 programs have publicly available data as of December 2025, and one still proprietary program is led by us. The 33 programs among these 37 that contain spectral windows in FDM (Frequency Division Mode) are the focus of this work and cover Bands 3 to 7, while the three remaining programs still in their proprietary time cover Bands 8 and 9. The majority of programs (20) contain only observations with the Main Array (12m array), while five programs use only the Atacama Compact Array (ACA or 7m array). Both arrays are utilized in eight programs, of which one program additionally uses Total Power observations (in Band 4). All program details are listed in Table~\ref{tab:ALMAprograms}, and a visual representation of the covered frequencies is shown in Fig.~\ref{fig:Observations}. Band 6 has been targeted by the most programs (18), followed by Bands 7 and 3 (10 and 8, resp.). Bands 4 and 5 are the least observed (4 and 1, resp.). In terms of frequency coverage, 65\% has been observed of the Band with the smallest tuneable bandwidth, Band 3, while only 5\% of Band 5 has been observed. The coverage fractions of Band 4, 6 and 7 are 31\%, 33\% and 21\%, respectively. The shortest total integration time at a given frequency is 8--15 minutes for the different Bands, except for Band 5, where the single program has an integration time of 1.2 hours. The longest total integration times are ~4 hours in Bands 3, 4 and 7 and 18 hours in Band 6. An overview of the distribution of the angular resolution, velocity resolution, point-source sensitivity and surface-brightness sensitivity across spectral windows for each ALMA Band is presented in the Supplementary Material (Fig. S1).  

In this work, we primarily used the pipeline generated ``product'' images provided in the ALMA archive. When those are not available, we used images generated by the Additional Representative Images for Legacy (ARI-L) project \citep{Massardi2021} or requested images to be produced through the ALMA User-Defined Imaging (AUDI) service as part of The Science Ready Data Products (SRDP) Initiative from the National Radio Astronomy Observatory (NRAO) \citep{Lacy2020}. For AUDI imaging, we selected the default angular and spectral resolution of each spectral window, and used the image generated from the self-calibrated measurement set. For only a small number of datasets from the earliest cycles, images were not available through either of these three ways and we reduced the data manually, including phase and amplitude selfcalibration, using the appropriate CASA version.  The imaging method used for each dataset is listed in Table~\ref{tab:ALMAprograms}. For programs with observations taken in multiple configurations, only the individual datasets are presented, as the combined datasets are not archived.

% ...................................................................
% Figure - Overview Observations
% ...................................................................
% .....................................................................

% ...................................................................
\clearpage
\input{Tables/Dataset_overview}
% ...................................................................

To account for small differences in source position in different observations, we fit a 2D Gaussian to the continuum image of each dataset and use that central position as the source position. The location of the continuum peak varies over 0.43$''$ and 0.62$''$ in right ascension and declination, respectively, with the majority of observations within 0.2$''$ in both directions (except the ACA observations of program 2017.1.01375.S which has the continuum peak $\sim$1$''$ further north). This centering is thus essential to compare the high resolution datasets. In terms of flux calibration uncertainty between different datasets, \citet{Cacciapuoti2023} combined continuum observations toward L1527 using 27 ALMA datasets, and found rescaling factors ranging between 1-15\%. As discussed in Sect.~\ref{sec:Columndensities}, column density measurements of the same transition in different datasets range within a factor of $\sim$5 after correcting for emitting area, so the uncertainty on the column density is not dominated by the calibration uncertainty. The main caveats of using product images is that very weak lines may have gone undetected at the default weighting scheme employed by the automated pipeline for imaging, in spectral windows observed at a high spectral resolution, or without combining datasets in programs that utilize multiple antenna configurations.

% ...................................................................
%\clearpage
\input{Tables/Observations}
% ...................................................................

% ==================================================================
% RESULTS
% ==================================================================

\section{Results} \label{sec:Results}

% -----------------------------------------------------------------
% Overview detected molecules
% -----------------------------------------------------------------

\subsection{Detected molecular species} \label{sec:Detections}

Every spectral window was inspected for the presence of molecular lines through the creation of peak intensity maps (moment 8) and spectra integrated over apertures with different sizes and at different locations. Multi-configuration datasets were not combined and data were not re-imaged at lower spectral and/or angular resolution, so some weak features may have gone undetected. We count a species as detected when it has emission for at least one transition above the $>3\sigma$ level in at least more than one pixel in at least three velocity channels or spatially integrated emission above the $>3\sigma$ level, but in practice all detections are well above this level.  Overall, 23 different molecular species are detected, and 40 species when isotopologues are included. Of these, 29 are reported here for the first time toward L1527 based on ALMA observations (see Table~\ref{tab:Observations}), and two (H$_2$S and $^{34}$SO) were also not reported in the Nobeyama 45m survey \citep{Yoshida2019}. With seven atoms, CH$_3$CCH is the largest molecule detected. Carbon and oxygen-bearing molecules include isotopologues of CO, HCO$^+$, H$_2$CO, H$_2$CCO, and CH$_3$OH. Hydrocarbons include isotopologues of CCH, c-C$_3$H, c-C$_3$H$_2$, l-C$_3$H$_2$, C$_4$H, l-C$_4$H$_2$, and CH$_3$CCH. Nitrogen-bearing molecules include isotopologues of N$_2$H$^+$, CN, HCN, HNC, HC$_3$N, and HC$_5$N. HNCO is the only detected molecule containing carbon, oxygen and nitrogen. Sulfur-bearing molecules include isotopologues of CS, SO, SO$_2$, and CCS, while OCS and H$_2$CS are not detected. A tentative detection of OCS was reported in the combined FAUST datasets \citep{Zhang2024}. Other notable non-detections are water isotopologues, although only one strong HDO transition (at 225.896720 GHz) has been covered. SiO is not detected in individual datasets, but a detection was reported in the combined datasets of the eDisk Large Program \citep{vantHoff_eDisk}.

Of the less abundant isotopes, deuterated species are the most common and include N$_2$D$^+$, DCO$^+$, DCN, DNC, DC$_3$N, CCD, c-HCCCD, HDCO, and D$_2$CO. The $^{13}$C isotope has been detected in $^{13}$CO, H$^{13}$CO$^+$, H$^{13}$CN, HN$^{13}$C, and c-H$^{13}$CCCH. Oxygen isotopes have only been detected in C$^{18}$O, C$^{17}$O, and HC$^{18}$O$^+$. Sulfur and nitrogen isotopologues are even more rare, with only C$^{34}$S and $^{34}$SO, and HC$^{15}$N detected, respectively. In total, 159 unique transitions are detected, counting lines that are completely blended only once. Due to their hyperfine structure, CN, CCD, and CCH have the most transitions detected (19, 17 and 12, resp.), followed by H$_2$CO (10) and SO (nine). The detected transition with the lowest upper energy level was HN$^{13}$C 1--0 (4.2 K), which is only detected in absorption. The lowest energy transition detected in emission is N$_2$H$^+$ 1--0 (4.5 K). The highest upper energy level detected was for SO$_2$ (148 K), followed by H$_2$CO (99 K). Although the upper energy levels of detected transitions thus range between 4 and 148 K, the average and median values are 34 and 30 K, respectively. All detected transitions and nondetections used to derive upper limits are listed in the Supplementary Materials, and Table~\ref{tab:Observations} gives a global overview of the angular resolution at which each molecule is observed and detected. A brief overview of the transitions covered and detected per molecule is presented in the Supplementary Materials.

% ...................................................................
% Figure - Overview Protostellar Components
% ...................................................................
\begin{figure}
    \centering
    \vspace{-0.5cm}
    \includegraphics[width=0.90\linewidth]{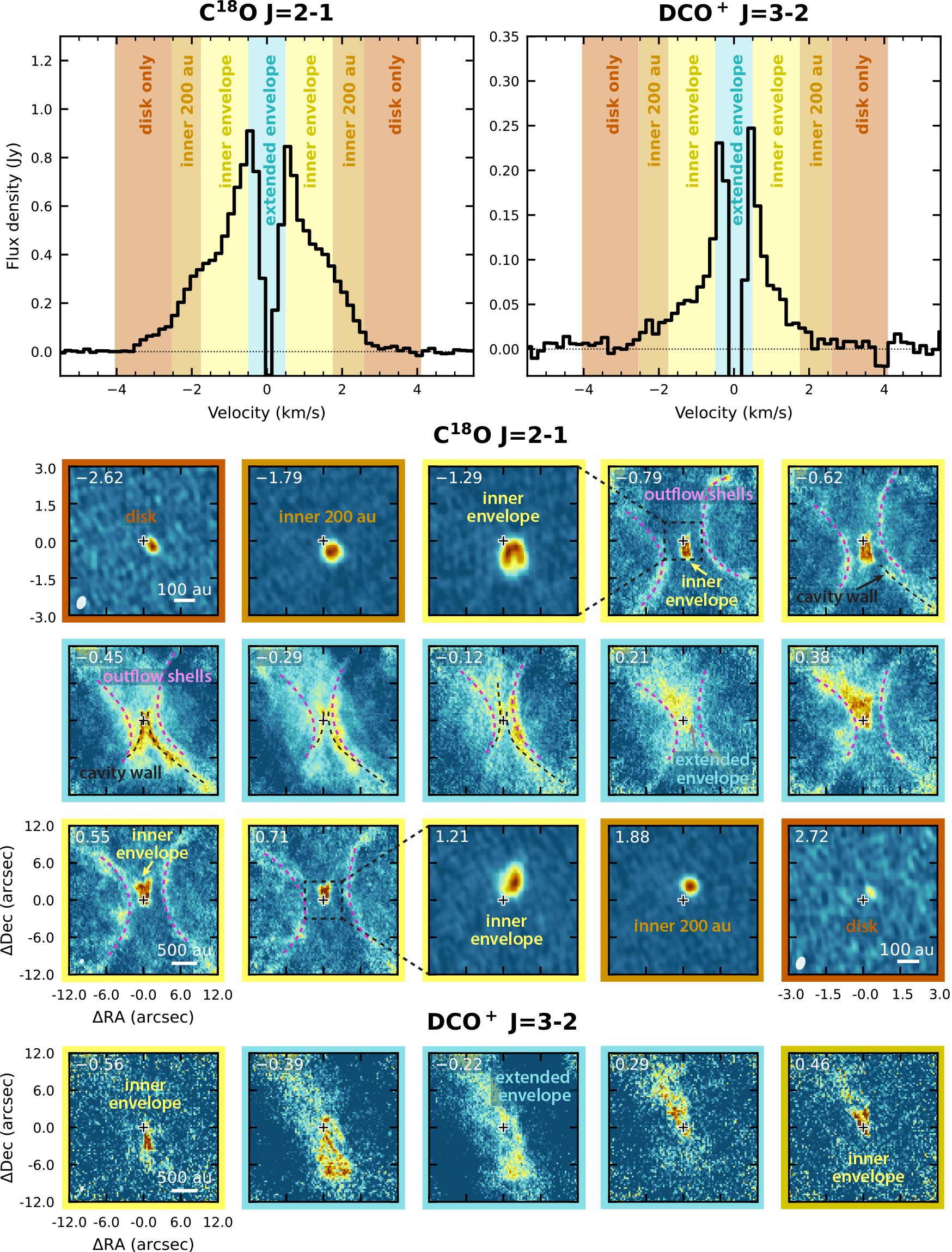}
    \vspace{-0.2cm}
    \caption{Different components of the protostellar system can be identified based on their velocity and morphology. Top panels: spectra of the C$^{18}$O $J=2-1$ and DCO$^+$ $J=3-2$ transitions extracted in circular apertures with a 1$''$ radius, with different shaded regions marking different components based on a 100 au Keplerian rotating disk within a rotating infalling envelope around a 0.45 $M_{\odot}$ star \citep{vantHoff2018}. Bottom panels: selected velocity channels (0.17 km s$^{-1}$ resolution) of the same transitions with different components annotated in the same color scheme as the top panels. The dashed contours are drawn by hand to guide the eye. The first three and last three C$^{18}$O panels have a different intensity stretch and spatial scale compared to the remaining panels. All velocities are with respect to the systemic velocity of 5.9 km s$^{-1}$. Data from 2018.1.01205.L.} 
    \label{fig:ProtostellarComponents}
\end{figure}
% .....................................................................

% -----------------------------------------------------------------
% Chemical structure 
% -----------------------------------------------------------------

\subsection{Chemical structure} \label{sec:Structure}

To determine the spatial origin of the different molecular emissions, we look for characteristic signatures of the protostellar components in the individual velocity channels. Using both the spatial extent and the velocity range of the detected emission, we categorize species as predominantly tracing the following components: the outflow, the cavity wall, the extended envelope, the extended and inner envelope, the inner envelope, the disk, and a southeast ``tail''. Figure~\ref{fig:ProtostellarComponents} present channelmaps and spatially integrated spectra for C$^{18}$O and DCO$^+$, with the different features that are used to characterize emission morphology and origin annotated. We define the extended envelope component as low-velocity emission ($|\Delta v| \lesssim$ 0.5~km~s$^{-1}$) as there is a clear change in morphology around $|\Delta v| \sim$0.5~km~s$^{-1}$, with the emission extending more than 7$''$ ($\sim$1000 au) to the north and south at the lowest velocities. The southeast tail is present at similar velocities, but extends $\sim40''$ to the southeast. Based on the 3D radiative transfer modeling of $^{13}$CO and C$^{18}$O emission done by \citet{vantHoff2018} using a Keplerian disk surrounded by a rotating-infalling envelope \citep{Ulrich1976,Cassen1981}, for a central stellar mass of 0.45 $M_{\odot}$ \citep{Aso2017,vantHoff_eDisk} velocity offsets $\leq$2.54 km s$^{-1}$ and $\geq$2.59 km s$^{-1}$ with respect to the systemic velocity (referred to as $|\Delta v| \gtrsim$ 2.5~km~s$^{-1}$ in the text for simplicity) will be free of envelope emission and contain only emission originating in the disk. Emission at intermediate velocities ($|\Delta v| \sim$0.5--2.5~km~s$^{-1}$) will be referred to as the inner envelope, and typically extends $\lesssim4''$ ($\sim$550 au) north and south of the source position. Some of this emission may originate in the outer disk, but more detailed modeling is required to make this distinction. Finally, we refer to narrow strips of emission on the envelope side of the $^{12}$CO outflow emission as the cavity wall, but a detailed kinematic analysis would be required to determine whether this material is infalling or outflowing.

\begin{figure}
    \centering
    \includegraphics[width=\linewidth]{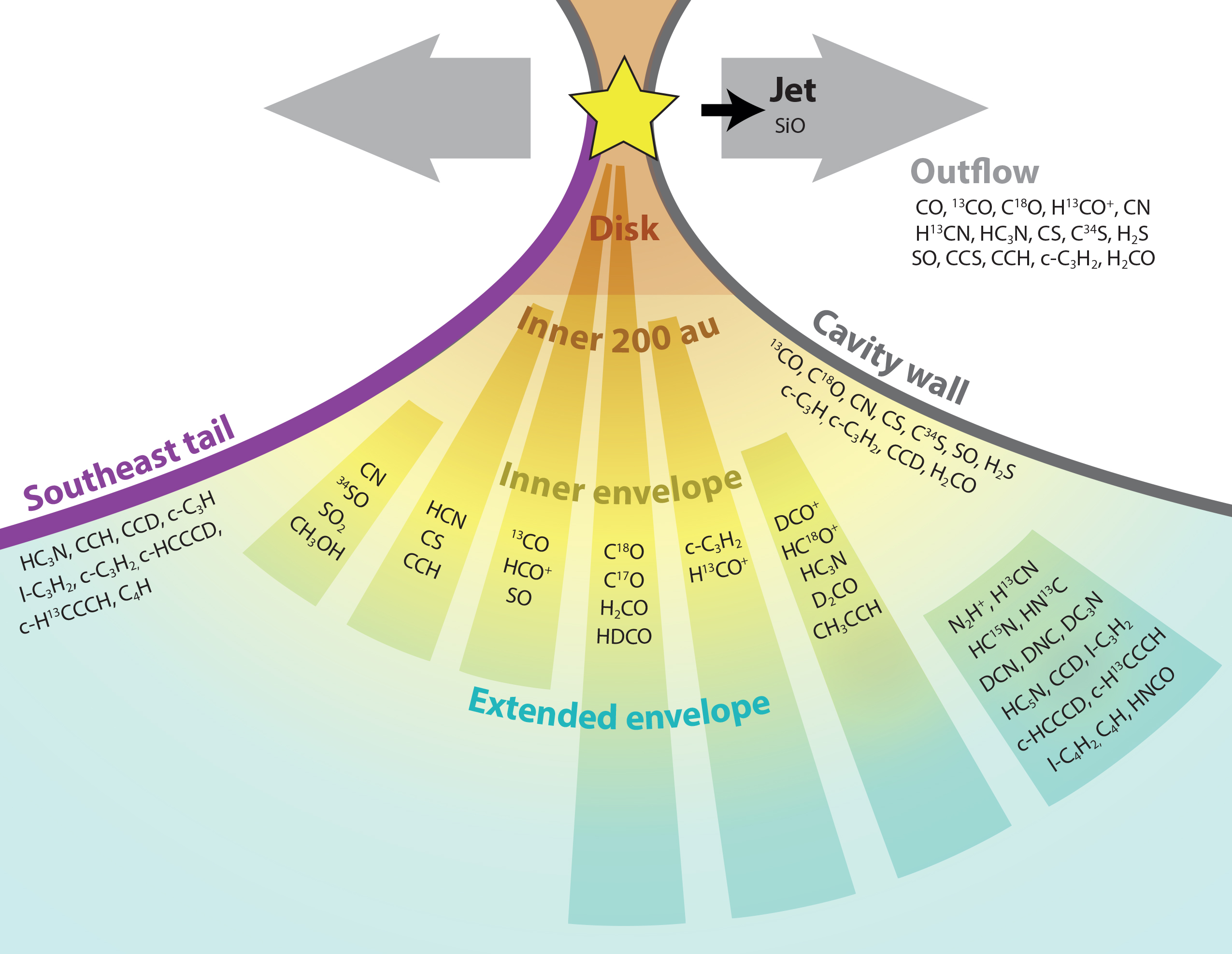}
    \caption{Schematic overview (not to scale) of the chemical structure of the inner $\sim$3500 au of the L1527 protostellar system based on ALMA observations.}
    \label{fig:Cartoon}
\end{figure}

Each component and its characteristics are described in more detail in the following sections (Sec.~\ref{sec:Outflow}--\ref{sec:Cavitywall}) and representative images (peak intensity maps) of the different molecular species are presented in Figs.~\ref{fig:Outflow}--\ref{fig:SmallScale}. Spatially integrated spectra of the weakest detections are shown in Fig.~\ref{fig:Spectra}, and a visual overview is shown in Fig.~\ref{fig:Cartoon}. We note that whether or not a molecule shows emission from a certain component can depend on observational parameters such as angular resolution, maximum recoverable scale, surface brightness sensitivity and point source sensitivity. This becomes evident as several transitions have been observed by multiple programs and some protostellar components are not visible in all observations. In addition, excitation effects may contribute to whether or not components are visible in the observed transitions. Disentangling excitation effects for different transitions of a molecule observed in different programs with different resolution and sensitivity is not trivial. We therefore focus here on the overall distribution of molecular species. Based on $^{13}$CO, C$^{18}$O and H$_2$CO observations, the temperature in the disk midplane ranges from $\sim$50 K at $\sim$10 au down to $\sim$20 K (the CO snowline) at $\sim$350 au (in the envelope), while the surface layers reach temperatures of $\sim$70-80 K \citep{vantHoff2018,vantHoff2020,vantHoff_eDisk}. The excitation conditions in the disk and inner envelope are thus ideal for transitions with upper-level energies of several tens of Kelvin.

% --- Outflow -----------------------------------------------------

\subsubsection{Outflow}\label{sec:Outflow}

\begin{figure}
    \centering
    \includegraphics[width=\linewidth]{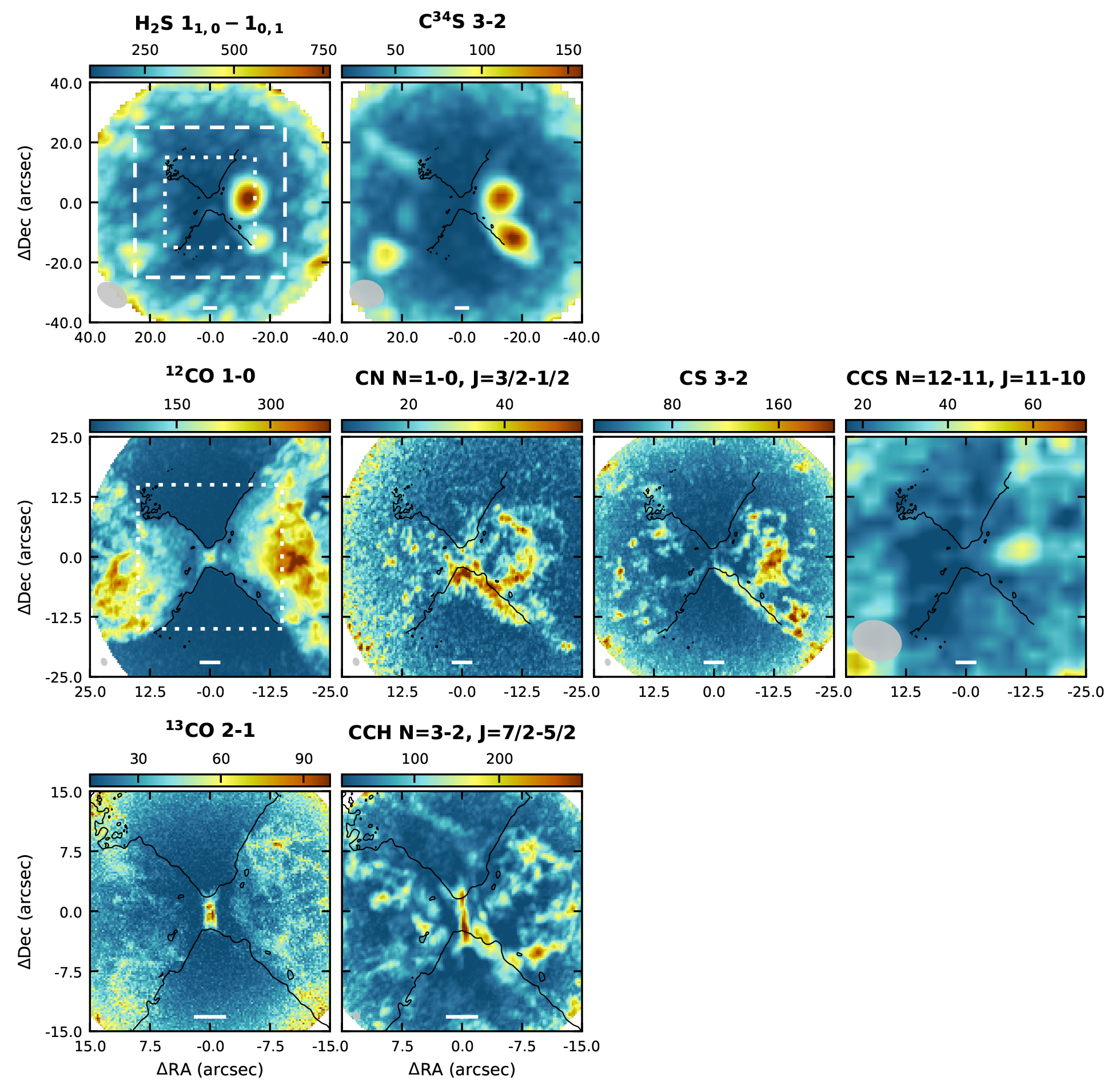}
    \caption{Peak intensity maps of molecules whose emission is dominated by the outflow. Images on different rows display a different angular scale, with the scale of the row(s) below indicated by a white box in the left most panel. The black contours mark the outflow cavity as traced by the peak intensity map of $^{12}$CO $J=1-0$ and correspond to 30 mJy beam$^{-1}$ (6 times the rms of the image cube). In each panel, the gray ellipse in the bottom left corner represents the synthesized beam and the white scalebar marks 500 au. The following datasets are used: 2011.0.00604.S (CCH), 2016.2.00171.S (C$^{34}$S, CCS), 2016.1.01203.S (CS), 2018.1.00375.S (H$_2$S), 2022.1.00131.S ($^{12}$CO, CN), and 2019.A.00034.S ($^{13}$CO).}
    \label{fig:Outflow}
\end{figure}

If present, outflows are often the most prominent and spectacular component of young protostellar systems and L1527 is no exception (Fig.~\ref{fig:Outflow}). The outflow is most clear in $^{12}$CO, the second most abundant molecule after H$_2$, with emission filling up and outlining the outflow cavity. With $^{13}$CO being less abundant, outflow emission is still present but it is not as strong as $^{12}$CO. CN, CS and CCH are particularly bright in the western outflow, and trace more detailed structures compared to the more uniform distribution of $^{12}$CO. Although not exactly the same, CN, CS and CCH roughly trace a similar structure. Their main difference is in the extent to which they trace the cavity wall: CN traces deeper into the envelope than CS and CCH, and CCH is brighter just inside the cavity outlined by $^{12}$CO. 

H$_2$S, C$^{34}$S and CCS are only detected at low spatial resolution ($\gtrsim 3''$). H$_2$S and C$^{34}$S show two bright unresolved peaks in the western outflow that cover the more detailed structure seen in CS and CN. While for H$_2$S the northwestern peak is clearly brighter than the southwestern, for C$^{34}$S the southwestern peak is slightly brighter. For CCS only the northwestern peak is visible. All three molecules also show an unresolved peak in the southeastern outflow as well as emission along the northeastern cavity wall (most clear for C$^{34}$S). Both regions fall outside the primary beam for the other molecules shown in Fig.~\ref{fig:Outflow}. A weak contribution of the outflow is also visible for C$^{18}$O, c-C$_3$H$_2$, H$_2$CO, H$^{13}$CO$^+$, H$^{13}$CN, HC$_3$N, and SO. A very compact jet ($<$ 28 au in diameter) was detected in SiO only 11 au west of the source in the combined datasets from the eDisk Large Program \citep{vantHoff_eDisk}. 

% ...................................................................
% Figure - Outflow
% ...................................................................
% ...................................................................

Overall, the molecules detected in the outflow are, as expected, species associated with the bulk of the gas (CO), dense gas (CS), and those formed in shocks (e.g., SO) and irradiated environments (e.g., CN). On the other hand, shock-sputtering products such as CH$_3$OH and other COMs are not detected. \citet{Feeney-Johansson2025} categorizes the outflow as observed in $^{12}$CO with the eDisk Large Program as a slow disk wind based on the conical structure in the velocity channel maps, with the opening angle decreasing at larger velocities. As shown by \citet{vantHoff_eDisk}, the outflow emission from $^{13}$CO and C$^{18}$O is dominated by shells that move outward with increasing velocity offsets. This appearance can be expected when a wide-angle wind blows into ambient material \citep[e.g.,][]{Shu1991,Li1996,Lee2000,Lee2001}. Such potentially wind-driven shells are also visible in CN, CS, H$^{13}$CO$^+$ and H$_2$CO. We may thus be seeing the disk wind in the most abundant gas tracer $^{12}$CO, while less abundant species are showing the material swept up (and/or shocked) by the wind. A more detailed analysis is required to confirm this, especially since a slow disk wind and wind-driven shell are hard to distinguish for an outflow in the plane of the sky \citep{Feeney-Johansson2025}.

% --- Extended envelope ------------------------------------------------ 

\subsubsection{Extended envelope}\label{sec:ExtendedEnvelope}

The cold-gas mass and the emitting area of the circumstellar material is dominated by the extended envelope. We would therefore expect all molecules to trace this material, unless 1) chemistry prevents a molecule from being abundant in cold ($\lesssim$20 K) gas, 2) the emission becomes optically thick on scales larger than the interferometric observations are sensitive to and is therefore resolved out, or 3) only high-energy transitions that are weak in ~20 K gas are observed.

In the most compact configuration of the 12m array, the MRS ranges between $\sim$8$''$ (Band 7) to $\sim$29$''$ (Band 3), and for the 7m array this increases to $\sim$19$''$ (Band 7) to $\sim$67$''$ (Band 3) \citep{ALMAtechnicalHandbook}. Thus, care must be taken when comparing the extent of emission between different observations. Nevertheless, the extended envelope component would dominate the emission at velocities close to the systemic velocity. Based upon visual inspection of the L1527 data, we loosely define extended envelope emission as emission detected at $|\Delta v| \lesssim$ 0.5 km s$^{-1}$ that is extending more than 6$''$ from the source position in the north-south direction (see the DCO$^+$ velocity channelmaps in Fig.~\ref{fig:ProtostellarComponents} as an example). Molecules that have only been detected at velocities close to the systemic velocity are predominantly nitrogen-bearing molecules (N$_2$H$^+$, H$^{13}$CN, HC$^{15}$N, DCN, HN$^{13}$C, HC$_3$N, HC$_5$N, HNCO) and hydrocarbons (c-H$^{13}$CCCH and CH$_3$CCH), as well as less-abundant isotopologues (HC$^{18}$O$^+$ and D$_2$CO). It is not trivial to distinguish between a linewidth that is truly narrow and one that appears narrow due to the signal-to-noise ratio, especially at low angular and/or spectral resolution. The presence of these molecules in the inner envelope and/or disk can therefore not be ruled out without a more detailed analysis and/or new observations. 

\begin{figure}
    \centering
    \includegraphics[width=\linewidth]{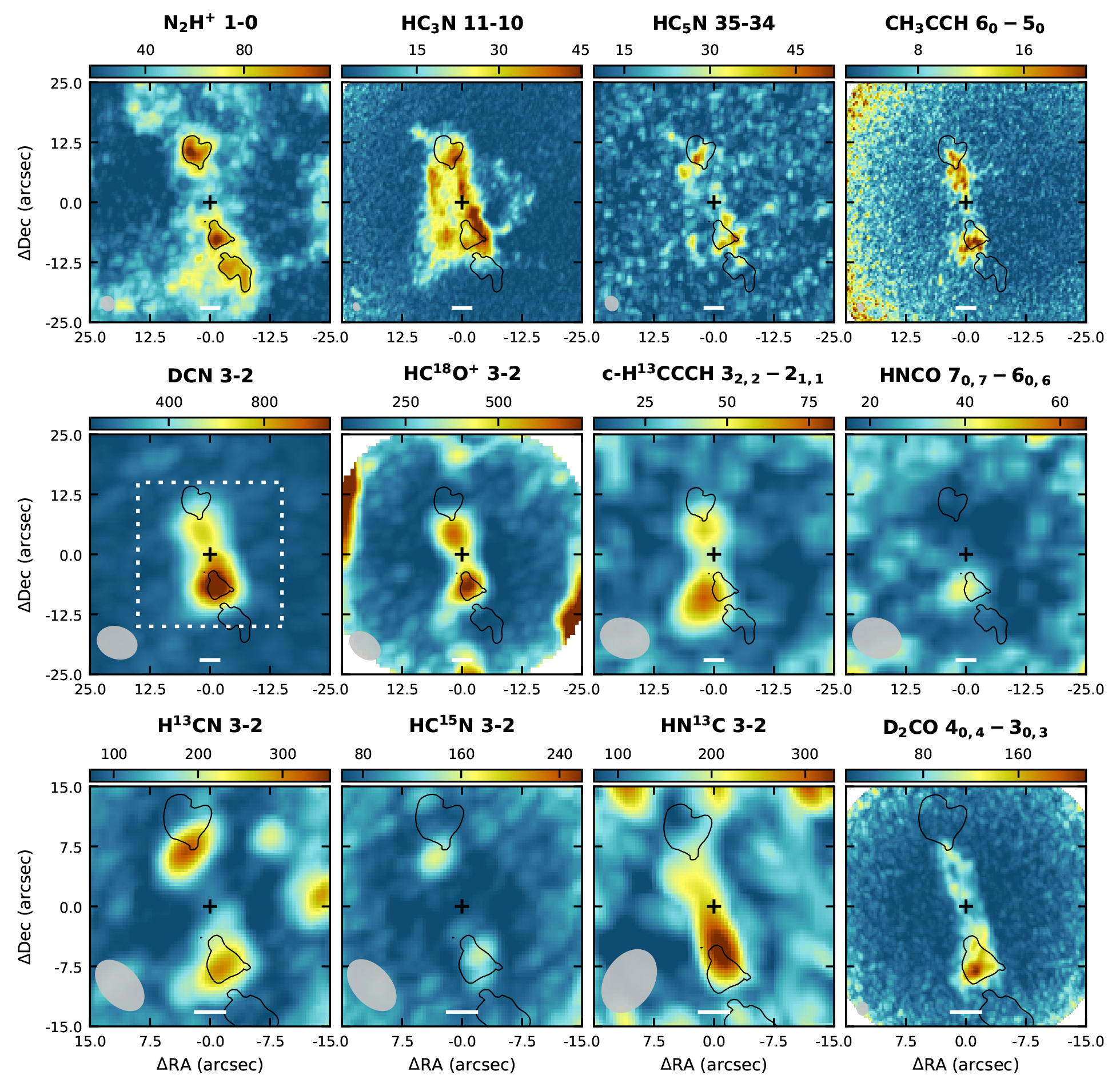}
    \caption{Peak intensity maps of molecules whose emission is dominated by the extended envelope. Images in the bottom row display a smaller angular scale, which is indicated by a white dotted box in the left most panel in the middle row. The black contours mark the bright N$_2$H$^+$ peaks as a reference for the spatial extent of the different molecules, and corresponds to 80 mJy beam$^{-1}$ (12 times the rms of the image cube). In each panel, the gray ellipse in the bottom left corner represents the synthesized beam and the white scalebar marks 500 au. The following datasets are used: 2013.1.01331.S (DCN), 2016.2.00171.S (c-H$^{13}$CCCH, HNCO), 2017.1.01375.S (HN$^{13}$C), 2018.1.00799.S (N$_2$H$^+$, HC$_5$N), 2018.1.01205.L (D$_2$CO), 2021.1.00536.S (HC$^{18}$O$^+$, H$^{13}$CN, HC$^{15}$N), and 2022.1.00131.S (HC$_3$N, CH$_3$CCH).}
    \label{fig:ExtendedEnvelope}
\end{figure}

At the observed angular resolutions and sensitivities, there appear two main morphologies that can be associated with the extended envelope emission (Fig.~\ref{fig:ExtendedEnvelope}): a two-lane structure in the north--south direction and two peaks offset from the source position along the major axis towards the north and south. In both cases, the emission extends to roughly 12$''$ ($\sim$1700 au) off source, except for N$_2$H$^+$, which is more extended. The two-lane structure is visible for HC$_3$N and HC$_5$N. The western lane is brighter than the eastern one, and at a smaller angular offset from the central source. The two-peak structure is visible for the other molecules, and in the vertical direction, the peaks roughly fall in between the HC$_3$N lanes. However, there are some differences among the molecules. N$_2$H$^+$ shows more extended emission in the south along with a third peak $\sim$15$''$ off source. The southern peak seen in HC$^{18}$O$^+$, H$^{13}$CN, HC$^{15}$N, DCN, HN$^{13}$C, D$_2$CO and CH$_3$CCH coincides with the first southern peak of N$_2$H$^+$ ($\sim$8$''$ off source). The northern peak of these molecules is closer to the central source ($\sim$7$''$) than the northern N$_2$H$^+$ peak ($\sim$11$''$). For c-H$^{13}$CCCH, the southern emission peaks more to the east.  For HC$^{18}$O$^+$, DCN, c-H$^{13}$CCCH and D$_2$CO, the southern peak is clearly stronger than the northern peak and for HNCO only the southern peak is visible. 

Other molecules primarily associated with the extended envelope are DNC, DC$_3$N, H$_2$CCO, and l-C$_4$H$_2$. DNC is only detected in absorption against the continuum, suggesting it is present in cold gas at systemic velocities. DC$_3$N, H$_2$CCO, and l-C$_4$H$_2$ are only detected in spatially integrated spectra (Fig.~\ref{fig:Spectra}), hinting at a more extended distribution. N$_2$D$^+$ has only been detected in Total Power observations and peaks at least 60$''$ north of the source, consistent with IRAM 30m observations \citep{Tobin_N2H+}.  While the IRAM 30m observations show a clear offset between N$_2$D$^+$ and N$_2$H$^+$, with N$_2$H$^+$ peaking only 35$''$ north of the source, the 2$''$-resolution ALMA mosaic displays a complex N$_2$H$^+$ emission morphology on large scales, so the details of the relative distributions remain unclear without higher resolution N$_2$D$^+$ images. 

% ...................................................................
% Figure - Extended envelope
% ...................................................................
% ...................................................................

% ...................................................................
% Figure - SE tail
% ...................................................................
\begin{figure}
    \centering
    \includegraphics[width=\linewidth]{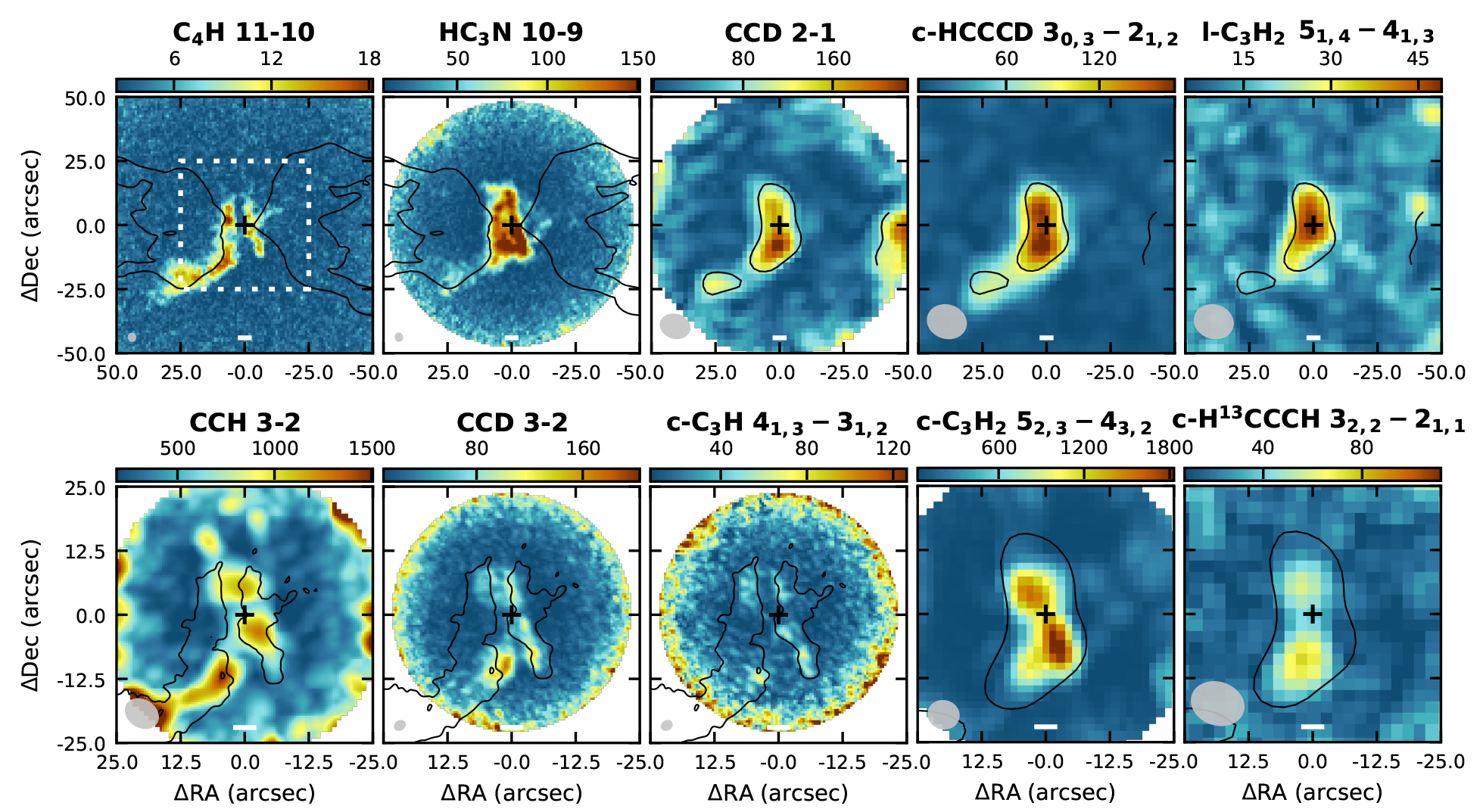}
    \caption{Peak intensity maps of molecules that have a strong contribution from a SE tail. Images in the bottom row display a smaller angular scale, which is indicated by a white dotted box in the left most panel in the top row. The black contours in the C$_4$H and HC$_3$N panels (first and second panel on the top row) outline the outflow cavity as seen in $^{12}$CO $J=2-1$ (from 2019.1.01063.S), and correspond to 10 Jy beam$^{-1}$ (90 times the rms of the image cube). The black contours in the CCH, CCD $N=3-2$ and c-C$_3$H panels (first, second and third panels on the bottom row) outline the C$_4$H emission shown in the top left panel (7 mJy beam$^{-1}$, i.e., 5 times the cube rms), and the black contours in the other panels outline the CCD $N=2-1$ emission shown in the second panel in the top row (80 mJy beam$^{-1}$, i.e., 10 times the cube rms). In each panel, the gray ellipse in the bottom left corner represents the synthesized beam and the white scalebar marks 500 au. The following datasets are used: 2013.1.01331.S (CCD $N=3-2$, c-C$_3$H), 2015.1.00261.S (HC$_3$N), 2016.2.00117.S (c-C$_3$H$_2$), 2016.2.00171.S (CCD $N=2-1$, c-H$^{13}$CCCH), 2017.1.01375.S (CCH), and 2018.1.00799.S (C$_4$H, c-HCCCD, l-C$_3$H$_2$).}
    \label{fig:SEtail}
\end{figure}
% ...................................................................

% --- Extended envelope and tail --------------------------------------- 

\subsubsection{Southeast tail}

In addition to the molecules described in the previous section, the emission of several other hydrocarbons is also restricted to low velocities ($|\Delta v| \lesssim$ 0.5 km s$^{-1}$), that is, CCD, c-C$_3$H, c-HCCCD, C$_4$H and l-C$_3$H$_2$. However, the most striking feature in their morphology is a tail of emission in the southeastern direction extending up to $\sim$40$''$ ($\sim$5600 au) away from the source position. 

Only C$_4$H emission has been detected at high enough angular resolution ($\sim$2$''$) and over a large enough FOV ($\gtrsim$50$''$) to display this morphology in detail. It shows a similar two-lane structure as seen in HC$_3$N, but with the emission extending much further in the SE direction. This tail is weakly visible for HC$_3$N. Comparing with lower resolution ($\sim$6$''$) CO observations that trace out to these large offsets shows that while the beginning of the tail (like the rest of the C$_4$H emission) lines the envelope side of the outflow, the end of the tail appears to fall inside the outflow cavity, but this could be a projection effect. 

ACA observations of CCD, c-HCCCD and l-C$_3$H$_2$ display a similar tail, although the l-C$_3$H$_2$ emission is not as strongly detected and the tail does not extend as far. At this resolution ($\gtrsim$10$''$), the two-lane structure is not visible, but the emission peaks closer to the source and more to the southeast than for the molecules discussed in the previous section. The c-H$^{13}$CCCH is weak, but the southern peak is toward the southeast as well, so the tail may not be visible at the low signal-to-noise ratio of the observations. CCD $N=3-2$ emission has also been detected at high angular resolution ($\sim$1.5$''$), but with a FOV too small ($\sim$20$''$) to see the full tail. Its morphology appears to be a blend between the two-lane (in the south) and two-peak structures (in the north), which is also the case for c-C$_3$H. However, the southeastern component of CCD ($\sim$12$''$ from source) is anti-correlated with C$_4$H and located closer to the midplane. Higher resolution observations with a large FOV are thus required to determine the exact distribution of hydrocarbons in this extended component.

While the main isotopologues CCH and c-C$_3$H$_2$ display higher velocity ($|\Delta v| \gtrsim 0.5$ km s$^{-1}$) emission, their peak intensity maps are dominated by the low velocity emission. CCH clearly traces the beginning of the tail, but the FOV is not large enough to assess the full extent. The tail appears more strongly for the weaker hyperfine components with lower Einstein A coefficient, probably because the stronger transitions suffer from resolved out emission at the lowest velocities. For c-C$_3$H$_2$, which also has resolved out emission, the tail is even more compact, but the emission displays a clear southeastern component at low resolution in addition to the southern peak.

% --- Extended and inner envelope ---------------------------------------

\subsubsection{Extended and inner envelope} \label{sec:Envelope}

\begin{figure}
    \centering
    \includegraphics[width=\linewidth]{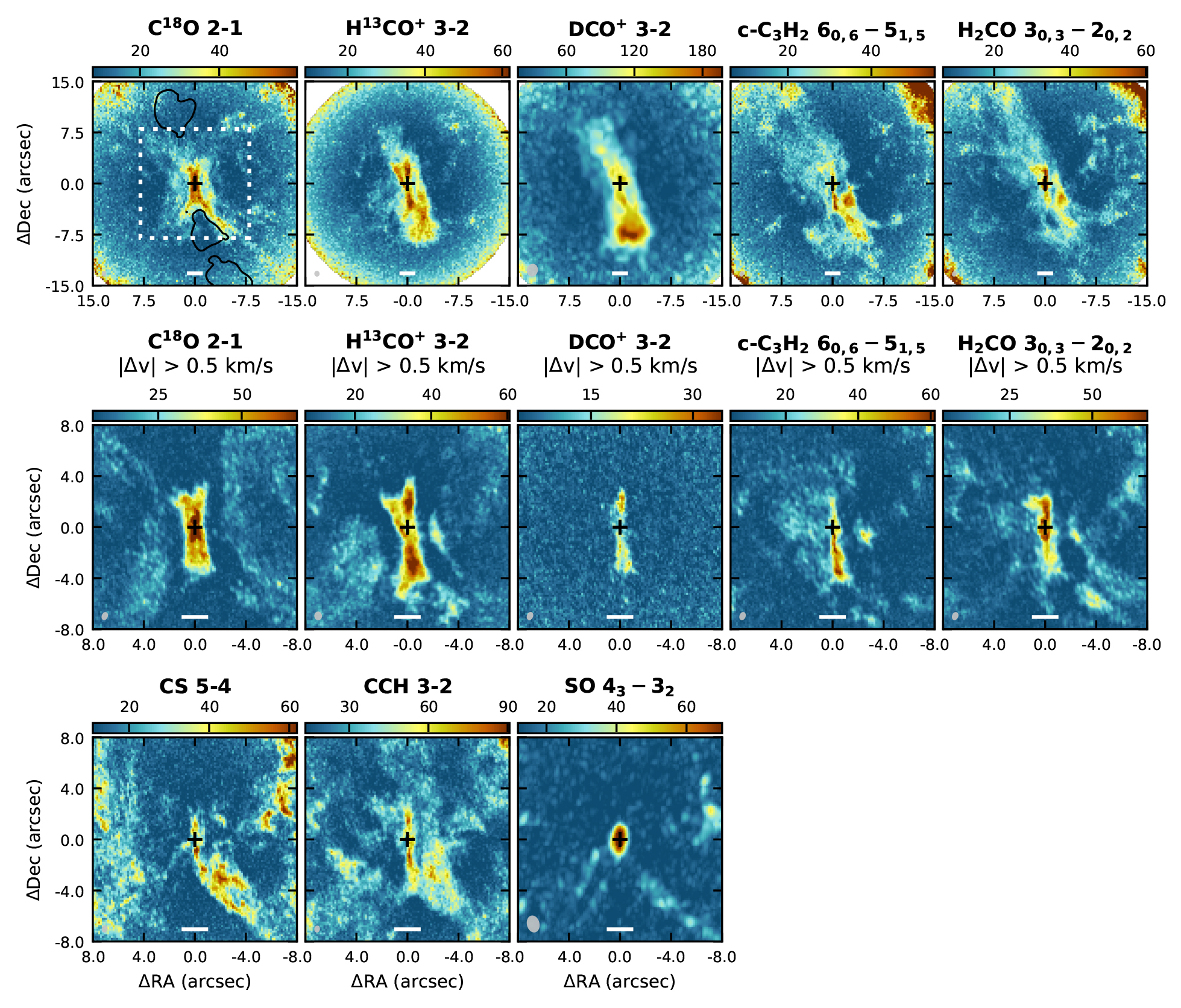}
    \caption{Peak intensity maps of molecules that have a strong contribution from the inner envelope. Images in the middle and bottom row display a smaller angular scale, which is indicated by a white dotted box in the left most panel in the top row. The middle row shows the same transitions as the top row, except only velocity offsets $>$0.5 km/s from the systemic velocity are used to create the peak intensity maps. The black contours in the C$^{18}$O panel mark the bright N$_2$H$^+$ peaks (as shown in Fig.~\ref{fig:ExtendedEnvelope}, corresponding to 80 mJy beam$^{-1}$). In each panel, the gray ellipse in the bottom left corner represents the synthesized beam and the white scalebar marks 250 au. The following datasets are used: 2016.1.01203.S (SO), 2018.1.01205.L (C$^{18}$O, DCO$^+$, c-C$_3$H$_2$, H$_2$CO, CS, CCH), and 2021.1.00536.S (H$^{13}$CO$^+$).}
    \label{fig:Envelope}
\end{figure}

Some of the more abundant isotopologues of the molecules detected only at low velocities ($|\Delta v| \lesssim 0.5$ km~s$^{-1}$; described in Sect.~\ref{sec:ExtendedEnvelope}) typically display the most complex morphologies due to emission arising over a larger velocity range ($|\Delta v| \lesssim$ 2.5 km s$^{-1}$) and originating from several of the protostellar components. In the existing observations, these molecules are C$^{18}$O, H$^{13}$CO$^+$, DCO$^+$, c-C$_3$H$_2$, H$_2$CO and HDCO. They are abundant enough to display resolved out emission around the systemic velocity, but their peak intensity maps are still dominated by extended envelope emission ($\sim$12$''$) at $v_{\rm{sys}}~\pm \sim$0.5 km s$^{-1}$. At intermediate to high angular resolution, a more detailed structure due to contributions from the envelope, outflow shells and cavity wall becomes apparent (Fig.~\ref{fig:Envelope}). With some small differences, H$^{13}$CO$^+$, DCO$^+$, c-C$_3$H$_2$, H$_2$CO, and HDCO display a similar north--south morphology extending roughly 12$''$ on the sky from the source position. The DCO$^+$, c-C$_3$H$_2$, and H$_2$CO emission is clearly stronger in the south, while H$^{13}$CO$^+$ and HDCO display a more even distribution. C$^{18}$O shows a distinct X-shape profile due to freeze out and emission arising only from the surface layers of the envelope. ACA images of these molecules (not shown) are more centrally peaked compared to the two-peak morphologies displayed by the extended envelope tracers discussed in Sect.~\ref{sec:ExtendedEnvelope}. This suggests that the less abundant isotopologues trace the largest cold gas column, which is the line of sight at larger angular offsets from the protostar that does not go through the warmer inner envelope and disk, while the more abundant isotopologues have a strong contribution from warmer gas closer to the source. 

% ...................................................................
% Figure - Envelope
% ...................................................................
% ...................................................................

To remove the contribution from the extended envelope and low-velocity outflow shells close to the source, we make peak intensity maps excluding the central 1 km s$^{-1}$ (except for HDCO, as the spectral resolution is too low). The resulting maps are then dominated by emission from the inner envelope ($\lesssim$~4$''$~$\approx$~550~au), which shows as a narrow band of emission almost exactly north--south. The most profound difference between these molecules is the vertical extent (i.e., east--west) of the emission. C$^{18}$O is most vertically extended ($\sim$2$''$ total) and the emission is brightest in higher layers compared to the midplane. H$^{13}$CO$^+$ is slightly less extended and peaks toward the midplane. DCO$^+$, c-C$_3$H$_2$, and H$_2$CO are about half as extended at similar angular resolution. This difference could be related to molecular abundances, sensitivity, or chemistry. For example, DCO$^+$ should be more abundant in cold gas and is therefore more likely to trace the midplane, while the lower abundance of H$_2$CO compared to C$^{18}$O makes it harder to observe H$_2$CO emission from the lower-density surface layers. As noted in earlier work, the distribution of c-C$_3$H$_2$ in the midplane is somewhat unexpected, as this molecule is often a tracer of UV-irradiated environments. Finally, C$^{18}$O and H$_2$CO show centrally peaked emission, while H$^{13}$CO$^+$, DCO$^+$, and c-C$_3$H$_2$ peak off source. This may be because no high velocity ($|\Delta v| \gtrsim$~2.5~km~s$^{-1}$) emission has been detected for those last three molecules.

% --- Inner envelope ---------------------------------------

\subsubsection{Inner envelope}

Several molecules lack emission at low velocities ($v_{\rm{sys}}~\pm \sim$0.5 km s$^{-1}$). For $^{13}$CO, HCO$^+$, and HCN this is due to low velocity emission being resolved out. This means an extended envelope component is present, in fact, so much so that the emission becomes optically thick on scales larger than probed, but thus not visible. For CS and C$^{17}$O, signs of resolved out emission are only weakly visible at the systemic velocity, so these molecules may be predominantly present in the inner regions of the envelope. The emission from $^{13}$CO, C$^{17}$O, HCO$^+$ and HCN is more compact than the inner envelope component ($|\Delta v | >$ 0.5~km~s$^{-1}$) of C$^{18}$O, H$^{13}$CO$^+$, DCO$^+$, c-C$_3$H$_2$ and H$_2$CO discussed in the previous section (Sect.~\ref{sec:Envelope}), extending only $\sim$1--2$''$ ($\sim$140--280 au) from the source along the major axis. This emission is associated with the inner envelope and potentially the disk, but without a detailed kinematic analysis these components cannot easily be distinguished. 

The $^{13}$CO emission is both radially and vertically the most extended, tracing the bulk of the gas, while C$^{17}$O traces the highest density regions and is therefore more compact. The extent of HCO$^+$ falls in between that of $^{13}$CO and C$^{17}$O. The radially more extended nature of H$^{13}$CO$^+$ (even when the inner 2 km~s$^{-1}$ are excluded; Fig~\ref{fig:SmallScale}) may be related to the higher optical depth of the main isotopologue and/or the fact that a lower-energy transition ($J=3-2$ versus $J=4-3$) has been observed for H$^{13}$CO$^+$ (see \citealt{vantHoff2022} for a more in-depth discussion). The HCN emission has a similar extent as C$^{17}$O, but forms a two-lane structure, indicating that the emission dominates in the surface layers while C$^{17}$O (and $^{13}$CO and HCO$^+$) traces down to the midplane. 

SO, $^{34}$SO, SO$_2$ and CH$_3$OH also display compact emission at intermediate velocities (Fig.~\ref{fig:SmallScale}), but no signs of resolved out emission. This suggests that they are predominantly present in the inner region. SO displays the same two-lane morphology as HCN, and the weak SO$_2$ emission hints at a similar distribution, indicating an origin in the surface layers. $^{34}$SO has only been detected at low resolution, but seems to be stronger in the north, just like SO$_2$. The weak CH$_3$OH emission is centrally peaked, but given the low signal-to-noise ratio and relatively large beam ($\sim$0.6$''$), the spatial origin is ambiguous. Observations of the CO isotopologues show that the continuum in Band 6 becomes optically thick in the inner $\sim$10 au and no line emission is detected at high velocities toward the source position (see also \citealt{vantHoff2018}). A centrally peaked morphology does therefore not directly point to an inner disk origin, but could be the result of emission originating at larger radii along the line of sight toward the source. Alternatively, CH$_3$OH may follow a similar (perhaps spatially more confined) distribution as SO and SO$_2$, with the two-lane structure washed out due to the larger beam and lower signal-to-noise ratio of the CH$_3$OH observations.   

CS also lacks extended low-velocity emission without strong signs of resolved out emission, but its morphology resembles the narrow elongated north--south structure seen for DCO$^+$, CCH and c-C$_3$H$_2$ (Figs.~\ref{fig:Envelope}), even when the central 2 km s$^{-1}$ are excluded (Fig.~\ref{fig:SmallScale}). This suggests that the bulk of the emission originates at larger radii and is more confined to the midplane compared to, for example, HCO$^+$, HCN and SO.

% --- Disk ---------------------------------------

\subsubsection{Disk}\label{sec:Disk}

\begin{figure}
    \centering
    \includegraphics[width=\linewidth]{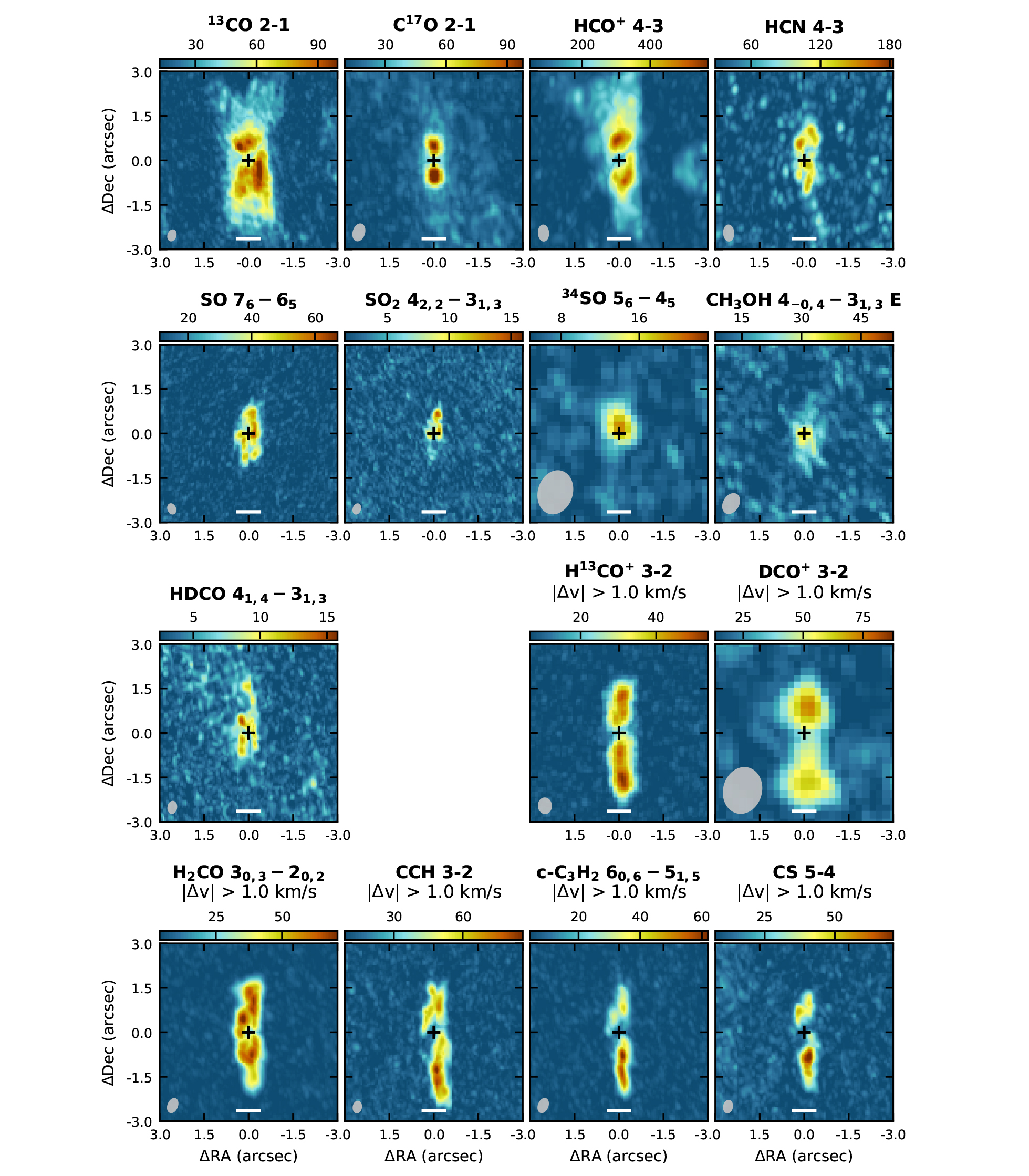}
    \caption{Peak intensity maps of molecules that are dominated by (first nine panels) or show strong emission (last six panels) on small angular scales. For the last six panels, only velocity offsets $>$1.0 km/s from the systemic velocity are used to create the peak intensity maps. In each panel, the gray ellipse in the bottom left corner represents the synthesized beam and the white scalebar marks 100 au. The following datasets are used: 2011.0.00604.S (CH$_3$OH), 2012.1.00346.S (HCO$^+$, HCN), 2013.1.00858.S (SO), 2017.1.01413.S (C$^{17}$O), 2018.1.01205.L ($^{34}$SO, HDCO, DCO$^+$, H$_2$CO, CCH, c-C$_3$H$_2$, CS), 2019.A.00034.S ($^{13}$CO, SO$_2$), and 2021.1.00536.S (H$^{13}$CO$^+$). }
    \label{fig:SmallScale}
\end{figure}

In the existing data, only four molecular species have emission detected out to high enough velocities to confidently confirm their presence in the disk: CO (through $^{13}$CO, C$^{18}$O and C$^{17}$O), HCO$^+$, H$_2$CO (and HDCO) and SO. All other species detected at intermediate velocities could very well be present in the disk, but a more detailed kinematic analysis and probably higher sensitivity observations would be required to confirm this. At high angular resolution, HDCO displays the same two lane morphology as SO, suggesting its emission also originates predominantly in the surface layers (Fig.~\ref{fig:SmallScale}). This structure is also present for the main isotopologue, although less clearly due to the envelope contribution, even when the central 2 km s$^{-1}$ are excluded (Fig.~\ref{fig:SmallScale}). C$^{17}$O and HCO$^+$ on the other hand, show less structure and the emission peaks north and south of the source, suggesting these species trace the full vertical extent of the disk. 

% ...................................................................
% Figure - Inner envelope and disk
% ...................................................................
% ...................................................................

% ...................................................................
% Figure - Spectra of weak detections
% ...................................................................
\begin{figure}
    \centering
    \includegraphics[width=\linewidth,trim={0cm 16.5cm 0cm 2cm},clip]{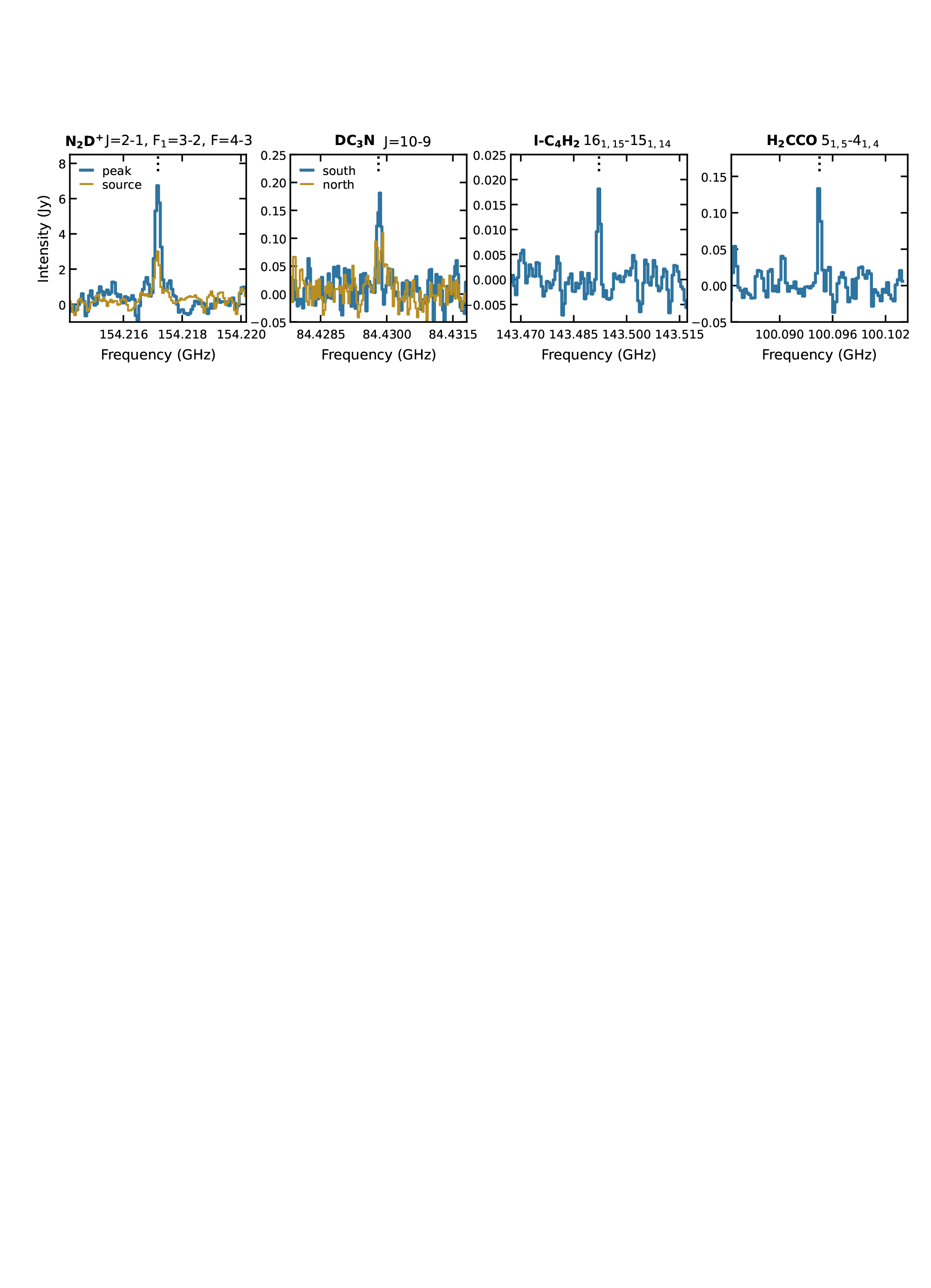}
    \caption{Spectra of molecules that are either only detected in Total Power observations (N$_2$D$^+$; project 2016.1.01541.S) or when spatially integrated (DC$_3$N, l-C$_4$H$_2$, and H$_2$CCO; projects 2015.1.00261.S, 2016.2.00171.S, and 2022.1.00131.S, respectively). The N$_2$D$^+$ spectra are extracted in pixels toward the N$_2$D$^+$ peak (blue) and the source position (orange). The DC$_3$N spectra are extracted in 8.0$''$ $\times$ 6.5$''$ apertures just north (blue) and south (orange) of the source position. The l-C$_4$H$_2$ and H$_2$CCO spectra are extracted in 5.5$''$ $\times$ 4.2$''$ and 18$''$ $\times$ 9$''$ apertures toward the source position, respectively. The vertical dotted line marks the rest frequency of each transition.}
    \label{fig:Spectra}
\end{figure}
% ...................................................................

% --- Cavity wall ---------------------------------------

\subsubsection{Cavity wall}\label{sec:Cavitywall}

Distinguishing the cavity wall, i.e., the infalling envelope surface along the outflow, from the outflowing material is not trivial, especially for an edge-on source where the outflow lies in the plane of the sky and therefore also has low velocities. We define the cavity wall here as a strip of material on the envelope side of the $^{12}$CO emission. A full ``cross'' morphology is visible for $^{13}$CO, C$^{18}$O, CS and weakly in SO (see also \citealt{vantHoff_eDisk,Liu2025}). However, in most cases, such structure is most clearly visible in the southwest (Figs.~\ref{fig:Outflow} and \ref{fig:Envelope}), due to the slight deviation from a perfect edge-on orientation (see \citealt{Oya2015}). Many of the molecules displaying outflow emission also show emission from the cavity wall (Fig.~\ref{fig:Outflow}). \citet{Liu2025} performed a detailed analysis of the SO emission, finding that it is likely tracing infalling material.

% -----------------------------------------------------------------
% Column densities
% -----------------------------------------------------------------

\subsection{Column densities}\label{sec:Columndensities}

To obtain disk-averaged column densities, we calculate the total integrated flux at velocities high enough to contain only disk emission. Based on the modeling done for L1527 by \citet{vantHoff2018}, velocity offsets $\leq$2.54 km s$^{-1}$ and $\geq$2.59 km s$^{-1}$ with respect to the systemic velocity are expected to be free of envelope emission. Since only a few molecules have emission detected out to these high velocities (Sect.~\ref{sec:Disk}), we also calculate the total integrated flux over $|\Delta v| \geq 1.75$ km s$^{-1}$, which should contain emission from the inner $\sim$200 au, which encompasses the disk (radius of $\sim$100--130 au; \citealt{vantHoff_eDisk}) and innermost envelope. This velocity range also corresponds to disk-only emission for the lower end (0.3 $M_{\odot}$) of the reported stellar masses in the literature \citep{Aso2017,vantHoff_eDisk}. The column density in the inner envelope is then calculated over $|\Delta v| \sim 0.5 - 1.75$ km s$^{-1}$, and in the extended envelope over $|\Delta v| \lesssim 0.5$ km s$^{-1}$. For the disk and inner 200 au components, we typically do not use ACA observations, as this compact component is often hard to make out in these datasets, especially when the integration time is short. 

% ...................................................................
\input{Tables/Components}
% ...................................................................

% ...................................................................
% Overview cartoon
% ...................................................................
% ...................................................................

In practice, we make integrated-intensity (moment-0) maps (using primary-beam corrected images) over these velocity ranges for blueshifted and redshifted emission separately (except for the extended envelope component), including channels with $\geq3\sigma$ emission. Given the wide range in spectral resolution, the actual velocity range used is different for each observation of a molecular line. We then estimate the rms in the moment-0 maps based on the rms per channel (in a spatial region similar as to from which the flux is extracted) and number of channels used to create the moment map, and extract the integrated flux in an ellipse encompassing the $2\sigma$ contour. The column density, $N$, is then calculated under the assumption of local thermodynamic equilibrium (LTE) using the partition functions, $Q$, and molecular spectroscopy data from the Cologne Database for Molecular Spectroscopy (CDMS; \citealt{Muller2001,Muller2005,Endres2016}) as follows:
\begin{equation}
   N = \frac{4\pi F\Delta\nu Q(T_{ex})}{\Omega A_{ul} hc g_u} e^{E_u/kT_{ex}}, 
\end{equation}
where $F\Delta\nu$ is the integrated flux density, $\Omega$ is the emitting area, $A_{ul}$ is the Einstein A coefficient, $g_u$ and $E_u$ are the degeneracy and energy of the upper level, respectively, and $T_{ex}$ is the excitation temperature. We adopt an excitation temperature of 50 K for the disk and inner 200 au \citep{vantHoff2018,vantHoff_eDisk}, and 20 K for the inner and extended envelope. The ratios of CH$_3$CCH lines observed on envelope scales are consistent with an excitation temperature of 20~K. 

% ...................................................................
% Figure - C18O column densities
% ...................................................................
\begin{figure}
    \centering
    \includegraphics[width=\linewidth]{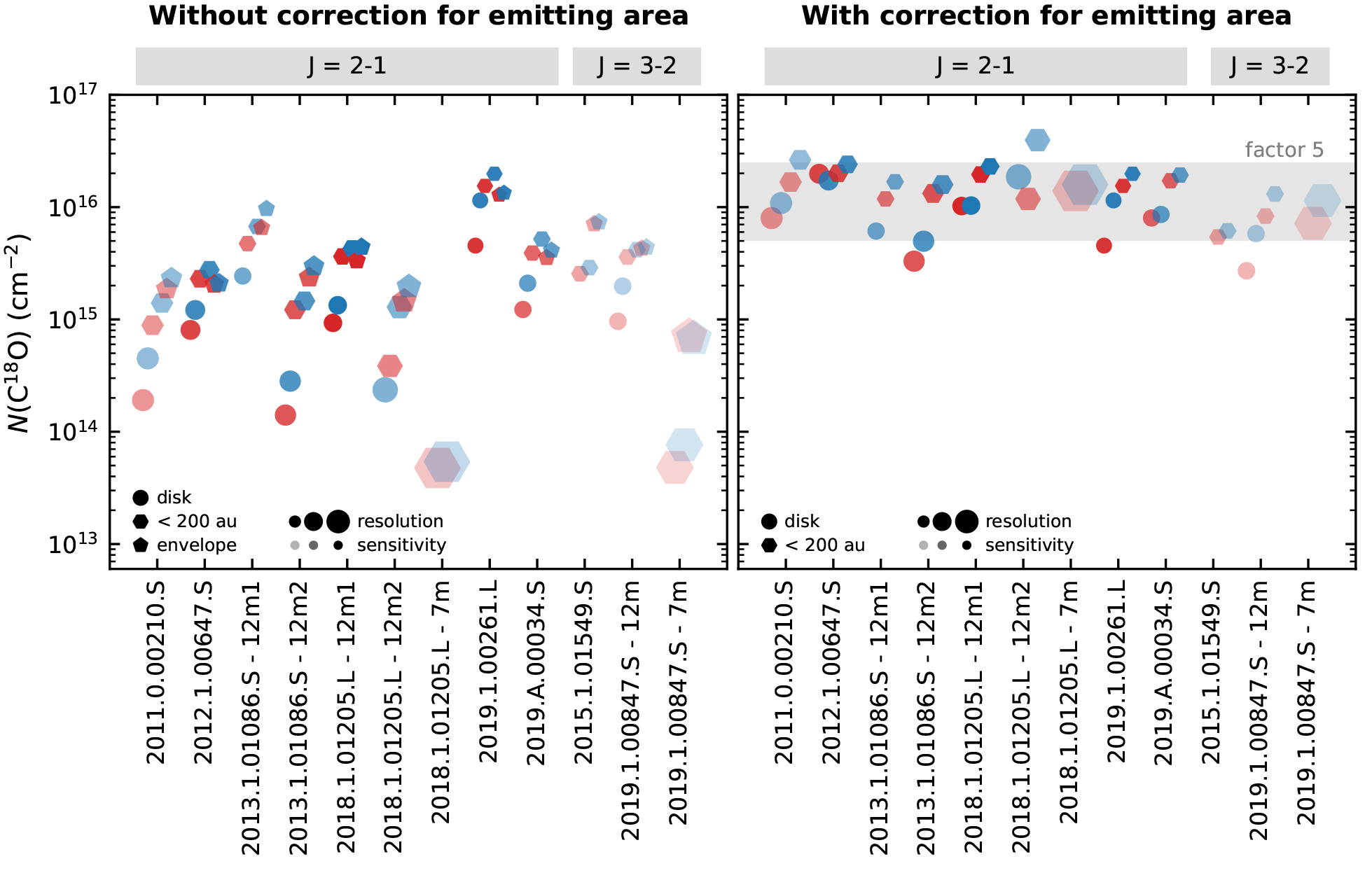}
    \caption{C$^{18}$O column densities calculated over different velocity ranges as proxy for different regions. In the left panel, column densities are calculated assuming the emission uniformly fills the aperture over which the flux is extracted. In the right panel, the emitting area is set to be the resolved emitting area of C$^{18}$O at 0.05$''$ resolution (program 2019.1.00261.L). Emission at velocity offsets $\Delta v \lesssim$~$-2.54$ km~s$^{-1}$ (blue circles) and $\Delta v \gtrsim$~2.59 km~s$^{-1}$ (red circles) originates solely from the disk. Emission at $|\Delta v| \gtrsim$~1.75 km~s$^{-1}$ (red and blue hexagons for redshifted and blueshifted emission, respectively) originates from the inner $\sim$200 au, that is, the disk and innermost part of the envelope, and emission at $|\Delta v| \approx 0.5-1.75$~km~s$^{-1}$ is dominated by the inner envelope (red and blue pentagons). The size of the markers represents the angular resolution of the observations, with larger markers indicating larger synthesized beam sizes (or poorer resolution). The opacity of the markers represents the sensitivity of the observations, with darker markers indicating higher sensitivity (or lower rms values). }
    \label{fig:C18OColumndensities}
\end{figure}
% ...................................................................

As C$^{18}$O $J=2-1$ is the most observed transition, all C$^{18}$O column densities are shown in Fig.~\ref{fig:C18OColumndensities}. The extended envelope component is excluded because of resolved-out emission around the systemic velocity. Within one dataset, the different velocity cuts result in column densities varying within a factor $\lesssim$ 10. Column densities extracted from redshifted emission are typically a bit smaller than those from the same blueshifted velocity offsets. Disk column densities are generally a factor of a few lower than those for the inner 200 au, consistent with a larger gas column when considering material out to larger radii. The envelope measurements are in turn often quite similar to the inner 200 au measurements, potentially because the density gradient becomes more shallow at larger radii and emission from much larger scales is resolved out. Taking into account all column density measurements from the different datasets, a spread of more than two orders of magnitude is found. There is no clear trend with sensitivity, but higher resolution observations typically result in higher column densities. Fig.~\ref{fig:C18OColumndensities} (right panel) clearly demonstrates that this effect is due to beam dilution, since setting the emitting area in the column density calculation to the resolved emitting area of the highest resolution (0.05$''$) observations (eDisk: 2019.1.00261.L) reduces the spread in column density to a factor of $\sim$5. In addition, the trend with angular resolution disappears. Column densities derived from disk-only velocities are still typically a bit smaller than those derived over a slightly larger velocity range, but the spread for both disk and inner 200 au measurements is about a factor of 5. This emitting-area correction is only applied for the disk and inner 200 au measurements, as for the more extended emission the MRS of the observations becomes more important. Part of the remaining spread may be related to the differences in spectral resolution between datasets and the consequently different velocity ranges used.

Assuming C$^{18}$O traces the bulk of the gas in the disk and inner envelope, and other molecules have the same distribution as C$^{18}$O, we adopt the emitting area derived from the resolved C$^{18}$O observations for all column density calculations. This results in a similar spread as seen for C$^{18}$O of about a factor of 5 for all molecules. The column density scales linearly with emitting area. In the existing datasets, emission arising predominantly from the surface layers of the disk and inner 200 au has an emitting area smaller by less than a factor of 2. As distributions become clearly different on envelope scales and the MRS becomes more important, we do not correct the emitting area for envelope column densities. Although this can result in spreads larger than a factor of 5 for some molecules, the extended emission is more readily spatially resolved so the effects of beam dilution are not as large as for compact emission. An additional source of uncertainty is the adopted excitation temperature. However, changing the excitation temperature within a reasonable range generally changes the column densities by less than a factor of 2. Specifically, adopting temperatures of 30 and 80 K, rather than 50 K, for the disk and inner 200 au changes individual column densities by factors of 0.6--1.7 and 0.5--1.0, respectively. However, a single temperature is a simplified assumption, which would impact molecules differently depending on their detailed spatial distribution. For the (extended) envelope, the temperature is likely more uniform, and the ratios of CH$_3$CCH lines are consistent with 20 K. Adopting temperatures of 10 and 30 K, instead of 20 K, generally changes individual column densities by factors of 0.4--1.6 and 0.7--1.8, respectively. Larger changes are found for a few of the higher energy transitions, with changes up to a factor of 5 for transitions with $E_{\rm{up}} \sim 60-80$ K at 30 K, and up to a factor of 10 for transitions with $E_{\rm{up}} \sim 45-80$ K at 10 K. For molecules with multiple transitions detected, some of the temperature uncertainty is likely mitigated by taking the median column density as some transitions would result in a higher and other in a lower column density at different temperatures.  Combining and propagating all different sources of uncertainty into one error estimate is non trivial, but taken together, a spread of a factor of $\sim$5 for a given molecule appears a reasonable estimate. 

\begin{figure}
    \centering
    \includegraphics[width=0.91\linewidth]{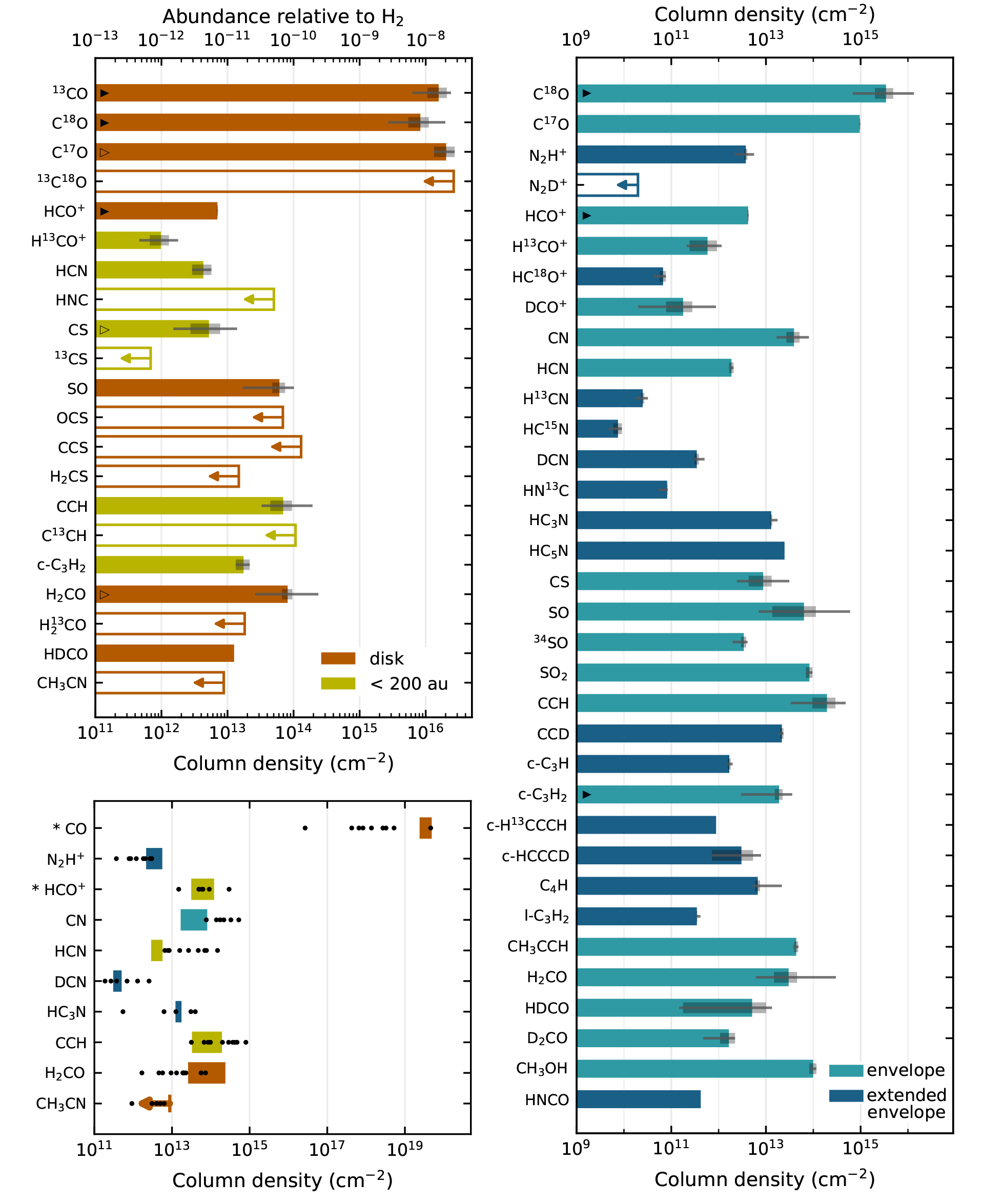}
    \vspace{-0.5cm}
    \caption{Median column densities of molecular species in L1527. The top left panel shows column densities for the disk ($|\Delta v| > 2.55$ km s$^{-1}$; orange) and inner 200 au ($|\Delta v| > 1.75$ km s$^{-1}$; yellow) after correcting the emitting area using the resolved C$^{18}$O emitting area. For species detected on both scales, only the disk component is shown, which is typically less than a factor 2 lower than the inner 200 au component. The top axis shows the abundance relative to H$_2$ for a H$_2$ column of $1 \times 10^{24}$ cm$^{-2}$ based on continuum emission. The right panel shows column densities for the envelope (light blue) and extended envelope components (dark blue). If both components are present, only the envelope component is shown. Black and open triangles indicate that the emission is likely or may be, respectively, (marginally) optically thick and the derived column densities are or may be lower limits. The bottom left panel shows a comparison with disk-averaged column densities in Class II protoplanetary disks (black points) from the review by \citet{Oberg2023}. For species marked with an asterisk, their column density in L1527 has been derived from an optically thin isotopologue.}
    \label{fig:Columndensities}
\end{figure}

The median column densities on small and large scales are shown in Fig.~\ref{fig:Columndensities}. For the small scales, the disk column density is shown when available, which is less than a factor 2 lower than the inner 200 au column density. For the large scales, most molecules are either detected in the envelope or in the extended envelope. In the few cases where the emission is present on both scales, the envelope column density is shown, which is within a factor of 2--3 from the extended envelope column density. For species that are not detected, upper limits are also calculated using the C$^{18}$O emitting area, except for N$_2$D$^+$, where the N$_2$H$^+$ aperture is used.

% ...................................................................
% Figure - Column densities
% ...................................................................
% ...................................................................

\subsubsection{Disk and inner most envelope}

On small scales, the CO isotopologues have, as expected, the highest column densities. All three detected isotopologues have columns on the order of $10^{16}$ cm$^{-2}$ with the highest values for C$^{17}$O, indicating at least the $^{13}$CO and C$^{18}$O emission is optically thick. Using isotope ratios of $^{12}$C/$^{13}$C = 68 \citep{Milam2005}, $^{16}$O/$^{18}$O = 560 and $^{16}$O/$^{17}$O = 1790 \citep{Wilson1994}, the C$^{17}$O/$^{13}$C$^{18}$O ratio is 21 for optically thin emission. The most stringent $^{13}$C$^{18}$O upper limit is similar to the largest C$^{17}$O column density, suggesting C$^{17}$O may indeed be optically thin and existing datasets are not sensitive enough to detect $^{13}$C$^{18}$O. If the C$^{17}$O emission is fully optically thin, this would then result in a total CO column of $\sim4\times10^{19}$ cm$^{-2}$. We estimate the CO abundance by calculating the dust mass from the eDisk observations following \citep{vantHoff_eDisk} over the resolved C$^{18}$O emitting area ($\sim$0.14--0.28$''$ radius for blueshifted and redshifted disk and inner 200 au components) and converting that into an H$_2$ column density. These apertures start 0.12$''$ north of the source and 0.09$''$ south of the source, avoiding the brightest continuum regions, so the emission is most likely optically thin. The resulting H$_2$ column densities range between $1-3 \times 10^{24}$ cm$^{-2}$ for the different apertures. This then suggests a CO abundance of $\sim 7 \times 10^{-6} - 6 \times 10^{-5}$, which is slightly lower than the canonical CO abundance of $10^{-4}$. However, given the typical spread of a factor of 5 in column densities between different datasets, and that C$^{17}$O is only present in two datasets, a canonical CO abundance cannot be ruled out. The $^{13}$C$^{18}$O upper limits are consistent with CO abundances less than $2-5 \times 10^{-4}$, and thus not stringent enough to directly rule out a canonical CO abundance. Additional caveats are in the uncertainty of converting the continuum flux into a dust mass (although the C$^{18}$O emitting areas do not overlap with the optically thick inner dust disk), converting the dust mass into a gas mass assuming a fixed gas-to-dust ratio (100), and the relative distributions of H$_2$ and CO. A more detailed analysis is thus required to derive the spatially resolved CO abundance, but based on a global analysis and typical assumptions a canonical CO abundance cannot be ruled out and there is no immediate evidence for a strong depletion, consistent with earlier results for other young disks \citep{Harsono2014,Bergner2019,vantHoff2020,Zhang2020}. 

After the CO isotopologues, the largest column densities are found for SO, CCH and H$_2$CO, at values just below $10^{14}$ cm$^{-2}$, two orders of magnitude less than for C$^{17}$O, or at abundances of a few times $\sim$$10^{-11}$ with respect to H$_2$. The H$_2$$^{13}$CO upper limit suggests a H$_2$CO column only 5--10 times higher than observed, indicating H$_2$CO is at best somewhat optically thick. The upper limit for C$^{13}$CH is comparable to the main isotopologue's column and does therefore not place stringent constraints on the optical depth. Column densities of order $10^{13}$ cm$^{-2}$ are found for c-C$_3$H$_2$ and HDCO, and column densities of several times $10^{12}$ cm$^{-2}$ are seen for HCO$^+$, HCN and CS. On scales of the inner 200 au, the H$^{13}$CO$^+$ column density is $\sim$ 20 times lower than that for HCO$^{+}$, indicating HCO$^{+}$ is optically thick, and actually in abundance similar to CCH, H$_2$CO and SO (assuming they are optically thin). The $^{13}$CS upper limit also allows for slightly larger CS abundance. The upper limits for HNC, OCS and CCS are of the order 10$^{14}$ cm$^{-2}$, and those for H$_2$CS and CH$_3$CN are about 10 times more stringent. 

\subsubsection{Envelope scales}

The column densities calculated over low velocities are typically lower than the corresponding column densities on disk and inner 200 au scales, except for SO and c-C$_3$H$_2$, for which those columns are similar, and CS and CCH, whose column densities are slightly higher on larger scales. The column densities on larger scales are more difficult to convert into abundances as C$^{17}$O may suffer from some resolved out emission and freeze out. Indeed, assuming a similar C$^{17}$O abundance as on small scales results in abundances of other molecules of roughly a factor 10 higher on larger scales compared to smaller scales. Nevertheless, the most abundant molecule on envelope scales ($|\Delta v| \sim 0.5-1.75$ km s$^{-1}$) after the CO isotopologues is CCH with a median column density of a few times $10^{14}$ cm$^{-2}$. Slightly lower columns ($10^{13}$--$10^{14}$ cm$^{-2}$) are found for CN, SO, SO$_2$, CCH, c-C$_3$H$_2$, CH$_3$CCH, H$_2$CO and CH$_3$OH, followed by HCO$^+$, HCN, $^{34}$SO, HDCO and D$_2$CO ($10^{12}$--$10^{13}$ cm$^{-2}$). The lowest column densities are found for H$^{13}$CO$^+$ and DCO$^+$ ($10^{11}$--$10^{12}$ cm$^{-2}$). 

On the most extended scales and lowest velocities ($|\Delta v| < 0.5$ km s$^{-1}$), the highest column densities are found for HC$_3$N, HC$_5$N and CCD (a few times $10^{13}$ cm$^{-2}$), followed by N$_2$H$^+$, c-HCCCD and C$_4$H ($10^{12}$--$10^{13}$ cm$^{-2}$). c-H$^{13}$CCCH, l-C$_3$H$_2$ and HNCO have column densities of $10^{11}$--$10^{12}$ cm$^{-2}$, followed by HC$^{18}$O$^+$, H$^{13}$CN and HN$^{13}$C ($10^{10}$--$10^{11}$ cm$^{-2}$). HC$^{15}$N has the lowest column just below $10^{10}$ cm$^{-2}$.

\subsubsection{Isotope ratios}

After CO, most isotopologues are detected for HCO$^+$. The HCO$^+$/H$^{13}$CO$^+$ ratio is 16--20 on disks scales and even lower on envelope scales, indicating that at least the HCO$^+$ emission is optically thick. Assuming H$^{13}$CO$^+$ is optically thin and a $^{12}$C/$^{13}$C ratio of 68 \citep{Milam2005}, the D/H ratio for HCO$^+$ in the inner envelope is of order 0.001--0.02. HC$^{18}$O$^+$ is predominantly detected in the more extended envelope, where H$^{13}$CO$^+$ suffers from resolved out emission. The D/H ratio in the extended envelope, assuming a $^{16}$O/$^{18}$O ratio of 560 \citep{Wilson1994} is 0.004--0.01. 

The $^{13}$C isotope has also been detected for HCN, HNC and c-C$_3$H$_2$. The c-C$_3$H$_2$/c-H$^{13}$CCCH ratio of less than 10 in the extended envelope suggests the main isotopologue is optically thick. For the N-bearing species the $^{12}$C/$^{13}$C ratio cannot be assessed, because HCN and H$^{13}$CN have very different distributions and the main HNC isotopologue has not been detected. The nondetections of $^{13}$CS, C$^{13}$CH and H$_2^{13}$CO are not stringent enough to constrain the optical depth of the main isotopologues. Less abundant oxygen isotopes are only detected with CO and HCO$^+$, and $^{15}$N is only detected in HC$^{15}$N. In the extended envelope, the $^{14}$N/$^{15}$N ratio based on H$^{13}$CN/HC$^{15}$N is 220--320, where the higher value is for the redshifted side. For sulfur, the $^{34}$S isotope is detected only in $^{34}$SO. On envelope scales, the ratio of the median column densities is consistent with the ISM $^{32}$S/$^{34}$S ratio of 22 \citep{Wilson1994}.  

The most detected isotope is deuterium. On small scales, the only deuterated species is HDCO, which suggests a D/H ratio of 0.06--0.16 based on the H$_2$CO emission and a lower limit to the D/H ratio of 0.005--0.02 based on the H$_2^{13}$CO upper limit. In the envelope, D/H ratios of $\sim$0.2 and $\sim$0.08 are suggested by HDCO/H$_2$CO and D$_2$CO/H$_2$CO, respectively. In the extended envelope, HDCO/D$_2$CO point to a high D/H ratio of $\gtrsim 0.6$ in both 12m and ACA observations. The H$_2$CO D/H calculations take the statistics of the two hydrogen positions into account \citep{Manigand2019}. D/H ratios in the extended envelope are 0.01--0.1 for c-C$_3$H$_2$ and 0.2--0.3 for DCN, in both cases using the $^{13}$C isotopologue and a $^{12}$C/$^{13}$C ratio of 68. Although CCH and CCD are detected in the envelope and extended envelope, respectively, their column density ratio suggest a D/H ratio of $\sim$0.1. Finally, the N$_2$D$^+$ upper limit in the extended envelope where N$_2$H$^+$ is detected suggests an upper limit to the D/H ratio of $\lesssim 0.01$, which is less stringent than the single-dish upper limit of 0.001 \citep{Tobin_N2H+}. Overall, the D/H ratio seems to be roughly on the order of $10^{-2}-10^{-1}$ for most detected species, except for D$_2$CO, which suggests a higher D/H ratio.

% ==================================================================
% DISCUSSION
% ==================================================================

\section{Discussion} \label{sec:Discussion}

% -----------------------------------------------------------------
% L1527 observations 
% -----------------------------------------------------------------

\subsection{Comparison with previous work on L1527}

\subsubsection{Carbon chains}
Based on single-dish observations, the main chemical characteristic of L1527 is a rich carbon-chain chemistry that is typically associated with prestellar cores \citep{Sakai2008}. The rotational temperature and linewidth pointed to an origin in the inner envelope, which was confirmed by spatially resolved observations with the Plateau de Bure Interferometer (PdBI) showing emission out to $\sim$20$''$ \citep{Sakai2010}. As carbon-chains formed in the early stages of molecular clouds are predicted to freeze out or get destroyed during the later stages of starless cores, their presence in L1527 was attributed to the thermal desorption of CH$_4$ near the protostar in combination with a short collapse time after the prestellar core phase \citep{Sakai2008,Sakai2013}. Chemical modeling confirmed that carbon chains can form following the sublimation of CH$_4$ \citep[e.g.,][]{Aikawa2008,Hassel2008}. On the other hand, the detection of different spatial distributions between c-C$_3$H$_2$ and CH$_3$OH toward prestellar cores suggests that differences in illumination by the interstellar radiation field, where carbon chains form in regions exposed to FUV, are driving distinct chemistries that may then be inherited by the protostellar stage \citep{Spezzano2016,Spezzano2020}. The frequent detection of CCH and c-C$_3$H$_2$ in outflow cavity walls in high-resolution observations is consistent with a photodissociation-driven formation of hydrocarbons in protostellar systems \citep[e.g.,][]{Zhang2018,Tychoniec2021}, although hydrocarbons have also been associated with the protostellar envelope in some sources \citep{Murillo2018}. 

Subarcsecond-resolution observations showed that on scales of a few arcseconds the CCH and c-C$_3$H$_2$ emission in L1527 was in fact dominated by the inner envelope  with clear signs of larger scale emission being resolved out \citep{Sakai2014_Nature,Sakai2014,Sakai2017,Oya2015,vantHoff_eDisk}. However, as presented here, observations with a larger field of view show a strong contribution along the southeastern outflow cavity up to 25$''$ (3500 au) south of the source position (Fig.~\ref{fig:SEtail}). A similar structure is seen for the detected isotopologues, C$_4$H, l-C$_3$H$_2$ and HC$_3$N, but not for other species, such as D$_2$CO, in similar datasets. In addition, CCH and c-C$_3$H$_2$ are also strongly detected in the western outflow cavity, a few arcseconds off source. The sulfur-bearing carbon chain CCS is only detected here in the outflow. A bright spot is present near the tip of the southeastern tail, but its true nature is more uncertain as it falls at the edge of the FOV. Given that the outflow direction is along the plane of the sky, a kinematic analysis to determine whether the carbon-chain tail is infalling or outflowing is challenging. Based on a comparison with $^{12}$CO emission (Fig.~\ref{fig:SEtail}), the part closest to the protostar falls inside the envelope, while the outer part either falls within the cavity or represents more extended envelope emission that lies in front of the cavity along our line of sight. A detailed analysis of SO emission along the outflow cavity wall by \citet{Liu2025} suggests that this material is undergoing downward-spiraling infall on scales of $\lesssim10''$. Regardless of the details that remain to be constrained, the carbon chains in L1527 seem thus, at least in part, the result of localized photochemistry and/or shock chemistry. 

Additional evidence presented for a short collapse phase in L1527 was D/H ratios in carbon-chain species of a few percent (2--7\%; \citealt{Sakai2009}), compared to higher ratios in hot cores (e.g., 18\% toward IRAS 16293; \citealt{vanDishoeck1995}). For the extended envelope component at systemic velocities, we derive similarly low D/H ratios for c-C$_3$H$_2$ and CCH (1--10\%). However, consistent with earlier IRAM 30m single-dish observations toward L1527, as well as hot cores, we find higher D/H ratios for H$_2$CO \citep[e.g.,][]{Parise2006} and also HCN. A higher D/H ratio may signal heavy CO depletion and a longer prestellar core phase, which would mean contradictory results between carbon chains and organics. Recent JWST observations of water ice in L1527 point in the same direction as the gas-phase organics. Analysis revealed ice HDO/H$_2$O ratios to be similar to gas-phase HDO/H$_2$O ratios in other low-mass protostellar envelopes in comparably isolated star-forming regions \citep{Slavicinska2025}. In contrast, this ratio is a factor $\sim$4--7 higher compared to gas-phase measurements toward protostars in more clustered environments. This difference in gaseous water D/H ratios has been suggested to be the result of either longer or colder prestellar stages in more isolated regions \citep{Jensen2019,Jensen2021}. Alternatively, different D/H ratios may indicate different formation times or gas-phase versus ice formation. Hints on the gas-phase processes at play for deuterium chemistry are obtained from the here confirmed spatial offset of N$_2$H$^+$ and N$_2$D$^+$, which indicates a region where CO is frozen out (and hence N$_2$H$^+$ is abundant) but the H$_2$D$^+$ abundance is low (and hence N$_2$D$^+$ is low), either due to a relatively warm CO freeze-out temperature or a H$_2$ ortho-to-para gradient \citep{Emprechtinger2009,Tobin_N2H+}. Gas-phase deuterium fractionation occurring in the inner few 1000 au of the envelope should thus be driven by CH$_2$D$^+$.

\subsubsection{The disk--envelope interface}

A second chemical characteristic observed toward L1527 is a change in composition at the interface between the disk and envelope. This was first reported with early ALMA observations for SO, where the emission seems to trace a ring of material around the disk--envelope boundary, potentially due to an accretion shock  \citep{Sakai2014_Nature,Ohashi2014}. Subsequent SO observations reveal a more complex morphology, including emission from the inner disk, as well as disk and envelope surface layers and/or the outflow cavity wall \citep{vantHoff_eDisk,Liu2025}. So far, no other molecular species has been found to have a similar distribution with respect to the disk--envelope interface. SO$_2$ is probably the most likely candidate \citep{ArturdelaVillarmois2023}, but so far only weak emission has been detected (see also \citealt{Liu2025}). Early ALMA results did suggest an absence of, for example, c-C$_3$H$_2$, CCH and CS in the disk because no emission was detected at high velocities \citep{Sakai2014_Nature,Sakai2014}. 

Among all molecules detected with ALMA as presented here, CCH and CS are the only ones were slightly higher column densities are derived on larger scales, suggesting that these molecules may indeed be most abundant in the more extended envelope. Nitrogen-bearing molecules are also mostly detected on larger scales only, but this may be a sensitivity issue combined with a cosmic lower nitrogen abundance compared to carbon and oxygen. Based on molecules that have been observed by many programs (such as C$^{18}$O), high-velocity disk emission can easily be missed if the sensitivity is not high enough, and most molecules do not have observations sensitive enough to detect emission out to large velocity offsets. A detailed analysis of the velocity profiles would therefore be required to determine whether more narrow line profiles and the absence of higher velocity line wings are truly indicative of an absence of disk emission or simply due to the small emitting area in combination with too low sensitivity.

% -----------------------------------------------------------------
% JWST
% -----------------------------------------------------------------

\subsection{Comparison with JWST observations}

\begin{figure}
    \centering
    \includegraphics[width=\linewidth]{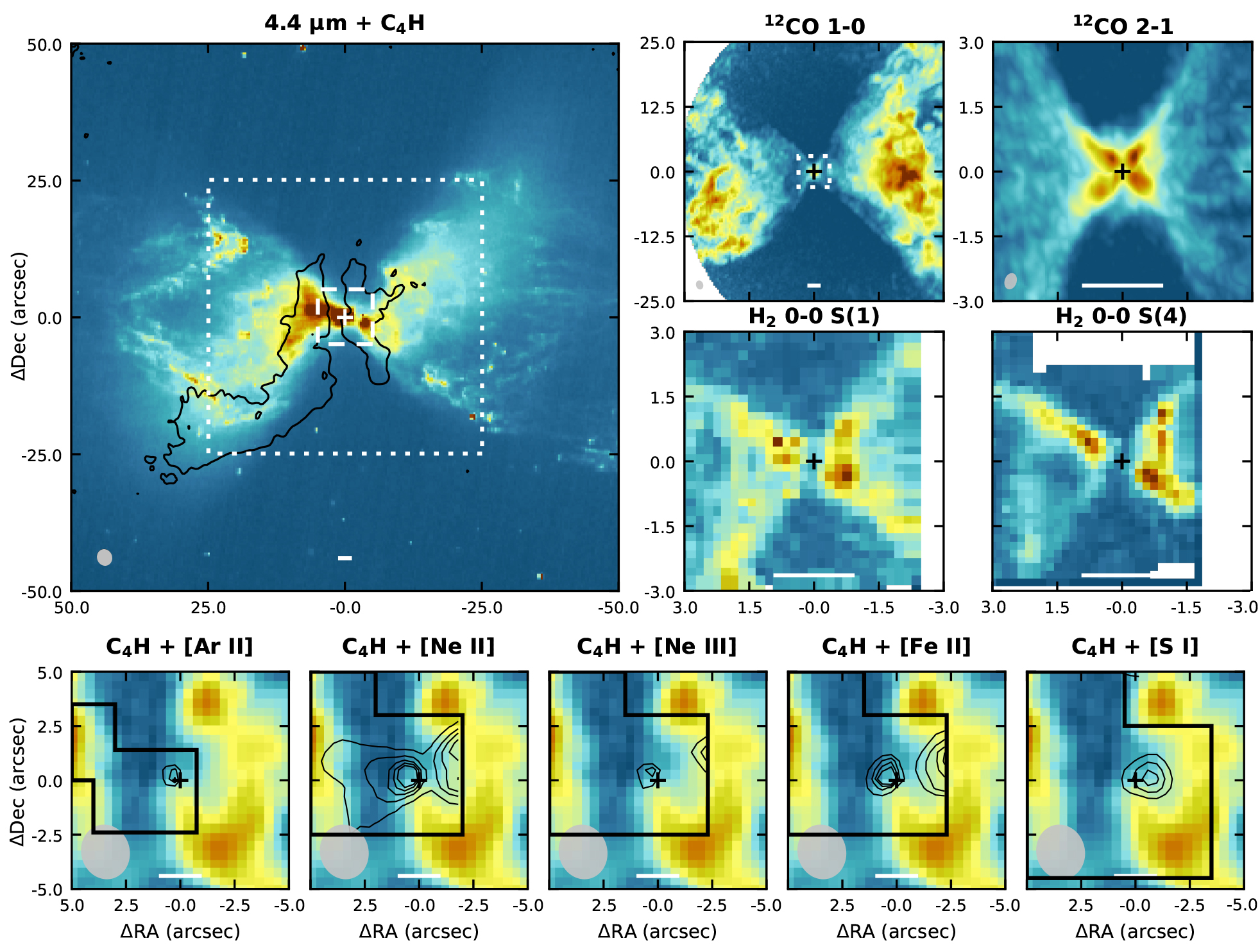}
    \caption{Comparison between select ALMA and JWST observations of L1527. The top left panel shows the JWST NIRCam 4.4 $\mu$m image with the ALMA C$_4$H peak intensity map overlaid in black contours corresponding to 7 mJy beam$^{-1}$, i.e., 5 times the rms of the image cube. The dotted white contours mark the image size of the top middle panel and the dashed white contours mark the size of the bottom panels. The top middle and right panels show ALMA peak intensity maps of $^{12}$CO $1-0$ and $2-1$, respectively. The dashed white contours in the top middle panel mark the image size of the top right and middle row panels. The middle row panels show JWST MIRI integrated intensity maps of H$_2$ $0-0$ S(1) and S(4). The bottom panels show the ALMA C$_4$H peak intensity maps with JWST MIRI integrated intensity maps of [Ar II], [Ne II], [Ne III], [Fe II] and [S I] overlaid in black contours starting at 3$\sigma$ and in steps of 2$\sigma$ except for [Ne II] where the steps are 5$\sigma$. The thick black line outlines the MIRI footprint. The gray ellipses depict the ALMA synthesized beam and the white scalebars mark 250 au in each panel. The following datasets are used: 2018.1.00799.S (C$_4$H), 2019.A.00034.S (CO $2-1$), 2022.1.00131.S (CO $1-0$), PID 2739 (4.4 $\mu$m), PID 1290 ([Ar II], [Ne II], [Ne III], [Fe II] and [S I]).}
    \label{fig:JWST}
\end{figure}

L1527 has been targeted by the NIRCam, NIRSpec, and MIRI/MRS and MIRI/Imager instruments onboard JWST. The stunning large-scale NIRCam image of the outflow has already been compared to ALMA maps of $^{12}$CO by \cite{vantHoff_eDisk}, and Fig.~\ref{fig:JWST} shows a comparison with C$_4$H. H$_2$, the most abundant molecule in the gas phase, is undetectable with ALMA due to the lack of a permanent dipole moment. Quadrupole rotational transitions detectable with JWST are an invaluable resource to observe the warm gas. \cite{Devaraj.vanDishoeck.ea2026} present MIRI observations of those lines, at scales of $\sim$ 500 au (see also Fig.~\ref{fig:JWST}). It has been shown that the H$_2$ S(1) line often exhibits a close resemblance to low-velocity CO emission \citep{Dutta2025, Narang.Tyagi.ea2026}. In L1527, the CO emission peaks at a significant distance from the source, beyond the MIRI/MRS FOV (Figs.~\ref{fig:Outflow} and \ref{fig:JWST}). At scales traced by MIRI, the emission morphology of H$_2$ shows indeed a clear component along the outflow cavity walls, but with a slight asymmetry: along the northwest--southeast direction the H$_2$ emission coincides with the edge of the CO emission, while along the northeast--southwest direction the H$_2$ emission falls within the cavity, i.e., is located closer to the outflow axis. The H$_2$ emission is brighter in the northeast to southwest direction, which likely results from shocks at the outflow cavity wall. 

% ...................................................................
% Figure - JWST 
% ...................................................................
% ...................................................................

\cite{Devaraj.vanDishoeck.ea2026} interpreted the weak emission from the coldest H$_2$ S(1) line as a sign of little contribution from cold gas, with H$_2$ emission being dominated by shocked gas at least close to the source. Based on shock models, H$_2$ is dominated by J-type shocks, with UV radiation contributing to gas heating. More sensitive NIRSpec observations of H$_2$ presented in \cite{BlakeDrechsler.Tobin.ea2026} also pick up a horizontal ridge close to the source cospatial with, for example, H$^{13}$CO$^+$ and c-C$_3$H$_2$ (Fig. \ref{fig:SmallScale}). Importantly, no collimated jet is detected in H$_2$, consistent with the absence of a high-velocity molecular jet in ALMA data, and H$_2$ is tracing a wide-angle wind with a fixed opening angle. H$_2$ rotational temperature maps revealed a temperature gradient with hotter southeast and northwest sides of the flow \citep{Devaraj.vanDishoeck.ea2026}, such an X-shape asymmetry could be related to precession, and a similar asymmetry in scattered light variable with time has been tracked by \citep{Cook.Tobin.ea2019} and attributed to inner disk misalignment. 

The ionized jet is, however, detected for the first time in mid-IR tracers [Ar II], [Ne II], and [Ne III] \citep{Devaraj.vanDishoeck.ea2026}, where it is detected as a distinct high-velocity component, exclusively on the eastern side of the outflow. This is consistent with the result from \cite{BlakeDrechsler.Tobin.ea2026} on hydrogen recombination lines, which point to asymmetric accretion with a higher accretion luminosity on the eastern side, and OH emission, which is stronger to the east, indicative of a stronger UV-field in that direction. The presence of small carbon chains such as CCH and C$_4$H and along the southeast ridge of the outflow cavity (Fig.~\ref{fig:SEtail}) seems thus associated with the presence of UV-irradiated gas. In addition to the jet, broad, low-velocity emission is detected for the ions with a morphology similar to the wide-angle wind traced by H$_2$. The velocity offsets of the wind are consistent with the western side being blueshifted and the eastern side redshifted, consistent with, for example, a SiO blueshifted jet detected on the western side \citep{vantHoff_eDisk}. The emission of ions with high ionization potentials, such as [Ar II], [Ne II], and [Ne III], is detected equally bright on the northwest and southeast sides of the outflow, in contrast to tracers such as H$_2$CO and CS (Fig.\ref{fig:Outflow}). Although the C$_4$H emission on small angular scales peaks just too far west to fall within the MIRI FOV, the ion emission peaks further off source in the west than in the east, increasing in brightness toward the C$_4$H (Fig.~\ref{fig:JWST}). In contrast, neutral sulfur peaks close to the source in the western cavity and displays an anti-correlation with the C$_4$H emission. \citet{Devaraj.vanDishoeck.ea2026} interpreted the differences between the neutral and ionized species as the ionized material being further away from the source in the west, which again would suggest that C$_4$H is enhanced in ionized gas. 

Overall, JWST and ALMA observations are highly complementary, with JWST revealing a warm H$_2$ wind also observed with CO in ALMA, but also showing some distinct features like an ionized jet, missed by ALMA due to a lack of ionized and atomic lines at the sub-mm range. Higher-energy lines of CO and SiO in Bands 8 and 9 could possibly reveal if a hotter molecular component exists in the L1527 jet. JWST also provides support for energetic radiation contributing to the hydrocarbon emission observed with ALMA and provides additional clues on the geometry of the system, showing, through wind and jet emission, that the eastern cavity is facing away from the observer, contrary to what would be expected from the interpretation of scattered-light emission.

% -----------------------------------------------------------------
% Other young disks 
% -----------------------------------------------------------------

\subsection{Comparison with other young disks}

Recently, \citet{Sharma2025} presented the first systematic comparison of the chemical structure of a large number (19) of young disks at high angular resolution using data from the eDisk Large Program. This work compared emission from nine molecular species (including three CO isotopologues) toward 19 protostellar disks, including L1527. Although some species trace predominantly the same structure in all systems (such as the molecular outflow in $^{12}$CO), other species such as SO and H$_2$CO display a variety of morphologies that are, at least without an in-depth analysis of each source, not easily associated with a particular protostellar component. Therefore, it is not straightforward to say whether the chemical structure of L1527 is typical or not. Nevertheless, within the eDisk sample, L1527 is among the sources with the highest number of molecular line detections. The notable exception is CH$_3$OH, which is detected toward $\sim$50\% of Class 0 sources in the survey. Moreover, although c-C$_{3}$H$_2$ was frequently observed to trace the cavity wall, L1527 was not the only source with a dominant envelope contribution. 

Another large comparison, although not as uniform, was made by \citet{Tychoniec2021}, who compared the chemical composition of 16 protostars using different ALMA programs. Overall, they discussed 17 molecular species, including isotopologues and five complex organics. Their conclusions included that the protostellar envelope is well traced by C$^{18}$O, DCO$^+$ and N$_2$D$^+$. L1527 seems in that regard an exception as N$_2$D$^+$ has only been detected at a large angular offset. Furthermore, although L1527 has a tiny blueshifted jet \citep{vantHoff_eDisk}, it is only visible in SiO and not in other tracers such as SO and H$_2$CO. \citet{Tychoniec2021} observed CCH, c-C$_3$H$_2$ and CN to often trace the cavity wall. Among these, CN displays the cavity wall most prominent in L1527, although predominantly toward the southwest. Finally, H$_2$S, SO, and OCS were often detected in the inner envelope, while in L1527, H$_2$S is only seen in the outflow and OCS has only been tentatively detected \citep{Zhang2024}. 

Smaller surveys of protostellar disk chemistry, in terms of number of sources and molecular species, have been performed by \citet{ArturdelaVillarmois2019} and \citet{Garufi2021}. The first study revealed strong SO$_2$ emission in 5 out of 10 protostars in Ophiuchus, which all have luminosities $\gtrsim3L_{\odot}$. The slightly lower luminosity of L1527 may then explain why only weak SO$_2$ emission has been detected. The second study showed very similar distributions for H$_2$CO and CS toward six protostars in Taurus. As shown in Fig.~\ref{fig:SmallScale}, that is not the case in L1527, where CS emission originates in the envelope, while H$_2$CO has a strong disk contribution. On the other hand, \citet{Garufi2021} detected H$_2$CS toward only two of the six disks, so the nondetection of H$_2$CS in the L1527 disk is not outstanding. 

Overall, the limited number of studies of the protostellar disk chemical structure revealed clear commonalities and trends for some molecular species, but also highly diverse morphologies for other molecules. So far, L1527 does not stand out as either very typical nor very different, but a large enough sample both in terms of sources and molecules, quantitative measurements and a detailed enough understanding of the protostellar chemistry are probably still lacking to make more precise and nuanced statements about the diversity of the chemical conditions in these systems.

% -----------------------------------------------------------------
% Protoplanetary disks 
% -----------------------------------------------------------------

\subsection{Comparison with protoplanetary disks}

In a recent review, \citet{Oberg2023} compiled an overview of disk-averaged column densities (inner 100 au) for nine small molecules (CO, HCO$^+$, N$_2$H$^+$, CN, HCN, DCN, HC$_3$N, CCH and H$_2$CO) in Class II protoplanetary disks. Figure~\ref{fig:Columndensities} presents a comparison of those results with the column densities toward L1527. The total CO column in the L1527 disk derived from C$^{17}$O falls on the high end of the more than two orders of magnitude spread of CO columns in the Class II disks. This would be in alignment with CO being a factor 10--100 depleted in more mature disks, while its abundance is closer to canonical in younger disks \citep{Bergner2019,vantHoff2020,Zhang2020,Zhang2021}. The H$_2$CO column in the L1527 disk also falls on the high end of the Class II range, while the HCN column over the inner 200 au in L1527 lies just below the columns derived for Class II disks. When looking at the radial column density profiles in the individual five disks observed with the MAPS Large Program \citep{Guzman2021}, the HCN column in the inner 100 au is either similar to or larger than the H$_2$CO column by up to two orders of magnitude. In the outer disk, the HCN column typically drops below the H$_2$CO column. In L1527, the spatially averaged column of H$_2$CO is $\sim$10 times higher than that of HCN. Although a more in-depth analysis and potentially deeper observations are required to confirm this, the current nondetection of HCN at the highest velocities (in contrast to H$_2$CO) would be consistent with a different distribution of these two species in L1527 compared to (the MAPS) Class II disks.   

The HCO$^+$ and CCH columns extracted over the inner 200 au are very similar to the columns in Class II disks, while CN, DCN, N$_2$H$^+$, and HC$_3$N are only detected in the L1527 envelope. For species with detections on small and large scales, the column densities are typically a factor of a few smaller for the disk, so under that assumption, the L1527 column densities of DCN, N$_2$H$^+$, and HC$_3$N appear similar to those in Class II disks, while CN appears to be lower in L1527. The CH$_3$CN upper limit on disk scales is on the high end of the Class II column densities, so deeper observations are required to make a meaningful comparison. 

One of the results of the MAPS Large Program was that the inner 100 au was rich in organics such as HC$_3$N and CH$_3$CN \citep{Ilee2021}. Toward L1527, HC$_3$N is only detected in the envelope and CH$_3$CN is not detected. This may be consistent with the scenario outlined by \citet{Ilee2021}, who show chemical models that include dynamical processes such as radial drift and vertical mixing are required to explain the high column densities. In that view, L1527 may not be old enough for a large enough supply of organics to have been delivered to the inner disk.  

Another difference compared to Class II disks, is that those mature disks are typically observed to be bright in CS emission, while SO detections are rarer, pointing to these disks being dominated by a carbon-rich chemistry (C/O $>$ 1; e.g., \citealt{Dutrey2011,Semenov2018,LeGal2021,Riviere-Marichalar2026}), although this is not a uniform result \citep{Huang2024} and spatially resolved observations have also found changes in the C/O ratio throughout the disk \citep{Booth2023,Keyte2023}. Even though the CS column density derived for the inner 200 au toward L1527 is similar to that found in Class II disks ($10^{12}-10^{13}$ cm$^{-2}$; \citealt{Law2026}), the SO column is an order of magnitude higher. For comparison, lower limits for the CS/SO column density ratios of $> 4-14$ were found for the disks in the MAPS sample \citep{LeGal2021}. In L1527, SO emission has also been detected out to higher velocity offsets, while CS extends out to larger scales in the envelope.  

Overall, the composition of Class II disks often points to a carbon-rich chemistry based on CS/SO as discussed above and/or the abundance of hydrocarbons such as CCH \citep{Bergin2016,Miotello2019,Bosman2021}. In contrast, the L1527 disk appears more oxygen-rich, with high columns of SO and H$_2$CO, while CS and hydrocarbons are more prominent in the envelope. More in-depth analysis is required in combination with deeper observations to constrain column densities of more species on disk-scales, but these results may point to a transition from a carbon-dominated chemistry in the envelope to a more oxygen-rich chemistry in the young disk.

% ==================================================================
% CONCLUSION
% ==================================================================

\section{Conclusion} \label{sec:Conclusion}

We have presented an overview of molecular line emission toward the young Class 0 protostar L1527 based on 33 publicly available ALMA programs with spectral windows in FDM mode, and calculated column densities for the detected species. The main results can be summarized as follows: 

\begin{itemize}
    \item In total, 40 molecules are detected, including 23 unique molecular species. Including isotopologues, 29 species are reported here for the first time in ALMA observations. The largest detected molecules are CH$_3$CCH, CH$_3$OH and l-C$_4$H$_2$. Notable nondetections include OCS, H$_2$CS, HDO and SiO, although OCS and SiO have been reported in datasets that combine observations in different configurations. Upper energy levels of detected transitions range between 4 and 148 K, with a mean value of 30 K. 
    
    \item In general, nitrogen-bearing molecules are predominantly detected on more extended envelope scales, potentially because existing observations are not equipped to detect nitrogen-bearing species on small scales where the emitting area is small given the cosmically lower abundance of nitrogen compared to carbon and oxygen. Sulfur-bearing molecules are also sparse on small scales, and instead often detected in the western outflow. Hydrocarbons display a distinct tail along the southeastern outflow cavity wall due to the high levels of UV irradiation in the eastern outflow lobe. For other species, the outflow cavity is brightest toward the southwest. 

    \item Many species display highly non-symmetric emission morphologies for the outflow, cavity wall and envelope components, indicating that the conditions in protostellar systems are far from spherically symmetric, even without streamers. 
        
    \item Most observations are not sensitive enough to detect emission without an envelope contribution based on kinematics, resulting in, so far, only $^{13}$CO, C$^{18}$O, C$^{17}$O, HCO$^+$, SO, H$_2$CO and HDCO established to unambiguously trace the disk. 

    \item Overall, the envelope appears characterized by a carbon-rich chemistry, which seems to transition into an oxygen-dominated chemistry in the disk. 
    
    \item The most important factor in deriving column densities is knowledge of the emitting area from spatially resolved observations. Using the emitting area of C$^{18}$O, column densities derived for a single species typically vary less than a factor of five between different datasets. 
 
    \item Confirming previous results, there is no evidence for a strong depletion of CO as typically observed toward Class II protoplanetary disks. 

\end{itemize}

Taken together, this work highlights the need for broadband observations at high angular and spectral resolution as well as high sensitivity in order to characterize the chemistry in embedded disks. Once such observations become available for more protostellar systems, we will be able to establish whether the characteristics of L1527 are typical for young disks. This will then allow an in-depth comparison with mature protoplanetary disks to constrain how the composition of the planet-forming material changes as planets are being formed.

\section*{Conflict of Interest Statement}
%All financial, commercial or other relationships that might be perceived by the academic community as representing a potential conflict of interest must be disclosed. If no such relationship exists, authors will be asked to confirm the following statement: 

The authors declare that the research was conducted in the absence of any commercial or financial relationships that could be construed as a potential conflict of interest.

\section*{Author Contributions}

MLRH performed the analysis of ALMA observations and wrote the manuscript. {\L}T performed the comparison between ALMA and JWST results and wrote the respective section of the manuscript. JJT and DH assisted with the data reduction of the ALMA data from early cycles, and contributed to the ALMA proposal led by MLRH for which the data are still proprietary. All authors provided input on the manuscript.   

%The Author Contributions section is mandatory for all articles, including articles by sole authors. If an appropriate statement is not provided on submission, a standard one will be inserted during the production process. The Author Contributions statement must describe the contributions of individual authors referred to by their initials and, in doing so, all authors agree to be accountable for the content of the work. Please see  \href{https://www.frontiersin.org/about/policies-and-publication-ethics#AuthorshipAuthorResponsibilities}{here} for full authorship criteria.

\section*{Funding}
Part of this work was performed at Leiden Observatory, Leiden University. Astrochemistry in Leiden is supported by the Netherlands Research School for Astronomy (NOVA). M.L.R.H. acknowledges support from a Huygens Fellowship from Leiden University for early phases of this work.
%Details of all funding sources should be provided, including grant numbers if applicable. Please ensure to add all necessary funding information, as after publication this is no longer possible.

\section*{Acknowledgments}
The authors would like to thank Ewine van Dishoeck for many insightful discussions, and the anonymous referees for their supportive reviews. This paper makes use of the following ALMA data: ADS/JAO.ALMA\#2011.0.00210.S, ADS/JAO.ALMA\#2011.0.00604.S, ADS/JAO.ALMA\#2012.1.00193.S, ADS/JAO.ALMA\#2012.1.00346.S, ADS/JAO.ALMA\#2012.1.00647.S, ADS/JAO.ALMA\#2013.1.00858.S, ADS/JAO.ALMA\#2013.1.01086.S, ADS/JAO.ALMA\#2013.1.01331.S, ADS/JAO.ALMA\#2015.1.00261.S, ADS/JAO.ALMA\#2015.1.01549.S, ADS/JAO.ALMA\#2016.1.01203.S, ADS/JAO.ALMA\#2016.1.01245.S, ADS/JAO.ALMA\#2016.1.01541.S, ADS/JAO.ALMA\#2016.2.00117.S, ADS/JAO.ALMA\#2016.2.00171.S, ADS/JAO.ALMA\#2016.A.00011.S, ADS/JAO.ALMA\#2017.1.00509.S, ADS/JAO.ALMA\#2017.1.01375.S, ADS/JAO.ALMA\#2017.1.01413.S, ADS/JAO.ALMA\#2018.1.00375.S, ADS/JAO.ALMA\#2018.1.00799.S, ADS/JAO.ALMA\#2018.1.01205.L, ADS/JAO.ALMA\#2019.1.00261.L, ADS/JAO.ALMA\#2019.1.00847.S, ADS/JAO.ALMA\#2019.1.01063.S, ADS/JAO.ALMA\#2019.1.01695.S, ADS/JAO.ALMA\#2019.A.00034.S, ADS/JAO.ALMA\#2021.1.00536.S, ADS/JAO.ALMA\#2022.1.00131.S, ADS/JAO.ALMA\#2022.1.01357.S, ADS/JAO.ALMA\#2023.1.00169.S, ADS/JAO.ALMA\#2023.1.00439.S, and ADS/JAO.ALMA\#2023.1.00592.S. ALMA is a partnership of ESO (representing its member states), NSF (USA) and NINS (Japan), together with NRC (Canada), MOST and ASIAA (Taiwan), and KASI (Republic of Korea), in cooperation with the Republic of Chile. The Joint ALMA Observatory is operated by ESO, AUI/NRAO and NAOJ. The National Radio Astronomy Observatory is a facility of the National Science Foundation operated under cooperative agreement by Associated Universities, Inc. This work is based on observations made with the NASA/ESA/CSA James Webb Space Telescope. The data were obtained from the Mikulski Archive for Space Telescopes at the Space Telescope Science Institute, which is operated by the Association of Universities for Research in Astronomy, Inc., under NASA contract NAS 5-03127 for JWST. These observations are associated with programs PID 1290 (can be obtained from https://doi.org/10.17909/5dw8-cs47) and PID 2739.

% \section*{Supplemental Data}
%  \href{http://home.frontiersin.org/about/author-guidelines#SupplementaryMaterial}{Supplementary Material} should be uploaded separately on submission, if there are Supplementary Figures, please include the caption in the same file as the figure. LaTeX Supplementary Material templates can be found in the Frontiers LaTeX folder.

\section*{Data Availability Statement}
The datasets analyzed for this study can be found in the ALMA archive (product images and ARI-L images) and NRAO archive (product images and AUDI images). The JWST data can be downloaded from the MAST archive.

% Please see the availability of data guidelines for more information, at https://www.frontiersin.org/about/author-guidelines#AvailabilityofData

\bibliographystyle{Frontiers-Harvard} %  Many Frontiers journals use the Harvard referencing system (Author-date), to find the style and resources for the journal you are submitting to: https://zendesk.frontiersin.org/hc/en-us/articles/360017860337-Frontiers-Reference-Styles-by-Journal. For Humanities and Social Sciences articles please include page numbers in the in-text citations 
\bibliography{bibliography}

%%% Make sure to upload the bib file along with the tex file and PDF
%%% Please see the test.bib file for some examples of references

%\section*{Figure captions}

%%% Please be aware that for original research articles we only permit a combined number of 15 figures and tables, one figure with multiple subfigures will count as only one figure.
%%% Use this if adding the figures directly in the mansucript, if so, please remember to also upload the files when submitting your article
%%% There is no need for adding the file termination, as long as you indicate where the file is saved. In the examples below the files (logo1.eps and logos.eps) are in the Frontiers LaTeX folder
%%% If using *.tif files convert them to .jpg or .png
%%%  NB logo1.eps is required in the path in order to correctly compile front page header %%%

% \begin{figure}[h!]
% \begin{center}
% \includegraphics[width=10cm]{logo1}% This is a *.eps file
% \end{center}
% \caption{ Enter the caption for your figure here.  Repeat as  necessary for each of your figures}\label{fig:1}
% \end{figure}

%%% If you don't add the figures in the LaTeX files, please upload them when submitting the article.
%%% Frontiers will add the figures at the end of the provisional pdf automatically
%%% The use of LaTeX coding to draw Diagrams/Figures/Structures should be avoided. They should be external callouts including graphics.

\newpage

\section*{Supplementary Materials}

\section{ALMA data}

%A brief overview of the ALMA datasets used in this work is listed in Table~\ref{tab:ALMAprograms}. The angular resolution, total frequency range and integration time are extracted directly from the ALMA Archive. 

\begin{figure}
    \centering
    \includegraphics[width=\linewidth,trim={0cm 11cm 0cm 1cm},clip]{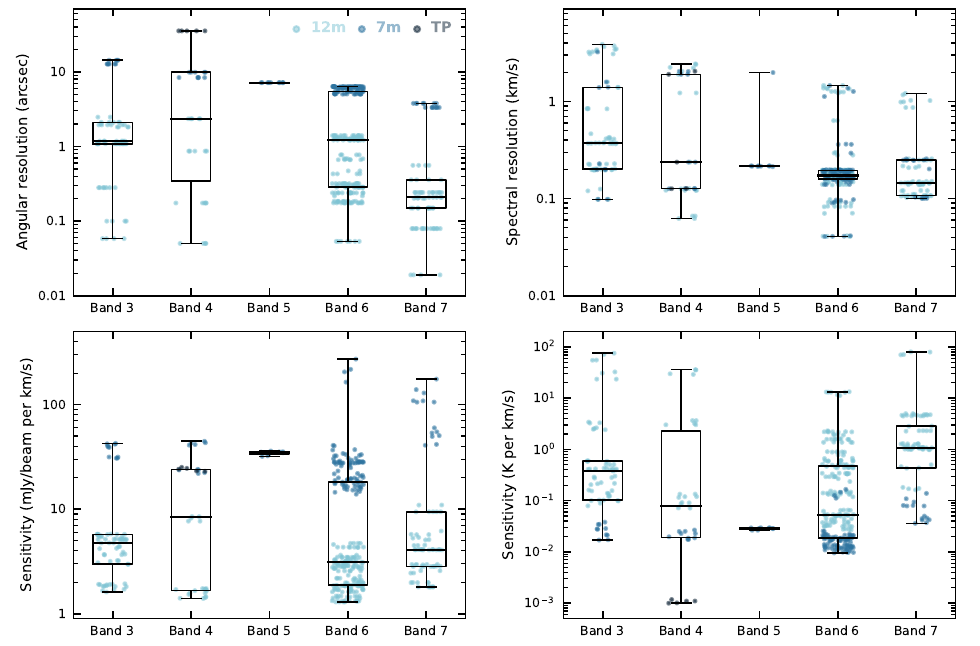}
    \caption{Distribution of angular resolution (top left), spectral resolution (top right), point-source sensitivity (bottom left) and surface-brightness sensitivity (bottom right) among the spectral windows used in this work per ALMA Band. The boxes extend from the first quartile to the third quartile of the data, with a line at the median. The whiskers extend from the box to the farthest data points. All data points are plotted as scattered circles, with light blue circles for 12m observations, blue circles for 7m observations and dark blue circles for Total Power (TP) observations (only present in Band 4).  }
    \label{fig:spwProperties}
\end{figure}

An overview of the distribution of the angular resolution, velocity resolution, point-source sensitivity and surface-brightness sensitivity across spectral windows for each ALMA Band is presented in Fig.~\ref{fig:spwProperties}.

\section{Observed molecular lines and column densities}

An overview of all detected molecular transitions and their column densities per dataset is listed in Table~\ref{tab:columndensity_full}. Transitions used to derive upper limits are included as well. The median column density for each molecular species is listed in Table~\ref{tab:columndensity_summary_disk_inner200} for disk and inner 200 au scales, and in Table~\ref{tab:columndensity_summary_envelope_extended} for envelope and extended envelope scales. A brief description of the available observations for each molecule is provided below.

% --- CO -------------------------------------------------------------

\subsection{CO isotopologues}

$^{12}$CO -- For the main isotopologue, the $J=1-0$ and $J=2-1$ transitions have been observed, where the $J=2-1$ transition is included in many programs covering resolutions ranging between 0.05$''$ and 6.1$''$. The $^{12}$CO emission is dominated by the east-west outflow and sharply outlines the outflow cavity wall. The western outflow is brighter than the eastern, and on both sides the emission peaks about 20$''$ from the protostar. This has two clear effects on the observed morphology depending on the spatial resolution, size of the primary beam, and largest angular scale. At low spatial resolution (and large field of view), this manifests as a ``bow-tie'' morphology with a faint east-west elongated central component with the bright semi-conical outflows at $\sim$6$''$ offsets, giving the impression of a narrow opening angle at small scales and a larger opening angle at larger scales. In high-resolution images with a smaller field of view, this results in the emission being dominated on scales of ~1$''$.

$^{13}$CO and C$^{18}$O -- For $^{13}$CO and C$^{18}$O, the $J=2-1$ and $J=3-2$ transitions have been observed by several programs, spanning angular resolutions between 0.05$''$ and 6.2$''$. As discussed in detail by van 't Hoff et al. (2023), the $J=2-1$ emission from both isotopologues is dominated by the north-south oriented disk and inner envelope, but they also show the cavity wall and outflow, including outward-moving shells that are not visible for $^{12}$CO. While the outflow emission is more pronounced for $^{13}$CO, the envelope is traced out to larger angular offsets by C$^{18}$O ($\sim3-4''$). The envelope emission for $^{13}$CO in particular appears asymmetric, with bright emission extending further to the south ($\sim$2$''$) than to the north ($\sim$1$''$). van 't Hoff et al. (2023) postulated that the X-shaped morphology of C$^{18}$O is due to CO freeze out in the deeper layers, resulting in emission only from the warmer surface layers. For both isotopologues, the contribution from larger-scale emission diminishes for the higher energy $J=3-2$ transition. However, these observations have a smaller FOV and more shallow observations of the $J=2-1$ lines also lack larger-scale emission, so this may not solely be an excitation effect.

% .....Figure Dataset statistics .........................
% ........................................................

As detailed by van 't Hoff et al. (2023), the difference between the $^{13}$CO and C$^{18}$O $J=2-1$ peak intensity maps is due to their different abundances. Due to its higher abundance, emission from $^{13}$CO is resolved out near the systemic velocity, while C$^{18}$O emission is detected in all low-velocity channels, resulting in emission from components with velocities close to systemic to be only visible in C$^{18}$O. In addition, $^{13}$CO has a stronger contribution from the inner envelope, while C$^{18}$O traces emission from the envelope out to larger scales where the $^{13}$CO emission already becomes optically thick and resolved out. This may also explain why the redshifted $^{13}$CO emission in the north is less extended along the major axis compared to the blueshifted emission in the south, as for the redshifted emission the outer envelope is in front of the disk along our line of sight, while the disk is in front of the envelope on the blueshifted side. 

C$^{17}$O -- For C$^{17}$O, the $J=1-0$ and $J=2-1$ transitions are each observed once. The weaker $J=1-0$ transition (even at higher sensitivity) displays solely compact emission, which most likely originates from the disk (i.e., highest density regions). In the $J=2-1$ transition an additional weak larger-scale envelope component is visible from northeast to southwest, which corresponds to the brightest regions in C$^{18}$O at low velocities ($v_{\rm{sys}} \pm 0.5$ km $^{-1}$). The different morphology of C$^{17}$O and C$^{18}$O can also be explained by abundance; due to its lower abundance, C$^{17}$O does not become optically thick as easily as C$^{18}$O and therefore shows better the larger scale structure at the lowest velocity offsets, while at the same time being only abundant enough to show the highest density regions on small scales. 

$^{13}$C$^{18}$O -- Only the $J=1-0$ transition has been observed for $^{13}$C$^{18}$O, albeit by several programs with high (0.06$''$), medium (2.0$''$) and low resolution (13$''$). However, the sensitivity is never high enough for a detection. 

% --- N2H+ and N2D+ -----------------------------------------------------

\subsection{N$_2$H$^+$ and N$_2$D$^+$}

N$_2$H$^+$ -- The N$_2$H$^+$ $J=1-0$ transition has been observed by several programs, but has only been detected at resolutions $> 1''$. The emission is elongated in the north-south direction, extending up to more than 50$''$ off source in mosaic observations. The emission peaks $\sim 8''$ off-source in the south and $\sim 10''$ off-source in the north, with only faint emission at smaller offsets. Emission is also present to the east and west at offsets of $>25''$. Over the inner region (1.8$''$ radius), the N$_2$H$^+$ hyperfine lines are very narrow with a FWHM of $\sim$0.2 km s$^{-1}$, compared to C$^{18}$O with a FWHM of $\sim$2 km s$^{-1}$. 

Since N$_2$H$^+$ is efficiently destroyed by CO gas, N$_2$H$^+$ is expected to only be present when CO is frozen out (Qi et al. 2013). Comparing the N$_2$H$^+$ emission to that of C$^{18}$O shows indeed an anti-correlation, confirming that the structure seen in C$^{18}$O is due to CO freeze out. This anti-correlation is most clearly seen in the south, probably because the redshifted C$^{18}$O emission in the north with the outer envelope closer to us is more affected by interferometric effects. The faint N$_2$H$^+$ emission in the inner $\sim8''$ is most likely coming from the outer envelope and projected at small angular offsets from the source. 

N$_2$D$^+$ -- For N$_2$D$^+$, both the $J=2-1$ and $J=3-2$ transitions are covered by numerous programs, but emission is only detected in Total Power observations (Fig.~\ref{fig:Spectra}). The N$_2$D$^+$ emission clearly peaks at least 60$''$ north of the source (at the edge of the FOV), consistent with IRAM 30m observations presented by Tobin et al. (2013). While the IRAM 30m observations show a clear offset between N$_2$D$^+$ and N$_2$H$^+$, with N$_2$H$^+$ peaking only 35$''$ north of the source, the 2$''$ ALMA mosaic displays a complex N$_2$H$^+$ emission morphology on large scales, so the details of the relative distributions remain unclear without higher resolution N$_2$D$^+$ images.

% --- HCO+  -------------------------------------------------------------

\subsection{HCO$^+$ isotopologues}

HCO$^+$ -- For the main isotopologue, only the $J=4-3$ transition has been observed (by a single program). Due to emission being resolved out over the central 1 km s$^{-1}$, we mainly see the disk and inner envelope. 

H$^{13}$CO$^+$ -- The $^{13}$C isotopologue has been observed more often and in both $J=1-0$ and $J=3-2$, although the $J=1-0$ transition has not been detected. Due to its lower abundance, H$^{13}$CO$^+$ suffers less from resolved-out emission and traces the same regions as C$^{18}$O at low velocities, including the envelope and outflow cavity. The X-shape seen in C$^{18}$O due to freeze out in the midplane is not present, as H$^{13}$CO$^+$ is not absent in the midplane, but its emission is less extended both radially and vertically. Finally, while the C$^{18}$O brightness is pretty uniformly distributed spatially outside of the bright inner $\sim1''$, H$^{13}$CO$^+$ is brightest at large angular offsets. Despite tracing the same regions, these differences make the peak intensity maps look quite distinct. In ACA observations, unresolved emission is also present in the outflow cavity. 

HC$^{18}$O$^+$ -- For HC$^{18}$O$^+$, $J=1-0$, $2-1$, $3-2$ and $4-3$ are observed, but only $J=2-1$ and $J=3-2$ are detected. Emission is only visible in images at low spatial resolution (5-7$''$), but it appears similar in distribution to H$^{13}$CO$^+$, with the exception that the HC$^{18}$O$^+$ emission more clearly peaks to the north and south of the source. 

DCO$^+$ -- The DCO$^+$ $J=2-1$ and $J=3-2$ transitions are covered by several datasets over a large range of resolutions. The 0.3$''$ resolution images show that DCO$^+$ traces overall the same regions as H$^{13}$CO$^+$, except that the outflow cavity is not visible and the emission does not extend to as high velocities. At 1-2$''$ resolution, DCO$^+$ is clearly brightest in the southernmost part ($\sim8''$ off source). 

% --- CN -------------------------------------------------------------

\subsection{CN}

For CN, hyperfine transitions from $N=1-0$, $N=2-1$, and $N=3-2$ are detected. The emission predominantly traces outflowing material and the southwestern cavity wall. As for $^{12}$CO, the western outflow is brighter than the eastern outflow. The brightest hyperfine lines also display a compact component close to the source along the major axis. This emission is detected at intermediate velocity offsets and thus likely originates in the inner envelope or maybe the outer disk. The inner envelope contribution appears to increase relative to the outflow and cavity wall components for higher energy transitions, but this may be, at least in part, related to the smaller primary beam at higher frequencies. The $^{13}$C and $^{15}$N isotopologues are not detected.

% --- HCN -------------------------------------------------------------

\subsection{HCN isotopologues}

HCN -- Only the $J=4-3$ transition has been observed (only once) for the main isotopologue. The line is only weakly detected, and displays strong redshifted absorption and very compact emission ($\lesssim1''$), suggesting that we are only seeing emission from the inner envelope or maybe the outer disk. 

H$^{13}$CN and HC$^{15}$N -- For H$^{13}$CN, the $J=3-2$ and $J=4-3$ transitions are observed and for HC$^{15}$N the $J=1-0$ is also observed. In both cases, only the $J=3-2$ transition has been detected and emission is only visible in the ACA images (5$''$ resolution), although spatially integrated emission is detected down to 0.2$''$ resolution. The images display two unresolved emission peaks at $\sim7''$ north and $\sim8''$ south of the source position, coinciding with the N$_2$H$^+$ peak in the south. H$^{13}$CN also shows unresolved blueshifted emission in the western outflow cavity $\sim14''$ off source. It is hard to constrain in the ACA data whether a compact component as seen for the main isotopologue is present, but it seems that the extended envelope, which is resolved out for HCN, is visible with the isotopologues, while HCN itself is abundant enough to show the more compact emission on smaller scales and at higher velocities.

DCN -- For DCN, the $J=2-1$ and $J=3-2$ transitions are detected, while the $J=4-3$ transition (only observed at 0.15$''$ resolution) is not. Although only weakly visible at intermediate resolution (2.3$''$), the DCN emission clearly extends from the northern H$^{13}$CN/HC$^{15}$N peak to the southern peak, which may reflect a higher abundance of DCN. Unlike the carbon and nitrogen isotopologues, the southern tip is much brighter than the northern peak. 
 
% --- HNC -------------------------------------------------------------

\subsection{HNC isotopologues}

The main isotopologue ($J=1-0$) is only covered by one program at very high resolution (0.06$''$) and not detected. HN$^{13}$C $J=3-2$ is weakly detected at 5$''$ resolution and has a similar distribution as the HCN isotopologues. The $J=2-1$ transition is present in absorption against the continuum in 0.1$''$ observations. DNC is only observed in the $J=2-1$ transition at very high angular resolution (0.04$''$), where it is detected in absorption against the continuum. The HNC isotopologues thus trace the large-scale envelope.

% --- Cyanopolyynes ----------------------------------------------------

\subsection{Cyanopolyynes}

HC$_3$N -- For HC$_3$N, the $J=10-9$, $J=11-10$, $J=17-16$, $J=24-23$, $J=27-26$ and $J=29-18$ transitions have been covered, but only the first three, with $E_{\rm{up}} < 70$ K, are detected. The lines are narrow (FWHM $\sim$0.8 km s$^{-1}$), and in contrast to most other species, the emission is strongest around the systemic velocity and thus most likely originates in the (extended) envelope. The peak intensity map shows two bright lanes of emission in the north-south direction; one just west of the source position, and the other at a larger offset ($\sim5''$) to the east. A weak tail of emission along the southeastern outflow cavity is also visible. 

DC$_3$N -- One transition ($J=10-9$) is detected for DC$_3$N, but only when spatially integrating over a few arcseconds. The emission is strongest south of the protostar.   

HC$_5$N -- For HC$_5$N, the $J=35-34$ transition is detected. Despite the relatively high upper-level energy of 80 K, the weak emission traces the brightest regions seen in HC$_3$N and thus most likely originates in the envelope. 

% ---  CCH -------------------------------------------------------------

\subsection{CCH and isotopologues}

CCH -- For CCH, hyperfine lines from the $N=3-2$ and $N=4-3$ transitions are detected, while weaker lines from $N=1-0$ are not. The emission has a clear compact elongated component originating in the inner envelope and maybe the outer disk. Emission from the cavity wall, outflowing shells and material further out in the cavity is also present. At lower angular resolution, a tail extending along the southeastern cavity wall becomes the most striking feature. 

$^{13}$CCH and C$^{13}$CH -- No $^{13}$CCH transitions are observed, and while several $N=1-0$, $N=2-1$ and $N=4-3$ transitions from C$^{13}$CH are covered, none are detected. This may have to do with the fact that the observations either have high angular resolution (0.1$''$ or 0.02$''$) or low velocity resolution (2 km s$^{-1}$); neither is favorable for detecting weak large-scale emission. 

CCD -- Hyperfine lines from the $N=2-1$ and $N=3-2$ transitions are detected for the deuterated isotopologue, while no $N=4-3$ transitions (only observed at 0.15$''$) are detected. The emission morphology is strikingly different from that of CCH at comparable resolution. Where CCH seems to trace the midplane of the envelope out to angular offsets of $\sim3''$, CCD does not have this compact component at intermediate velocities. Instead, the emission originates at larger vertical offsets from the midplane, likely the envelope surface layers and cavity wall, and extends out to angular offsets ($\sim10''$) in north and south direction. Comparing to the CCH ACA observations, the bright CCD regions anti-correlate with the bright CCH regions. In ACA observations of CCD, the emission is seen to extend even further to the southeast ($\sim35''$).

% --- c-C3H2 -------------------------------------------------------------

\subsection{c-C$_3$H$_2$ isotopologues}

c-C$_3$H$_2$ -- Multiple c-C$_3$H$_2$ transitions have been detected with upper-level energies ranging between 10 and 96 K. On scales of a few arcseconds, the emission morphology is similar to that of CCH with a vertically narrow envelope component. c-C$_3$H$_2$, though, shows a much stronger contribution from the envelope on larger scales, and fainter outflow emission. At intermediate angular resolution ($\sim1''$), c-C$_3$H$_2$ therefore more closely resembles HC$_3$N, although it is not as extended toward the east, potentially because c-C$_3$H$_2$ suffers more from resolved out emission. In ACA images, a southeast protrusion is visible. 

c-H$^{13}$CCCH -- Several c-H$^{13}$CCCH transitions are detected at $>1''$ resolution, although emission is only visible in images with $>5''$ resolution. The morphology resembles that of the main isotopologue at low resolution and the emission seems thus dominated by the larger-scale envelope. 

c-HCCCD -- Several transitions have been detected for the deuterated isotopologue, predominantly in low-resolution datasets. The lines are narrow ($\lesssim$ 1 km s$^{-1}$) and the emission is characterized by a strong contribution from the southeast tail extending about $\sim35''$ from the source. 

% --- Other hydrocarbons ------------------------------------------------

\subsection{Other hydrocarbons}

C$_4$H -- Multiple C$_4$H transitions are detected ($E_{\rm{up}}$ $\sim$ 25--80 K), but at a highest resolution of $\sim1''$. The lines appear narrow ($<$ 1 km s$^{-1}$) and the distinct morphology with two vertical lanes of emission sandwiching a dark midplane is created near the systemic velocity, suggesting an origin in the surface layers of the (larger-scale) envelope. The southeast tail is also clearly present. No isotopologues are detected. 

l-C$_3$H$_2$ -- Two transitions of l-C$_3$H$_2$ are detected in three datasets, but the emission is only visible in images with resolutions larger than a few arcseconds. All these datasets have low spectral resolution, but the narrow linewidth and emission morphology is consistent with an envelope component. The start of the southeast tail seen in other hydrocarbons is also present. 

l-C$_4$H$_2$ -- One transition is detected for l-C$_4$H$_2$ in one dataset at 8$''$ resolution. The narrow line is clearly detected in the spectrum, but appears only weakly in the image as a centrally unresolved blob. 

CH$_3$CCH -- The $K=0,1$ transitions are detected for the $J=10-9$ ladder, and the $K=0,1,2$ transitions are detected for the $J=6-5$ and $J=9-8$  ladders. The lines are very narrow (0.6 km s$^{-1}$) and show a bright region north and south of the source with very faint emission at angular offsets $< 3.5''$, indicating an origin in the extended envelope. 

% --- H2CO -------------------------------------------------------------

\subsection{H$_2$CO isotopologues}

H$_2$CO -- Multiple H$_2$CO transitions are detected at resolutions from 0.05$''$ to 6.2$''$ with upper-level energies ranging between 21 and 100 K. The emission is dominated by the disk and envelope, but there is also an outflow and extended envelope component. In datasets with lower surface-brightness sensitivity and/or smaller LAS, emission from the large-scale envelope is only weakly detected, and the disk and inner envelope dominate the emission morphology. At the highest velocities, the emission originates predominantly in the surface layers.  

H$_2^{13}$CO -- Three H$_2^{13}$CO transitions are covered, each by a different datasets, but none are detected. 

HDCO - For HDCO, two transitions from the $J=4-3$ ladder are observed and detected. The detailed distribution is hard to gauge because the strong 12m detections have low ($>$ 1 km s$^{-1}$) spectral resolution, but overall, the morphology is similar to that of H$_2$CO, although less extended radially and in particular vertically. HDCO also lacks an outflow component. At high angular resolution, the compact ($<1''$) component dominates and shows emission from surface layers. Emission is detected at high enough redshifted velocities in one dataset to confirm the presence of HDCO in the disk. 

D$_2$CO -- Several transitions from the $J=4-3$ and $J=6-5$ ladder are covered for the doubly deuterated isotopologue, but only three $J=4-3$ transitions are detected and the emission is mostly visible in only the $>1''$ resolution images. In contrast to the other isotopologues, there is only very faint compact emission and the emission seems to arise predominantly in the larger scale envelope, where the emission is brightest in the southernmost tip. This difference is very clear at high resolution, where D$_2$CO emission is still coming from large scales while HDCO only shows a compact component.

% --- CS -------------------------------------------------------------

\subsection{CS isotopologues}

CS -- For CS, the $J=3-2$, $J=5-4$ and $J=7-6$ transitions have been detected, while the $J=2-1$ transition (only observed at 0.1$''$) has not. The emission is dominated by the outflow, including outflowing shells, and the cavity walls (in particular in the southwest), but also has a more compact component. No emission is present at high enough velocities to confirm a disk origin. The relative strength of the compact component increases with higher $J$ transitions, but the $J=7-6$ transition has only been observed at high angular resolution resulting in a FOV too small to see the large-scale outflow. While there is some continuum oversubtraction at central velocities, there is emission present in all velocity channels. This points to a limited contribution from the extended envelope.

$^{13}$CS -- The $^{13}$CS isotopologue ($J=2-1$, $J=3-2$ and $J=5-4$) has been observed by multiple programs at angular resolutions ranging between 0.06$''$ and 14$''$, but has not been detected. 

C$^{34}$S -- The more abundant isotopologue C$^{34}$S has been observed twice, resulting in a detection of the $J=3-2$ transition at low resolution (8.4$''$) and a nondetection of the $J=6-5$ transition at high resolution (0.15$''$). The emission is dominated by the three bright unresolved peaks in the outflow, as well as the northeastern cavity wall at large angular offsets.

% --- SO -------------------------------------------------------------

\subsection{SO isotopologues}

SO -- Many SO transitions have been detected by many programs at resolutions ranging from 0.02$''$ to 6.22$''$. Detected upper-level energies range from 9 K to 81 K. The emission morphology is dominated by a compact component and detected at high enough velocities to constrain a disk contribution. All transitions observed at high-enough angular resolution resolve the compact component into a double lane structure on scales of $\lesssim 1-2''$. In addition, observations at high sensitivity show a strong contribution from the outflow cavity, which is most prominent in the south. Localized outflow emission is also present in these datasets. 

$^{34}$SO -- Two $^{34}$SO transitions are detected from the eight that have been observed, and both show unresolved (at 1$''$ resolution) compact emission at low velocities (probably due to low SNR), suggesting an envelope or potentially disk origin.

% --- Other S-bearing molecules -------------------------------------------------------------

\subsection{Other S-bearing molecules}

SO$_2$ -- While many transitions with $E_{\rm{up}}$ $<$ 100 K and $A_{ij}$ $> 10^{-4}$ s$^{-1}$ have been covered, only 1 transition (with $E_{\rm{up}}$ = 19 K and $A_{ij}$ = $7.7\times10^{-5}$ s$^{-1}$) has been detected. The emission is compact and shows a two-lane structure similar to SO. Given the weak detection, a disk origin cannot be ruled out. The $10_{1,9} - 10_{0,10}$ transition is tentatively detected in 1$''$ observations when spatially integrated over an 1.5$''$ region. 

H$_2$S -- For H$_2$S, two transitions have been observed (28 K and 84 K), of which only the lower energy one ($1_{1,0}-1_{0,1}$) has been detected. The emission displays two bright unresolved peaks (at 7.1$''$ resolution) in the western outflow cavity and a fainter peak in the eastern cavity. There is also a weak more central peak visible that is located $\sim3.7''$ south of the source position, and below the compact central component of $^{12}$CO seen at low resolution. This offset corresponds to the angular offset at which C$^{18}$O shows signs of freeze-out in the midplane. 

OCS -- Several OCS transitions are covered with $E_{\rm{up}}$ ranging between 16 and 237 K, but none are detected. Zhang et al. (2024) report a tentative detection in the combined datasets from the FAUST Large Program. 

CCS -- Emission from two transitions is weakly detected toward the brightest H$_2$S peak in the western outflow. 

H$_2$CS -- While many transitions have been covered at angular resolutions from 0.04$''$ to 14$''$, no H$_2$CS emission has been detected except for an unresolved peak about 1$'$ north of the source in a mosaic observation. The detected transition is $3_{1,2}-2_{1,1}$ with an upper-level energy of 23 K, and the emission is located where the western $^{12}$CO outflow cavity bends from a northwest direction to an almost pure west direction.

% --- H2O -------------------------------------------------------------

\subsection{H$_2$O isotopologues}

For water, only the HDO $3_{1,2}-2_{2,1}$ transition at 225 GHz has been observed. While this line is often strongly detected in protostellar envelopes, it is not detected here.

% --- SiO -------------------------------------------------------------

\subsection{SiO}

For SiO the $J=2-1$, $5-4$, $6-5$ and $7-6$ are observed, but only the $J=5-4$ transition is detected in the deep combined 0.05$''$ and 0.182$''$ observations from the eDisk Large Program program (van 't Hoff et al. 2023). The emission is unresolved and traces a compact jet component directly west of the source.

% --- HNCO -------------------------------------------------------------

\subsection{HNCO}

Of several transitions covered, only the $7_{0,7} - 6_{0,6}$ transition ($E_{\rm{up}}$ = 30 K) has been detected for HNCO. The Total Power observations display extended structure with emission peaking 18$''$ northwest of the source. The same transition is also detected in deep ACA observations, although there the emission is faintly visible $\sim8''$ south of the protostar, coinciding with the N$_2$H$^+$ peak. The higher energy (73 K) transitions $7_{1,7} - 6_{1,6}$ and $7_{1,6} - 6_{1,5}$ are not detected in the same datasets. Overall, this suggests that the HNCO emission originates in the larger-scale envelope.

% --- Complex organics ---------------------------------------------------

\subsection{Complex organics}

CH$_3$OH -- There are dozens of CH$_3$OH transitions covered that have $E_{\rm{up}}$ $<$ 100 K and $A_{ij}$ $> 10^{-5}$ s$^{-1}$, but only two transitions are weakly detected. The $4_{-0,4}-3_{1,3}$ E transition ($E_{\rm{up}}$ = 36 K) is detected at 0.6$''$ resolution, and displays 2-3$\sigma$ centrally peaked emission at intermediate velocities. The $4_{-2,3}-3_{-1,2}$ E transition ($E_{\rm{up}}$ = 45 K) is tentatively detected in ACA observations, but only when spatially integrated over a beam-sized area with a center $\sim5''$ southwest of the source position. Both transitions have similar parameters, and while the $4_{-0,4}-3_{1,3}$ E transition is expected to be $2-3$ times stronger at $20-200$ K, it is not clear why their spatial distribution should be different. Given that the $4_{-2,3}-3_{-1,2}$ E transition is not visible in individual velocity channels for the only 6-min integration ACA observations, we consider this detection more uncertain. Sakai et al. (2014) reported a 2$\sigma$ detection of the $7_{-1,7} - 6_{-1,6}$ E transition, but that is not clear from our images. 

Given the low SNR of the best detection, it is hard to constrain the spatial origin of the emission. Nevertheless, the centrally peaked nature of the emission, counterintuitively, rules out an inner disk origin as all molecules detected at high velocities show emission at an offset from the source position due to optically thick dust. The intermediate velocity range, compact size and relatively low upper-level energy of the $4_{-0,4}-3_{1,3}$ E transition of 45 K suggest an origin at moderate radii in the inner envelope and we only detect the largest column toward the protostar. Observations of CH$_3$OH transitions that will be stronger at temperatures of $20-200$ K than those currently covered (which fall in the range between 300 and 350 GHz) are required to constrain the CH$_3$OH distribution. 

CH$_3$CN -- Four $J$-ladders have been covered for the most abundant nitrogen-bearing complex organic, but no transitions are detected. Two transitions, however, are observed at $< 0.1''$ resolution. Only the $5_0-4_0$ transition has been observed at $> 1''$ resolution, but is also not detected in Total Power observations, although the integration time was only half that of the HNCO observations. Existing observations thus do not place strong constraints on the presence or absence of CH$_3$CN.

% ... Table All Column densities .........................
\clearpage
\input{Tables/Columndensity_Full}
\clearpage
% ........................................................

\clearpage
\input{Tables/Columndensity_Summary_SmallScale}
\clearpage

\clearpage
\input{Tables/Columndensity_Summary_LargeScale}
\clearpage

\end{document}

%% file: Tables/Dataset_overview.tex
% CHANGED THIS MANUALLY AFTER THE PYTHON SCRIPT
% Project code & PI & Band & Array & Resolution & Frequency & Integration & Imaging method \\
% & & & & ($''$) & (GHz) & (min) & \\

% Footnote also manually added 

{\scriptsize
\begin{longtable}{llp{0.5cm}p{1.1cm}p{1.5cm}p{3cm}p{1.5cm}l}
\caption{Overview of ALMA observations of L1527.} \label{tab:ALMAprograms} \\
\toprule
Project code & PI & Band & Array & Resolution & Frequency & Integration & Imaging method \\
& & & & ($''$) & (GHz) & (min) & \\
\midrule
\endfirsthead
\multicolumn{8}{c}{\textbf{Table \thetable}\ -- continued }\\
\toprule
Project code & PI & Band & Array & Resolution & Frequency & Integration & Imaging method \\
& & & & ($''$) & (GHz) & (min) & \\
\midrule
\endhead
\midrule
\multicolumn{8}{r}{Continued on next page} \\
\midrule
\endfoot
\bottomrule
\multicolumn{8}{l}{\textbf{Notes.} The listed frequency range is the start of the first and end of the last spectral window in FDM mode. The } \\
\multicolumn{8}{l}{exact frequency ranges covered by the spectral windows, and at what spectral resolution, are listed in the ALMA}\\
\multicolumn{8}{l}{Archive. An asterisk (*) in the ``Array'' column indicates mosaic obervations.}\\
\endlastfoot
2011.0.00210.S & Ohashi & 6 & 12m & 0.77 & 219.425--231.419 & 27.7 & manual \\
2011.0.00604.S & Sakai & 6 & 12m & 0.68 & 244.682--262.218 & 80.6 & manual \\
 &  & 7 & 12m & 0.56 & 338.086--351.978 & 70.6 & manual \\
2012.1.00193.S & Tobin & 6 & 12m & 0.22 & 258.948--260.309 & 122.5 & product \\
2012.1.00346.S & Evans & 7 & 12m & 0.36 & 342.776--356.861 & 10.6 & product, manual \\
2012.1.00647.S & Ohashi & 6 & 12m & 0.47 & 219.439--231.435 & 35.3 & product \\
2013.1.00858.S & Sakai & 6 & 12m & 0.24 & 244.902--264.294 & 39.8 & product, ARI-L \\
2013.1.01086.S & Koyamatsu & 6 & 12m & 0.18 & 219.528--231.347 & 24.2 & product, ARI-L \\
 &  & 6 & 12m & 0.66 & 219.528--231.347 & 24.2 & product, manual \\
2013.1.01331.S & Sakai & 6 & 12m & 1.32 & 216.323--235.216 & 20.2 & manual \\
 &  & 6 & ACA & 5.91 & 216.319--235.218 & 60.5 & manual \\
2015.1.00261.S & Ceccarelli & 3 & 12m & 2.47 & 84.397--88.194 & 35.3 & product, ARI-L \\
 &  & 3 & 12m & 2.00 & 90.859--91.094 & 16.6 & product, ARI-L \\
2015.1.01549.S & Ohashi & 7 & 12m & 0.20 & 328.971--340.739 & 14.1 & ARI-L \\
2016.1.01203.S & Oya & 4 & 12m & 0.17 & 137.879--150.525 & 42.0 & product, ARI-L \\
 &  & 4 & 12m & 0.87 & 137.879--150.525 & 36.3 & product \\
2016.1.01245.S & Cox & 3 & 12m & 2.10 & 99.241--114.338 & 15.1 & product, ARI-L \\
2016.1.01541.S & Harsono & 3 & 12m & 1.82 & 91.829--105.767 & 10.6 & product \\
 &  & 3 & ACA & 14.30 & 91.798--105.798 & 15.1 & product, ARI-L \\
 &  & 4 & 12m & 2.35 & 142.410--154.957 & 4.0 & product \\
 &  & 4 & ACA & 9.86 & 142.347--155.019 & 5.0 & product \\
 &  & 4 & TP & 35.35 & 142.347--155.019 & 35.3 & product \\
2016.2.00117.S & Yoshida & 6 & ACA & 5.04 & 230.471--249.297 & 449.1 & product \\
 &  & 7 & ACA & 3.33 & 351.636--364.304 & 46.4 & product \\
2016.2.00171.S & Harsono & 4 & ACA & 8.37 & 142.780--155.020 & 102.8 & product \\
2016.A.00011.S & Sakai & 7 & 12m & 0.08 & 337.729--351.988 & 59.9 & product, ARI-L \\
2017.1.00509.S & Sakai & 3 & 12m & 0.10 & 84.498--100.248 & 102.5 & AUDI \\
2017.1.01375.S & Yoshida & 6 & ACA & 4.74 & 259.495--262.287 & 8.6 & product, AUDI \\
2017.1.01413.S & van 't Hoff & 6 & 12m & 0.31 & 224.681--241.922 & 16.1 & product \\
2018.1.00375.S & Feng & 5 & ACA & 7.17 & 167.876--171.751 & 71.1 & product \\
2018.1.00799.S & Pineda & 3 & 12m * & 1.94 & 91.175--106.791 & 14.0 & product \\
 &  & 3 & ACA * & 12.72 & 91.173--106.793 & 87.3 & product \\
2018.1.01205.L & Yamamoto & 3 & 12m & 0.28 & 93.150--108.091 & 23.7 & product \\
 &  & 3 & 12m & 1.18 & 93.150--108.091 & 6.0 & product \\
 &  & 6 & 12m & 0.32 & 216.079--234.729 & 39.3 & product \\
 &  & 6 & 12m & 1.27 & 216.079--234.729 & 10.6 & product \\
 &  & 6 & ACA & 6.28 & 216.077--234.791 & 47.4 & product \\
 &  & 6 & 12m & 0.28 & 243.882--262.031 & 22.7 & product \\
 &  & 6 & 12m & 1.22 & 243.882--262.031 & 12.1 & product \\
 &  & 6 & ACA & 5.45 & 243.880--262.033 & 27.2 & product \\
2019.1.00261.L & Ohashi & 6 & 12m & 0.05 & 216.808--235.183 & 145.2 & product, AUDI \\
2019.1.00847.S & Sheehan & 7 & 12m & 0.21 & 329.218--343.122 & 13.3 & AUDI \\
 &  & 7 & ACA & 3.78 & 329.206--343.133 & 41.3 & AUDI \\
2019.1.01063.S & Sai & 6 & ACA * & 6.12 & 219.490--231.345 & 11.5 & product \\
2019.1.01695.S & Sakai & 4 & 12m & 0.04 & 137.962--153.732 & 79.2 & AUDI \\
 &  & 7 & 12m & 0.02 & 328.856--344.533 & 48.7 & AUDI \\
2019.A.00034.S & Tobin & 6 & 12m & 0.18 & 216.809--235.183 & 62.0 & product \\
2021.1.00536.S & Yen & 6 & 12m & 0.43 & 255.445--260.280 & 239.9 & product \\
 &  & 6 & ACA & 5.03 & 255.443--260.281 & 641.1 & product \\
2022.1.00131.S & Plunkett & 3 & 12m & 1.08 & 99.969--115.591 & 80.6 & product \\
 &  & 6 & ACA & 6.35 & 217.071--234.744 & 12.1 & product \\
2022.1.01357.S & Kwon & 3 & 12m & 0.06 & 89.580--105.413 & 67.1 & AUDI \\
2023.1.00169.S & van 't Hoff & 7 & 12m & 0.24 & 278.586--294.303 & 14.6 & product \\
2023.1.00439.S & van 't Hoff & 7 & 12m & 0.15 & 288.493--304.343 & 49.9 & product \\
2023.1.00592.S & van Gelder & 6 & 12m & 0.18 & 213.520--232.945 & 43.3 & product \\
 &  & 6 & 12m & 1.40 & 213.520--232.933 & 11.6 & product \\
 &  & 6 & ACA & 6.23 & 213.517--233.007 & 35.3 & product \\
\end{longtable}
}

%% file: Tables/Observations.tex
\begin{table}[p]
\centering
\tiny %\tiny
\setlength{\tabcolsep}{3pt}
\renewcommand{\arraystretch}{1.25}
\rotatebox{90}{%
\begin{minipage}{\textheight}
\centering
{\normalsize
\refstepcounter{table}\label{tab:Observations}
\textbf{Table~\thetable.} Overview of molecular line observations of L1527 obtained with ALMA.}
\vspace{0.8em}
\begin{tabular}{llllllll|llllllll}
\toprule
Molecule & $\leq$0.09$''$ & 0.09--0.4$''$ & 0.4--0.9$''$ & 1--2$''$ & 3--10$''$ & 10--15$''$ & $\sim$35$''$ & Molecule & $\leq$0.09$''$ & 0.09--0.4$''$ & 0.4--0.9$''$ & 1--2$''$ & 3--10$''$ & 10--15$''$ & $\sim$35$''$\\ 
    & (12m) & (12m) & (12m) & (12m) & (ACA) & (ACA) & (TP) & & (12m) & (12m) & (12m) & (12m) & (ACA) & (ACA) & (TP)\\
\midrule
CO & \checkmark & \checkmark~~(1,2,3) & \checkmark & \checkmark & \checkmark & --- & --- &
    SO & \checkmark & \checkmark (1,2,15,20,21) & \checkmark (5,16,19) & \checkmark & \checkmark & --- & --- \\
    
$^{13}$CO & \checkmark & \checkmark~~(1,2,4) & \checkmark & --- & \checkmark & --- & --- &
    $^{34}$SO & $\times$ & \checkmark & --- & \checkmark & $\times$ & --- & --- \\

C$^{18}$O & \checkmark & \checkmark~~(1,2,4) & \checkmark~~(5--11) & --- & \checkmark~~(11) & --- & --- &
    SO$_2$ & $\times$ & \checkmark (21) & $\times$ & \checkmark $^{\dagger}$ & $\times$ & --- & --- \\

C$^{17}$O & --- & \checkmark~~(12,13) & --- & \checkmark & --- & --- & --- &
    OCS & $\times$ & $\times$ & \checkmark $^{\star}$ (15) & $\times$ & $\times$ & --- & ---  \\
    
$^{13}$C$^{18}$O & $\times$ & --- & --- & $\times$ & --- & $\times$ & --- &
    CCS & $\times$ & $\times$ & --- & $\times$ & \checkmark & --- & $\times$ \\

HDO & --- & $\times$ (12) & --- & --- & --- & --- & --- & 
    H$_2$CS & $\times$ & $\times$ & $\times$ (15) & \checkmark & $\times$ & \checkmark & --- \\

HCO$^+$ & --- & --- & \checkmark~~(14) & --- & --- & --- & --- &
    CCH & \checkmark & \checkmark (20) & \checkmark (16) & \checkmark & \checkmark & --- & --- \\
    
H$^{13}$CO$^+$ & --- & \checkmark~~(14) & \checkmark & \checkmark & \checkmark & --- & --- &
    CCD & --- & $\times$ & --- & \checkmark & \checkmark & --- & \checkmark \\ 
    
HC$^{18}$O$^+$ & --- & $\times$ & \checkmark & --- & \checkmark & --- & --- &
    C$^{13}$CH & $\times$ & $\times$ & --- & --- & $\times$ & --- & --- \\
    
DCO$^+$ & --- & \checkmark & --- & \checkmark & \checkmark & --- & \checkmark &
     c-C$_3$H & --- & --- & --- & \checkmark & \checkmark & --- & --- \\
    
N$_2$H$^+$ & $\times$ & \checkmark & --- & \checkmark & --- & \checkmark & --- &
    c-C$_3$H$_2$ & \checkmark & \checkmark (1,2) & \checkmark (16,17,19) & \checkmark & \checkmark & --- & --- \\
    
N$_2$D$^+$ & --- & $\times$ & $\times$~~(4) & $\times$ & $\times$ & --- & \checkmark &
    c-HCCCD & $\times$ & $\times$ & --- & \checkmark & $\times$ & \checkmark & --- \\
    
CN & \checkmark & \checkmark~~(13) & --- & \checkmark & \checkmark & --- & --- & 
    c-H$^{13}$CCCH & $\times$ & $\times$ & --- & \checkmark $^{\dagger}$ & \checkmark & --- & \checkmark \\
    
HCN & --- & \checkmark & --- & --- & --- & --- & --- &
    c-HCC$^{13}$CH & $\times$ & $\times$ & $\times$ & $\times$ & $\times$ & --- & --- \\
    
H$^{13}$CN  & --- & \checkmark $^{\dagger}$ & \checkmark $^{\dagger}$ & \checkmark & \checkmark & --- & --- &
    C$_4$H & $\times$ & $\times$ & $\times$ & \checkmark & \checkmark & \checkmark & \checkmark \\

HC$^{15}$N & $\times$ & $\times$ & \checkmark $^{\dagger}$ & --- & \checkmark & --- & --- &
    l-C$_3$H$_2$ & $\times$ & $\times$ & --- & \checkmark $^{\dagger}$ & $\times$ & \checkmark & \checkmark \\
    
DCN & \checkmark $^{\ddagger}$ & \checkmark $^{\ddagger}$~~(1,2) & --- & \checkmark & \checkmark & --- & \checkmark &
    l-C$_4$H$_2$ & $\times$ & $\times$ & --- & $\times$ & \checkmark & --- & $\times$ \\
    
HNC & $\times$ & --- & --- & --- & --- & --- & ---  &
    CH$_3$CCH & $\times$ & $\times$ & --- & \checkmark & \checkmark & --- & \checkmark \\
    
HN$^{13}$C & --- & $\times$ & \checkmark & --- & \checkmark & --- & --- &
    H$_2$CO & \checkmark & \checkmark (1,2,12) & \checkmark (16) & \checkmark & \checkmark & --- & --- \\
    
DNC & \checkmark $^{\ddagger}$ & --- & --- & --- & --- & --- & --- &
    H$_2$$^{13}$CO & $\times$ & --- & $\times$ & --- & $\times$ & --- & --- \\
    
HC$_3$N & $\times$ & $\times$ & --- & \checkmark & \checkmark & --- & \checkmark &
    HDCO & --- & \checkmark & \checkmark & \checkmark & \checkmark & --- & --- \\
    
DC$_3$N & $\times$ & --- & --- & \checkmark $^{\dagger}$ & $\times$ & --- & $\times$ &
    D$_2$CO & $\times$ & \checkmark & $\times$ & \checkmark & \checkmark & --- & --- \\
    
HC$_5$N & $\times$ & $\times$ & --- & \checkmark & --- & \checkmark & --- &
    H$_2$CCO & --- & --- & --- & \checkmark & --- & --- & --- \\
    
CS & $\times$ & \checkmark~~(15) & \checkmark~~(16--18) & \checkmark & \checkmark & --- & --- &
    HNCO & $\times$ & $\times$ & --- & $\times$ & \checkmark & --- & \checkmark \\
    
$^{13}$CS & $\times$ & $\times$~~(15) & $\times$ & $\times$ & $\times$ & $\times$ & --- &
    CH$_3$OH & $\times$ & $\times$ (2,12) & \checkmark (16) & $\times$ & $\times$ & --- & $\times$ \\
    
C$^{34}$S & --- & $\times$ & --- & --- & \checkmark & --- & --- &
    CH$_3$CN & $\times$ & $\times$ & --- & $\times$ & --- & $\times$ & --- \\

H$_2$S & --- & $\times$ & --- & $\times$ & \checkmark & --- & --- &
    SiO & $\times$ & \checkmark $^{\star}$ (1,2) & --- & $\times$ & $\times$ & --- & --- \\
              
\bottomrule
{\bfseries Notes.} & \multicolumn{15}{l}{ A \checkmark indicates emission has been detected, a $\times$ indicates that a molecule has been observed but not detected, and --- indicates that a molecule is not} \\ 
& \multicolumn{15}{l}{observed on that angular scale. A $^{\dagger}$ indicates emission is only detected in a spatially integrated spectrum, a $^{\ddagger}$ marks detections in absorption against} \\
& \multicolumn{15}{l}{the continuum and a $^{\star}$ is used when emission has been presented in the literature in combined datasets. } \\ 

{\bfseries References.} & \multicolumn{15}{l}{ (1) \citealt{vantHoff_eDisk}, (2) \citealt{Sharma2025}, (3) \citealt{Liu2025}, (4) \citealt{vantHoff2018}, (5) \citealt{Ohashi2014}, (6) \citealt{Flores-Rivera2021},} \\
& \multicolumn{15}{l}{  (7) \citealt{Terebey2025}, (8) \citealt{Diep2016}, (9) \citealt{Tuan-Anh2016}, (10) \citealt{Aso2017}, (11) \citealt{Sai2023}, (12) \citealt{vantHoff2020}, } \\
& \multicolumn{15}{l}{ (13) \citealt{Tychoniec2021}, (14) \citealt{vantHoff2022}, (15) \citealt{Zhang2024}, (16) \citealt{Sakai2014}, (17) \citealt{Oya2015}, (18) \citealt{Oya2022},  }\\
& \multicolumn{15}{l}{  (19) \citealt{Sakai2014_Nature}, (20) \citealt{Sakai2017}, (21) \citealt{Liu2025}.}
 
\end{tabular}
\end{minipage}
}
\end{table}

%% file: Tables/Components.tex
\begin{table}[p]
\centering
\scriptsize
\setlength{\tabcolsep}{3pt}
\renewcommand{\arraystretch}{1.25}
% {\normalsize
% \refstepcounter{table}\label{tab:columndensity_summary}
% \textbf{Table~\thetable.} Column densities and upper limits toward L1527. Values are median $\pm$ the mean absolute deviation (MAD) \
%             with number of measurements in parentheses. For a single measurment, the listed uncertainty is the 1$\sigma$ uncertainty. \
%             For molecules that were not detected, the most stringent upper limit is listed.
% }
\caption{Molecular tracers of physical components in L1527.}
%\vspace{0.8em}
\begin{tabular}{lcccccc}
\toprule
Molecule & Disk & Inner Envelope & Extended Envelope & Cavity Wall & Outflow & Other \\
& $|\Delta v| \geq 2.55$ km s$^{-1}$ & $|\Delta v| \geq 0.5$ km s$^{-1}$ & $|\Delta v| < 0.5$ km s$^{-1}$ &\\
\midrule
CO              & $\times$ & $\times$ & $\times$ & --- & \checkmark\checkmark & ---  \\
$^{13}$CO       & \checkmark\checkmark & \checkmark\checkmark & $\times$ & \checkmark & \checkmark\checkmark & ---  \\
C$^{18}$O       & \checkmark\checkmark & \checkmark\checkmark & \checkmark\checkmark & \checkmark\checkmark & \checkmark & ---  \\
C$^{17}$O       & \checkmark\checkmark & \checkmark & \checkmark & --- & --- & ---  \\
HCO$^+$         & \checkmark\checkmark  & \checkmark\checkmark  & $\times$ & --- & --- & ---   \\
H$^{13}$CO$^+$  & --- & \checkmark\checkmark & $\times$\checkmark & --- & \checkmark & ---  \\
HC$^{18}$O$^+$  & --- & \checkmark & \checkmark\checkmark & --- & --- &   \\
DCO$^+$         & --- & \checkmark\checkmark & \checkmark\checkmark & ---- & --- & ---  \\
N$_2$H$^+$      & --- & --- & \checkmark\checkmark & --- & --- & ---  \\
N$_2$D$^+$      & --- & --- & --- & --- & --- & N offset  \\
CN              & --- & \checkmark & --- & \checkmark & \checkmark\checkmark & ---  \\
HCN             & --- & \checkmark\checkmark & --- & --- & --- & ---   \\
H$^{13}$CN      & --- & --- & \checkmark\checkmark & --- & \checkmark & ---  \\
HC$^{15}$N      & --- & --- & \checkmark\checkmark & --- & --- & ---  \\
DCN             & --- & --- & \checkmark\checkmark & --- & --- & ---  \\
HN$^{13}$C      & --- & --- & \checkmark\checkmark & --- & --- & ---  \\
DNC             & --- & --- & \checkmark & --- & --- & ---  \\
HC$_3$N         & --- & \checkmark & \checkmark\checkmark & --- & \checkmark & SE tail  \\
DC$_3$N         & --- & --- & \checkmark & --- & --- & ---  \\
HC$_5$N         & --- & --- & \checkmark\checkmark & --- & --- & ---  \\
CS              & --- & \checkmark\checkmark & --- & \checkmark & \checkmark & ---   \\
C$^{34}$S       & --- & --- & --- & \checkmark & \checkmark &  --- \\
H$_2$S          & --- & --- & --- & \checkmark & \checkmark &  --- \\
SO              & \checkmark & \checkmark\checkmark & --- & \checkmark & \checkmark & ---  \\
$^{34}$SO       & --- & \checkmark\checkmark & --- & --- & --- & ---   \\
SO$_2$          & --- & \checkmark\checkmark & --- & --- & --- & ---   \\
CCS             & --- & --- & --- & --- & \checkmark & ---  \\
H$_2$CS         & --- & --- & --- & --- & --- & N offset  \\
CCH             & --- & \checkmark\checkmark & $\times$ & \checkmark & \checkmark & SE tail  \\
CCD             & --- & --- & \checkmark\checkmark & \checkmark & -- & SE tail   \\

c-C$_3$H    & --- & --- & \checkmark & \checkmark & --- & SE tail  \\

c-C$_3$H$_2$    & --- & \checkmark\checkmark & \checkmark\checkmark & \checkmark & \checkmark & SE tail  \\
c-HCCCCD        & --- & --- & \checkmark\checkmark & --- & --- & SE tail \\
c-H$^{13}$CCCCH & --- & --- & \checkmark\checkmark & --- & --- & SE tail \\
C$_4$H          & --- & --- & \checkmark\checkmark & --- & --- & SE tail \\
l-C$_3$H$_2$      & --- & --- & \checkmark\checkmark & --- & --- & SE tail \\
l-C$_4$H$_2$    & --- & --- & \checkmark & --- & --- & --- \\
CH$_3$CCH       & --- & \checkmark\checkmark & \checkmark\checkmark & --- & --- & ---  \\
H$_2$CO         & \checkmark\checkmark & \checkmark\checkmark & \checkmark\checkmark & \checkmark & \checkmark & ---  \\
HDCO            & \checkmark & \checkmark\checkmark & \checkmark\checkmark & --- & --- & ---  \\
D$_2$CO         & --- & \checkmark & \checkmark\checkmark & --- & --- & ---   \\
H$_2$CCO        & --- & --- & \checkmark & --- & --- & --- \\
HNCO            & --- & --- & \checkmark & --- & --- & ---  \\
CH$_3$OH        & --- & \checkmark & --- & --- & --- & ---  \\           
\bottomrule
%\vspace{0.0001cm}
\end{tabular}
~~~~~~~~~~~~\textbf{Notes.} Checkmarks indicate that emission from this component has been detected. Two checkmarks (\checkmark\checkmark) are used to mark strong and dominant component(s) for a species, while one checkmark (\checkmark) is used for weak components.
A $\times$ indicates that a component is not detected due to emission being resolved out, and --- indicates a component is not detected.
\end{table}

%% file: Tables/Columndensity_Full.tex
%% Removed -12m1 etc from dataset manually 
%% Empty rows manually added so one transition is not split over two pages
%% 13C18O is somehow not included by the script - manually added
%% Manually remove transition name when multiple datasets are present. 
%% Some things do not go correctly with the script, seems when there is only extended envelope?

\begin{landscape}
\footnotesize
\setlength{\tabcolsep}{1.5pt}
\renewcommand{\arraystretch}{1.2}
\setlength{\LTleft}{-0.7cm} % shift table closer to footer
\begin{longtable}{lllllllll}
\caption{Column densities and upper limits toward L1527.}\label{tab:columndensity_full} \\
\toprule
Molecule & Transition & Frequency & Dataset & Resolution & Disk & $<200$ au & Envelope & Extended envelope \\
\midrule
\endfirsthead
\multicolumn{9}{l}{\textit{Table \thetable\ continued.}} \\
\toprule
Molecule & Transition & Frequency & Dataset & Resolution & Disk & $<200$ au & Envelope & Extended envelope \\
\midrule
\endhead
\midrule
\multicolumn{9}{r}{Continued on next page} \\
\endfoot
\bottomrule
\endlastfoot
C$_4$H & 10-9, 21/2-19/2 & 95.105389 & 2018.1.01205.L & 1.18$^{\prime\prime}$ & \dots & \dots & \dots & T: $(2.2{\pm}0.1) {\times} 10^{13}$ [470$^{\prime\prime2}$] \\
C$_4$H & 10-9, 19/2-17/2 & 95.188949 & 2018.1.01205.L & 1.18$^{\prime\prime}$ & \dots & \dots & \dots & T: $(2.1{\pm}0.1) {\times} 10^{13}$ [470$^{\prime\prime2}$] \\
C$_4$H & 11-10, 23/2-21/2 & 104.666573 & 2016.1.01541.S & 14.30$^{\prime\prime}$ & \dots & \dots & \dots & T: $(6.7{\pm}0.5) {\times} 10^{12}$ [701$^{\prime\prime2}$] \\
 &  &  &  & 1.82$^{\prime\prime}$ & \dots & \dots & \dots & T: $(5.9{\pm}0.5) {\times} 10^{12}$ [501$^{\prime\prime2}$] \\
 &  &  & 2018.1.00799.S & 12.72$^{\prime\prime}$ & \dots & \dots & \dots & T: $(9.6{\pm}0.3) {\times} 10^{12}$ [1244$^{\prime\prime2}$] \\
 &  &  &  & 1.94$^{\prime\prime}$ & \dots & \dots & \dots & T: $(1.1{\pm}0.1) {\times} 10^{13}$ [699$^{\prime\prime2}$] \\
C$_4$H & 11-10, 21/2-19/2 & 104.70511 & 2016.1.01541.S & 14.30$^{\prime\prime}$ & \dots & \dots & \dots & T: $(6.3{\pm}0.5) {\times} 10^{12}$ [701$^{\prime\prime2}$] \\
 &  &  &  & 1.82$^{\prime\prime}$ & \dots & \dots & \dots & T: $(6.1{\pm}0.6) {\times} 10^{12}$ [501$^{\prime\prime2}$] \\
\end{longtable}
{\footnotesize
\noindent
%\vspace{-1cm}
\textit{Note.} Frequencies are listed in GHz, and column densities in cm$^{-2}$. The area used for the column density calculation is listed in square brackets after the column density. For the disk and inner 200 au, column densities are calculated assuming the emission uniformly fills the aperture used for flux extraction (first entry) as well as assuming that the emitting area is equal to the resolved C$^{18}$O emitting area from 2019.1.00261.L (eDisk Large Program) (second entry). In both cases, an excitation temperature of 50 K is adopted. For the envelope and extended envelope, the emission is assumed to uniformly fill the aperture used for flux extraction, and a temperature of 20 K is adopted. No optical depth correction has been performed and some column densities may therefore be only lower limits. See Sect. 3.3 of the main text for more details. For the disk, inner 200 au, and envelope components, ``R'' and ``B'' denote the red- and blueshifted components, respectively. For the extended envelope, ``N'', ``S'', and ``T'' denote the north, south, and total components. Empty entries are shown as $\dots$. Upper limits are indicated with a leading $<$ symbol.
\par}
\end{landscape}

%% file: Tables/Columndensity_Summary_SmallScale.tex
%%% ORDER OF SPECIES IS CHANGED MANUALLY AFTER RUNNING THE PYTHON SCRIPT. 

\begin{table}[p]
\centering
\footnotesize
\setlength{\tabcolsep}{4pt}
\renewcommand{\arraystretch}{1.25}
\rotatebox{90}{%
\begin{minipage}{\textheight}
\centering
\newsavebox{\columndensitytablebox}
\sbox{\columndensitytablebox}{%
\begin{tabular}{lcccc}
\toprule
Molecule & \multicolumn{2}{c}{Disk} & \multicolumn{2}{c}{$<200$ au} \\
\cmidrule(lr){2-3}\cmidrule(lr){4-5}
  & Red & Blue & Red & Blue \\
\midrule
c-C$_3$H$_2$ & \dots & \dots & $(1.3 \pm 0.0) \times 10^{13}\,(1)$ & $(2.2 \pm 0.0) \times 10^{13}\,(1)$ \\
CCH & \dots & \dots & $(7.0 \pm 3.8) \times 10^{13}\,(7)$ & $(6.8 \pm 0.6) \times 10^{13}\,(3)$ \\
C$^{13}$CH & \dots & \dots & $<1.1 \times 10^{14}\,(3)$ & $<1.1 \times 10^{14}\,(3)$ \\
CCS & $<1.8 \times 10^{14}\,(16)$ & $<1.3 \times 10^{14}\,(16)$ & $<7.1 \times 10^{13}\,(16)$ & $<7.1 \times 10^{13}\,(16)$ \\
CH$_3$CN & $<1.5 \times 10^{13}\,(4)$ & $<8.8 \times 10^{12}\,(4)$ & $<4.2 \times 10^{12}\,(4)$ & $<4.2 \times 10^{12}\,(4)$ \\
$^{13}$CO & $(1.5 \pm 0.4) \times 10^{16}\,(6)$ & $(1.7 \pm 0.5) \times 10^{16}\,(6)$ & $(2.5 \pm 0.5) \times 10^{16}\,(6)$ & $(3.0 \pm 0.4) \times 10^{16}\,(6)$ \\
C$^{18}$O & $(8.0 \pm 3.4) \times 10^{15}\,(7)$ & $(1.0 \pm 0.4) \times 10^{16}\,(9)$ & $(1.4 \pm 0.3) \times 10^{16}\,(12)$ & $(1.8 \pm 0.5) \times 10^{16}\,(12)$ \\
C$^{17}$O & $(1.3 \pm 0.0) \times 10^{16}\,(1)$ & $(2.7 \pm 0.0) \times 10^{16}\,(1)$ & $(3.9 \pm 2.4) \times 10^{16}\,(2)$ & $(2.0 \pm 0.2) \times 10^{16}\,(2)$ \\
$^{13}$C$^{18}$O & $<4.6 \times 10^{16}\,(5)$ & $<2.7 \times 10^{16}\,(5)$ & $<1.2 \times 10^{16}\,(5)$ & $<1.3 \times 10^{16}\,(5)$ \\
CS & \dots & \dots & $(4.9 \pm 2.3) \times 10^{12}\,(6)$ & $(7.0 \pm 2.1) \times 10^{12}\,(4)$ \\
$^{13}$CS & \dots & \dots & $<6.9 \times 10^{11}\,(9)$ & $<1.2 \times 10^{12}\,(9)$ \\
HCN & \dots & \dots & $(5.7 \pm 0.0) \times 10^{12}\,(1)$ & $(2.9 \pm 0.0) \times 10^{12}\,(1)$ \\
HCO$^+$ & $(7.2 \pm 0.0) \times 10^{12}\,(1)$ & $(6.9 \pm 0.0) \times 10^{12}\,(1)$ & $(1.5 \pm 0.0) \times 10^{13}\,(1)$ & $(1.8 \pm 0.0) \times 10^{13}\,(1)$ \\
H$^{13}$CO$^+$ & \dots & \dots & $(6.7 \pm 1.8) \times 10^{11}\,(4)$ & $(1.1 \pm 0.1) \times 10^{12}\,(4)$ \\
H$_2$CO & $(1.0 \pm 0.2) \times 10^{14}\,(4)$ & $(8.1 \pm 0.7) \times 10^{13}\,(5)$ & $(1.1 \pm 0.5) \times 10^{14}\,(10)$ & $(1.0 \pm 0.6) \times 10^{14}\,(10)$ \\
H$_2$$^{13}$CO & $<2.0 \times 10^{13}\,(1)$ & $<1.8 \times 10^{13}\,(1)$ & $<8.4 \times 10^{12}\,(2)$ & $<1.0 \times 10^{13}\,(2)$ \\
HDCO & $(1.3 \pm 0.0) \times 10^{13}\,(1)$ & \dots & $(2.9 \pm 0.2) \times 10^{13}\,(3)$ & $(3.2 \pm 2.2) \times 10^{13}\,(3)$ \\
H$_2$CS & $<1.6 \times 10^{13}\,(11)$ & $<1.5 \times 10^{13}\,(11)$ & $<8.2 \times 10^{12}\,(11)$ & $<9.1 \times 10^{12}\,(11)$ \\
HNC & \dots & \dots & $<5.0 \times 10^{13}\,(1)$ & $<5.0 \times 10^{13}\,(1)$ \\
OCS & $<8.4 \times 10^{13}\,(6)$ & $<6.9 \times 10^{13}\,(6)$ & $<4.1 \times 10^{13}\,(6)$ & $<4.7 \times 10^{13}\,(6)$ \\
SO & $(2.6 \pm 0.3) \times 10^{13}\,(4)$ & $(6.7 \pm 0.7) \times 10^{13}\,(7)$ & $(6.3 \pm 2.7) \times 10^{13}\,(16)$ & $(8.4 \pm 3.5) \times 10^{13}\,(19)$ \\
\bottomrule
\end{tabular}%
}
{\normalsize
\refstepcounter{table}\label{tab:columndensity_summary_disk_inner200}
\noindent
\parbox[t]{\wd\columndensitytablebox}{\textbf{Table~\thetable.} Column densities and upper limits toward L1527 on scales of the disk and inner 200 au.}
\par}
\vspace{0.8em}
\noindent\usebox{\columndensitytablebox}
\par\vspace{0.6em}
{\footnotesize
\noindent
\parbox[t]{\wd\columndensitytablebox}{\textit{Note.} Molecules are ordered alphabetically by main isotopologue. Values (in cm$^{-2}$) are median $\pm$ the mean absolute deviation (MAD) with number of measurements in parentheses. For a single measurement, the listed uncertainty is the 1$\sigma$ uncertainty. For molecules that were not detected, the most stringent upper limit is listed.}
\par}
\end{minipage}
}
\end{table}

%% file: Tables/Columndensity_Summary_LargeScale.tex
%%% ORDER OF SPECIES IS CHANGED MANUALLY AFTER RUNNING THE PYTHON SCRIPT. 

\begin{table}[p]
\centering
\footnotesize
\setlength{\tabcolsep}{4pt}
\renewcommand{\arraystretch}{1.25}
\rotatebox{90}{%
\begin{minipage}{\textheight}
\centering
\newsavebox{\columndensitytableboxlarge}
\sbox{\columndensitytableboxlarge}{%
\begin{tabular}{lccccc}
\toprule
Molecule & \multicolumn{2}{c}{Envelope} & \multicolumn{3}{c}{Extended envelope} \\
\cmidrule(lr){2-3}\cmidrule(lr){4-6}
  & Red & Blue & North & South & Total \\
\midrule
C$_4$H & \dots & \dots & \dots & \dots & $(6.7 \pm 0.8) \times 10^{12}\,(16)$ \\
c-C$_3$H & \dots & \dots & $(1.6 \pm 0.1) \times 10^{12}\,(2)$ & $(1.8 \pm 0.2) \times 10^{12}\,(2)$ & \dots \\
c-C$_3$H$_2$ & $(2.0 \pm 0.1) \times 10^{13}\,(6)$ & $(1.8 \pm 0.6) \times 10^{13}\,(8)$ & \dots & \dots & $(4.2 \pm 1.5) \times 10^{12}\,(3)$ \\
c-H$^{13}$CCCH & \dots & \dots & \dots & \dots & $(8.8 \pm 0.0) \times 10^{11}\,(1)$ \\
c-HCCCD & \dots & \dots & $(7.1 \pm 0.0) \times 10^{12}\,(1)$ & $(7.4 \pm 0.0) \times 10^{12}\,(1)$ & $(2.4 \pm 0.7) \times 10^{12}\,(5)$ \\
l-C$_3$H$_2$ & \dots & \dots & \dots & \dots & $(3.5 \pm 0.1) \times 10^{11}\,(3)$ \\
CCH & $(1.8 \pm 0.9) \times 10^{14}\,(9)$ & $(2.1 \pm 1.3) \times 10^{14}\,(11)$ & \dots & \dots & \dots \\
C$^{13}$CH & $<7.3 \times 10^{13}\,(3)$ & $<7.3 \times 10^{12}\,(3)$ & \dots & \dots & \dots \\
CCD & \dots & \dots & \dots & $(2.3 \pm 0.1) \times 10^{13}\,(2)$ & $(1.9 \pm 0.0) \times 10^{13}\,(1)$ \\
CH$_3$CCH & $(4.3 \pm 0.6) \times 10^{13}\,(2)$ & $(4.4 \pm 0.4) \times 10^{13}\,(2)$ & \dots & \dots & \dots \\
CH$_3$OH & $(8.2 \pm 0.0) \times 10^{13}\,(1)$ & $(1.2 \pm 0.0) \times 10^{14}\,(1)$ & \dots & \dots & \dots \\
CN & $(4.4 \pm 1.2) \times 10^{13}\,(11)$ & $(3.7 \pm 1.2) \times 10^{13}\,(11)$ & \dots & \dots & \dots \\
C$^{18}$O & $(3.3 \pm 1.5) \times 10^{15}\,(11)$ & $(4.1 \pm 2.0) \times 10^{15}\,(11)$ & \dots & \dots & \dots \\
C$^{17}$O & $(9.3 \pm 0.0) \times 10^{14}\,(1)$ & $(9.8 \pm 0.0) \times 10^{14}\,(1)$ & \dots & \dots & \dots \\
CS & $(8.6 \pm 5.6) \times 10^{12}\,(7)$ & $(9.8 \pm 3.7) \times 10^{12}\,(7)$ & \dots & \dots & \dots \\
HC$_5$N & \dots & \dots & \dots & \dots & $(2.5 \pm 0.0) \times 10^{13}\,(1)$ \\
HC$_3$N & \dots & \dots & \dots & \dots & $(1.3 \pm 0.1) \times 10^{13}\,(6)$ \\
HCN & $(1.7 \pm 0.0) \times 10^{12}\,(1)$ & $(2.1 \pm 0.0) \times 10^{12}\,(1)$ & \dots & \dots & \dots \\
H$^{13}$CN & \dots & \dots & $(2.2 \pm 0.4) \times 10^{10}\,(2)$ & $(2.9 \pm 0.3) \times 10^{10}\,(2)$ & $(2.3 \pm 0.0) \times 10^{10}\,(2)$ \\
HC$^{15}$N & \dots & \dots & $(4.7 \pm 0.0) \times 10^{9}\,(1)$ & $(9.0 \pm 0.0) \times 10^{9}\,(1)$ & $(7.5 \pm 0.0) \times 10^{9}\,(1)$ \\
DCN & \dots & \dots & $(3.8 \pm 0.0) \times 10^{11}\,(1)$ & $(5.0 \pm 0.0) \times 10^{11}\,(1)$ & $(3.1 \pm 0.0) \times 10^{11}\,(2)$ \\
HCO$^+$ & $(4.1 \pm 0.0) \times 10^{12}\,(1)$ & $(4.3 \pm 0.0) \times 10^{12}\,(1)$ & \dots & \dots & \dots \\
H$^{13}$CO$^+$ & $(4.6 \pm 2.2) \times 10^{11}\,(5)$ & $(6.4 \pm 4.3) \times 10^{11}\,(5)$ & \dots & \dots & \dots \\
HC$^{18}$O$^+$ & \dots & \dots & $(5.9 \pm 1.7) \times 10^{10}\,(2)$ & $(6.1 \pm 1.6) \times 10^{10}\,(2)$ & $(6.7 \pm 0.3) \times 10^{10}\,(3)$ \\
DCO$^+$ & $(1.7 \pm 0.9) \times 10^{11}\,(4)$ & $(1.9 \pm 1.0) \times 10^{11}\,(4)$ & $(2.4 \pm 0.5) \times 10^{11}\,(4)$ & $(3.0 \pm 1.0) \times 10^{11}\,(4)$ & $(2.6 \pm 0.6) \times 10^{11}\,(4)$ \\
H$_2$CO & $(3.0 \pm 1.4) \times 10^{13}\,(10)$ & $(3.5 \pm 1.9) \times 10^{13}\,(10)$ & \dots & \dots & \dots \\
HDCO & $(5.1 \pm 4.9) \times 10^{12}\,(2)$ & $(6.8 \pm 6.6) \times 10^{12}\,(2)$ & $(2.2 \pm 0.7) \times 10^{12}\,(3)$ & $(2.2 \pm 0.4) \times 10^{12}\,(3)$ & $(1.6 \pm 0.3) \times 10^{12}\,(4)$ \\
D$_2$CO & $(1.3 \pm 0.9) \times 10^{12}\,(2)$ & $(1.6 \pm 0.0) \times 10^{12}\,(1)$ & $(9.8 \pm 3.0) \times 10^{11}\,(10)$ & $(1.7 \pm 0.7) \times 10^{12}\,(10)$ & $(9.6 \pm 3.3) \times 10^{11}\,(11)$ \\
HN$^{13}$C & \dots & \dots & $(8.4 \pm 0.0) \times 10^{10}\,(1)$ & $(8.2 \pm 0.0) \times 10^{10}\,(1)$ & $(6.8 \pm 1.4) \times 10^{10}\,(2)$ \\
HNCO & \dots & \dots & \dots & $(4.2 \pm 0.0) \times 10^{11}\,(1)$ & \dots \\
SO & $(5.8 \pm 4.3) \times 10^{13}\,(18)$ & $(7.2 \pm 5.4) \times 10^{13}\,(18)$ & \dots & \dots & \dots \\
$^{34}$SO & $(3.0 \pm 1.1) \times 10^{12}\,(2)$ & $(3.4 \pm 0.2) \times 10^{12}\,(2)$ & \dots & \dots & \dots \\
SO$_2$ & $(9.6 \pm 0.0) \times 10^{13}\,(1)$ & $(7.0 \pm 0.0) \times 10^{13}\,(1)$ & \dots & \dots & \dots \\
\bottomrule
\end{tabular}%
}
{\normalsize
\refstepcounter{table}\label{tab:columndensity_summary_envelope_extended}
\noindent
\parbox[t]{\wd\columndensitytableboxlarge}{\textbf{Table~\thetable.} Column densities toward L1527 on scales of the envelope and extended envelope.}
\par}
\vspace{0.8em}
\noindent\usebox{\columndensitytableboxlarge}
\par\vspace{0.6em}
{\footnotesize
\noindent
\parbox[t]{\wd\columndensitytableboxlarge}{\textit{Note.} Molecules are ordered alphabetically by main isotopologue. Values (in cm$^{-2}$) are median $\pm$ the mean absolute deviation (MAD) with number of measurements in parentheses. For a single measurement, the listed uncertainty is the 1$\sigma$ uncertainty. For molecules that were not detected, the most stringent upper limit is listed.}
\par}
\end{minipage}
}
\end{table}

%% file: Main-Suppl_ArXiv_final.bbl
\begin{thebibliography}{100}
\providecommand{\natexlab}[1]{#1}
\expandafter\ifx\csname urlstyle\endcsname\relax
  \providecommand{\doi}[1]{doi:\discretionary{}{}{}#1}\else
  \providecommand{\doi}{doi:\discretionary{}{}{}\begingroup \urlstyle{rm}\Url}\fi
\providecommand{\selectlanguage}[1]{\relax}
\providecommand{\bibAnnoteFile}[1]{%
  \IfFileExists{#1}{\begin{quotation}\noindent\textsc{Key:} #1\\
  \textsc{Annotation:}\ \input{#1}\end{quotation}}{}}
\providecommand{\bibAnnote}[2]{%
  \begin{quotation}\noindent\textsc{Key:} #1\\
  \textsc{Annotation:}\ #2\end{quotation}}

\bibitem[{{Aikawa} et~al.(2008){Aikawa}, {Wakelam}, {Garrod}, and {Herbst}}]{Aikawa2008}
{Aikawa}, Y., {Wakelam}, V., {Garrod}, R.~T., and {Herbst}, E. (2008).
\newblock {Molecular Evolution and Star Formation: From Prestellar Cores to Protostellar Cores}.
\newblock \emph{\apj} 674, 984--996.
\newblock \doi{10.1086/524096}
\bibAnnoteFile{Aikawa2008}

\bibitem[{{Armitage}(2024)}]{Armitage2024}
{Armitage}, P.~J. (2024).
\newblock {Planet formation theory: an overview}.
\newblock \emph{arXiv e-prints} , arXiv:2412.11064\doi{10.48550/arXiv.2412.11064}
\bibAnnoteFile{Armitage2024}

\bibitem[{{Artur de la Villarmois} et~al.(2023){Artur de la Villarmois}, {Guzm{\'a}n}, {Yang}, {Zhang}, and {Sakai}}]{ArturdelaVillarmois2023}
{Artur de la Villarmois}, E., {Guzm{\'a}n}, V.~V., {Yang}, Y.-L., {Zhang}, Y., and {Sakai}, N. (2023).
\newblock {The Perseus ALMA Chemical Survey (PEACHES). III. Sulfur-bearing species tracing accretion and ejection processes in young protostars}.
\newblock \emph{\aap} 678, A124.
\newblock \doi{10.1051/0004-6361/202346728}
\bibAnnoteFile{ArturdelaVillarmois2023}

\bibitem[{{Artur de la Villarmois} et~al.(2019){Artur de la Villarmois}, {J{\o}rgensen}, {Kristensen}, {Bergin}, {Harsono}, {Sakai} et~al.}]{ArturdelaVillarmois2019}
{Artur de la Villarmois}, E., {J{\o}rgensen}, J.~K., {Kristensen}, L.~E., {Bergin}, E.~A., {Harsono}, D., {Sakai}, N., et~al. (2019).
\newblock {Physical and chemical fingerprint of protostellar disc formation}.
\newblock \emph{\aap} 626, A71.
\newblock \doi{10.1051/0004-6361/201834877}
\bibAnnoteFile{ArturdelaVillarmois2019}

\bibitem[{{Arulanantham} et~al.(2025){Arulanantham}, {Salyk}, {Pontoppidan}, {Banzatti}, {Zhang}, {{\"O}berg} et~al.}]{Arulanatham2025}
{Arulanantham}, N., {Salyk}, C., {Pontoppidan}, K., {Banzatti}, A., {Zhang}, K., {{\"O}berg}, K., et~al. (2025).
\newblock {The JDISC Survey: Linking the Physics and Chemistry of Inner and Outer Protoplanetary Disk Zones}.
\newblock \emph{\aj} 170, 67.
\newblock \doi{10.3847/1538-3881/addd01}
\bibAnnoteFile{Arulanatham2025}

\bibitem[{{Aso} et~al.(2017){Aso}, {Ohashi}, {Aikawa}, {Machida}, {Saigo}, {Saito} et~al.}]{Aso2017}
{Aso}, Y., {Ohashi}, N., {Aikawa}, Y., {Machida}, M.~N., {Saigo}, K., {Saito}, M., et~al. (2017).
\newblock {ALMA Observations of the Protostar L1527 IRS: Probing Details of the Disk and the Envelope Structures}.
\newblock \emph{\apj} 849, 56.
\newblock \doi{10.3847/1538-4357/aa8264}
\bibAnnoteFile{Aso2017}

\bibitem[{{Bergin} et~al.(2016){Bergin}, {Du}, {Cleeves}, {Blake}, {Schwarz}, {Visser} et~al.}]{Bergin2016}
{Bergin}, E.~A., {Du}, F., {Cleeves}, L.~I., {Blake}, G.~A., {Schwarz}, K., {Visser}, R., et~al. (2016).
\newblock {Hydrocarbon Emission Rings in Protoplanetary Disks Induced by Dust Evolution}.
\newblock \emph{\apj} 831, 101.
\newblock \doi{10.3847/0004-637X/831/1/101}
\bibAnnoteFile{Bergin2016}

\bibitem[{{Bergner} et~al.(2019){Bergner}, {{\"O}berg}, {Bergin}, {Loomis}, {Pegues}, and {Qi}}]{Bergner2019}
{Bergner}, J.~B., {{\"O}berg}, K.~I., {Bergin}, E.~A., {Loomis}, R.~A., {Pegues}, J., and {Qi}, C. (2019).
\newblock {A Survey of C$_{2}$H, HCN, and C$^{18}$O in Protoplanetary Disks}.
\newblock \emph{\apj} 876, 25.
\newblock \doi{10.3847/1538-4357/ab141e}
\bibAnnoteFile{Bergner2019}

\bibitem[{{Booth} et~al.(2023){Booth}, {Law}, {Temmink}, {Leemker}, and {Mac{\'\i}as}}]{Booth2023}
{Booth}, A.~S., {Law}, C.~J., {Temmink}, M., {Leemker}, M., and {Mac{\'\i}as}, E. (2023).
\newblock {Tracing snowlines and C/O ratio in a planet-hosting disk. ALMA molecular line observations towards the HD 169142 disk}.
\newblock \emph{\aap} 678, A146.
\newblock \doi{10.1051/0004-6361/202346974}
\bibAnnoteFile{Booth2023}

\bibitem[{{Bosman} et~al.(2021){Bosman}, {Alarc{\'o}n}, {Bergin}, {Zhang}, {van't Hoff}, {{\"O}berg} et~al.}]{Bosman2021}
{Bosman}, A.~D., {Alarc{\'o}n}, F., {Bergin}, E.~A., {Zhang}, K., {van't Hoff}, M. L.~R., {{\"O}berg}, K.~I., et~al. (2021).
\newblock {Molecules with ALMA at Planet-forming Scales (MAPS). VII. Substellar O/H and C/H and Superstellar C/O in Planet-feeding Gas}.
\newblock \emph{\apjs} 257, 7.
\newblock \doi{10.3847/1538-4365/ac1435}
\bibAnnoteFile{Bosman2021}

\bibitem[{{Cacciapuoti} et~al.(2023){Cacciapuoti}, {Macias}, {Maury}, {Chandler}, {Sakai}, {Tychoniec} et~al.}]{Cacciapuoti2023}
{Cacciapuoti}, L., {Macias}, E., {Maury}, A.~J., {Chandler}, C.~J., {Sakai}, N., {Tychoniec}, {\L}., et~al. (2023).
\newblock {FAUST. IX. Multiband, multiscale dust study of L1527 IRS. Evidence for variations in dust properties within the envelope of a class 0/I young stellar object}.
\newblock \emph{\aap} 676, A4.
\newblock \doi{10.1051/0004-6361/202346204}
\bibAnnoteFile{Cacciapuoti2023}

\bibitem[{{Cassen} and {Moosman}(1981)}]{Cassen1981}
{Cassen}, P. and {Moosman}, A. (1981).
\newblock {On the formation of protostellar disks}.
\newblock \emph{\icarus} 48, 353--376.
\newblock \doi{10.1016/0019-1035(81)90051-8}
\bibAnnoteFile{Cassen1981}

\bibitem[{{Cook} et~al.(2019){Cook}, {Tobin}, {Skrutskie}, and {Nelson}}]{Cook.Tobin.ea2019}
{Cook}, B.~T., {Tobin}, J.~J., {Skrutskie}, M.~F., and {Nelson}, M.~J. (2019).
\newblock {Time variability in the bipolar scattered light nebula of L1527 IRS: a possible warped inner disk}.
\newblock \emph{\aap} 626, A51.
\newblock \doi{10.1051/0004-6361/201935419}
\bibAnnoteFile{Cook.Tobin.ea2019}

\bibitem[{{Devaraj} et~al.(2026){Devaraj}, {van Dishoeck}, {Ray}, {Tychoniec}, {Garatti}, {Francis} et~al.}]{Devaraj.vanDishoeck.ea2026}
{Devaraj}, R., {van Dishoeck}, E.~F., {Ray}, T.~P., {Tychoniec}, {\L}., {Garatti}, A. C.~o., {Francis}, L., et~al. (2026).
\newblock {JOYS: JWST MIRI/MRS spectra of the inner 500 au region of the L1527 IRS bipolar outflow}.
\newblock \emph{arXiv e-prints} , arXiv:2601.17820\doi{10.48550/arXiv.2601.17820}
\bibAnnoteFile{Devaraj.vanDishoeck.ea2026}

\bibitem[{{Diep} et~al.(2016){Diep}, {Phuong}, {Hoai}, {Nhung}, {Thao}, {Tuan-Anh} et~al.}]{Diep2016}
{Diep}, P.~N., {Phuong}, N.~T., {Hoai}, D.~T., {Nhung}, P.~T., {Thao}, N.~T., {Tuan-Anh}, P., et~al. (2016).
\newblock {Space reconstruction of the morphology and kinematics of axisymmetric radio sources}.
\newblock \emph{\mnras} 461, 4276--4294.
\newblock \doi{10.1093/mnras/stw1630}
\bibAnnoteFile{Diep2016}

\bibitem[{{Drechsler} et~al.(2026){Drechsler}, {Tobin}, {Sheehan}, {Looney}, {Megeath}, {Van Dishoeck} et~al.}]{BlakeDrechsler.Tobin.ea2026}
{Drechsler}, W.~B., {Tobin}, J.~J., {Sheehan}, P.~D., {Looney}, L.~W., {Megeath}, S.~T., {Van Dishoeck}, E.~F., et~al. (2026).
\newblock {Accretion onto the Embedded Protostar L1527 IRS: Insights from JWST NIRSpec and MIRI Observations}.
\newblock \emph{arXiv e-prints} , arXiv:2603.03575\doi{10.48550/arXiv.2603.03575}
\bibAnnoteFile{BlakeDrechsler.Tobin.ea2026}

\bibitem[{{Dutrey} et~al.(2011){Dutrey}, {Wakelam}, {Boehler}, {Guilloteau}, {Hersant}, {Semenov} et~al.}]{Dutrey2011}
{Dutrey}, A., {Wakelam}, V., {Boehler}, Y., {Guilloteau}, S., {Hersant}, F., {Semenov}, D., et~al. (2011).
\newblock {Chemistry in disks. V. Sulfur-bearing molecules in the protoplanetary disks surrounding LkCa15, MWC480, DM Tauri, and GO Tauri}.
\newblock \emph{\aap} 535, A104.
\newblock \doi{10.1051/0004-6361/201116931}
\bibAnnoteFile{Dutrey2011}

\bibitem[{Dutta(2025)}]{Dutta2025}
Dutta, S. (2025).
\newblock Molecular jets from an evolved protostar: Insights from jwst-alma synergy.
\newblock \emph{The Astrophysical Journal} 991, 45.
\newblock \doi{10.3847/1538-4357/adf8d6}
\bibAnnoteFile{Dutta2025}

\bibitem[{{Emprechtinger} et~al.(2009){Emprechtinger}, {Caselli}, {Volgenau}, {Stutzki}, and {Wiedner}}]{Emprechtinger2009}
{Emprechtinger}, M., {Caselli}, P., {Volgenau}, N.~H., {Stutzki}, J., and {Wiedner}, M.~C. (2009).
\newblock {The N\{2\}D\^+/N\{2\}H$^{+}$ ratio as an evolutionary tracer of Class 0 protostars}.
\newblock \emph{\aap} 493, 89--105.
\newblock \doi{10.1051/0004-6361:200810324}
\bibAnnoteFile{Emprechtinger2009}

\bibitem[{{Endres} et~al.(2016){Endres}, {Schlemmer}, {Schilke}, {Stutzki}, and {M{\"u}ller}}]{Endres2016}
{Endres}, C.~P., {Schlemmer}, S., {Schilke}, P., {Stutzki}, J., and {M{\"u}ller}, H. S.~P. (2016).
\newblock {The Cologne Database for Molecular Spectroscopy, CDMS, in the Virtual Atomic and Molecular Data Centre, VAMDC}.
\newblock \emph{Journal of Molecular Spectroscopy} 327, 95--104.
\newblock \doi{10.1016/j.jms.2016.03.005}
\bibAnnoteFile{Endres2016}

\bibitem[{{Favre} et~al.(2013){Favre}, {Cleeves}, {Bergin}, {Qi}, and {Blake}}]{Favre2013}
{Favre}, C., {Cleeves}, L.~I., {Bergin}, E.~A., {Qi}, C., and {Blake}, G.~A. (2013).
\newblock {A Significantly Low CO Abundance toward the TW Hya Protoplanetary Disk: A Path to Active Carbon Chemistry?}
\newblock \emph{\apjl} 776, L38.
\newblock \doi{10.1088/2041-8205/776/2/L38}
\bibAnnoteFile{Favre2013}

\bibitem[{{Feeney-Johansson} et~al.(2025){Feeney-Johansson}, {Aikawa}, {Takakuwa}, {Ohashi}, {Plunkett}, {Jorgensen} et~al.}]{Feeney-Johansson2025}
{Feeney-Johansson}, A., {Aikawa}, Y., {Takakuwa}, S., {Ohashi}, N., {Plunkett}, A., {Jorgensen}, J.~K., et~al. (2025).
\newblock {Early Planet Formation in Embedded Disks (eDisk). XIX. Structures of molecular outflows}.
\newblock \emph{arXiv e-prints} , arXiv:2512.21454\doi{10.48550/arXiv.2512.21454}
\bibAnnoteFile{Feeney-Johansson2025}

\bibitem[{{Flores-Rivera} et~al.(2021){Flores-Rivera}, {Terebey}, {Willacy}, {Isella}, {Turner}, and {Flock}}]{Flores-Rivera2021}
{Flores-Rivera}, L., {Terebey}, S., {Willacy}, K., {Isella}, A., {Turner}, N., and {Flock}, M. (2021).
\newblock {Physical and Chemical Structure of the Disk and Envelope of the Class 0/I Protostar L1527}.
\newblock \emph{\apj} 908, 108.
\newblock \doi{10.3847/1538-4357/abd1db}
\bibAnnoteFile{Flores-Rivera2021}

\bibitem[{{Garufi} et~al.(2021){Garufi}, {Podio}, {Codella}, {Fedele}, {Bianchi}, {Favre} et~al.}]{Garufi2021}
{Garufi}, A., {Podio}, L., {Codella}, C., {Fedele}, D., {Bianchi}, E., {Favre}, C., et~al. (2021).
\newblock {ALMA chemical survey of disk-outflow sources in Taurus (ALMA-DOT). V. Sample, overview, and demography of disk molecular emission}.
\newblock \emph{\aap} 645, A145.
\newblock \doi{10.1051/0004-6361/202039483}
\bibAnnoteFile{Garufi2021}

\bibitem[{{Grant} et~al.(2025){Grant}, {Temmink}, {van Dishoeck}, {Gasman}, {Arabhavi}, {Tabone} et~al.}]{Grant2025}
{Grant}, S.~L., {Temmink}, M., {van Dishoeck}, E.~F., {Gasman}, D., {Arabhavi}, A.~M., {Tabone}, B., et~al. (2025).
\newblock {MINDS: A transition from H$_{2}$O to C$_{2}$H$_{2}$ dominated disk spectra with decreasing stellar luminosity}.
\newblock \emph{\aap} 702, A126.
\newblock \doi{10.1051/0004-6361/202555862}
\bibAnnoteFile{Grant2025}

\bibitem[{{Guzm{\'a}n} et~al.(2021){Guzm{\'a}n}, {Bergner}, {Law}, {{\"O}berg}, {Walsh}, {Cataldi} et~al.}]{Guzman2021}
{Guzm{\'a}n}, V.~V., {Bergner}, J.~B., {Law}, C.~J., {{\"O}berg}, K.~I., {Walsh}, C., {Cataldi}, G., et~al. (2021).
\newblock {Molecules with ALMA at Planet-forming Scales (MAPS). VI. Distribution of the Small Organics HCN, C$_{2}$H, and H$_{2}$CO}.
\newblock \emph{\apjs} 257, 6.
\newblock \doi{10.3847/1538-4365/ac1440}
\bibAnnoteFile{Guzman2021}

\bibitem[{{Harsono} et~al.(2018){Harsono}, {Bjerkeli}, {van der Wiel}, {Ramsey}, {Maud}, {Kristensen} et~al.}]{Harsono2018}
{Harsono}, D., {Bjerkeli}, P., {van der Wiel}, M. H.~D., {Ramsey}, J.~P., {Maud}, L.~T., {Kristensen}, L.~E., et~al. (2018).
\newblock {Evidence for the start of planet formation in a young circumstellar disk}.
\newblock \emph{Nature Astronomy} 2, 646--651.
\newblock \doi{10.1038/s41550-018-0497-x}
\bibAnnoteFile{Harsono2018}

\bibitem[{{Harsono} et~al.(2014){Harsono}, {J{\o}rgensen}, {van Dishoeck}, {Hogerheijde}, {Bruderer}, {Persson} et~al.}]{Harsono2014}
{Harsono}, D., {J{\o}rgensen}, J.~K., {van Dishoeck}, E.~F., {Hogerheijde}, M.~R., {Bruderer}, S., {Persson}, M.~V., et~al. (2014).
\newblock {Rotationally-supported disks around Class I sources in Taurus: disk formation constraints}.
\newblock \emph{\aap} 562, A77.
\newblock \doi{10.1051/0004-6361/201322646}
\bibAnnoteFile{Harsono2014}

\bibitem[{{Hassel} et~al.(2008){Hassel}, {Herbst}, and {Garrod}}]{Hassel2008}
{Hassel}, G.~E., {Herbst}, E., and {Garrod}, R.~T. (2008).
\newblock {Modeling the Lukewarm Corino Phase: Is L1527 Unique?}
\newblock \emph{\apj} 681, 1385--1395.
\newblock \doi{10.1086/588185}
\bibAnnoteFile{Hassel2008}

\bibitem[{{Huang} et~al.(2024){Huang}, {Bergin}, {Le Gal}, {Andrews}, {Bae}, {Keyte} et~al.}]{Huang2024}
{Huang}, J., {Bergin}, E.~A., {Le Gal}, R., {Andrews}, S.~M., {Bae}, J., {Keyte}, L., et~al. (2024).
\newblock {Constraints on the Gas-phase C/O Ratio of DR Tau's Outer Disk from CS, SO, and C$_{2}$H Observations}.
\newblock \emph{\apj} 973, 135.
\newblock \doi{10.3847/1538-4357/ad6447}
\bibAnnoteFile{Huang2024}

\bibitem[{{Ikoma} and {Kobayashi}(2025)}]{Ikoma2025}
{Ikoma}, M. and {Kobayashi}, H. (2025).
\newblock {Formation of Giant Planets}.
\newblock \emph{\araa} 63, 217--258.
\newblock \doi{10.1146/annurev-astro-052722-094843}
\bibAnnoteFile{Ikoma2025}

\bibitem[{{Ilee} et~al.(2021){Ilee}, {Walsh}, {Booth}, {Aikawa}, {Andrews}, {Bae} et~al.}]{Ilee2021}
{Ilee}, J.~D., {Walsh}, C., {Booth}, A.~S., {Aikawa}, Y., {Andrews}, S.~M., {Bae}, J., et~al. (2021).
\newblock {Molecules with ALMA at Planet-forming Scales (MAPS). IX. Distribution and Properties of the Large Organic Molecules HC$_{3}$N, CH$_{3}$CN, and c-C$_{3}$H$_{2}$}.
\newblock \emph{\apjs} 257, 9.
\newblock \doi{10.3847/1538-4365/ac1441}
\bibAnnoteFile{Ilee2021}

\bibitem[{{Jensen} et~al.(2021){Jensen}, {J{\o}rgensen}, {Kristensen}, {Coutens}, {van Dishoeck}, {Furuya} et~al.}]{Jensen2021}
{Jensen}, S.~S., {J{\o}rgensen}, J.~K., {Kristensen}, L.~E., {Coutens}, A., {van Dishoeck}, E.~F., {Furuya}, K., et~al. (2021).
\newblock {ALMA observations of doubly deuterated water: inheritance of water from the prestellar environment}.
\newblock \emph{\aap} 650, A172.
\newblock \doi{10.1051/0004-6361/202140560}
\bibAnnoteFile{Jensen2021}

\bibitem[{{Jensen} et~al.(2019){Jensen}, {J{\o}rgensen}, {Kristensen}, {Furuya}, {Coutens}, {van Dishoeck} et~al.}]{Jensen2019}
{Jensen}, S.~S., {J{\o}rgensen}, J.~K., {Kristensen}, L.~E., {Furuya}, K., {Coutens}, A., {van Dishoeck}, E.~F., et~al. (2019).
\newblock {ALMA observations of water deuteration: a physical diagnostic of the formation of protostars}.
\newblock \emph{\aap} 631, A25.
\newblock \doi{10.1051/0004-6361/201936012}
\bibAnnoteFile{Jensen2019}

\bibitem[{{Kenyon} et~al.(1994){Kenyon}, {Dobrzycka}, and {Hartmann}}]{Kenyon1994}
{Kenyon}, S.~J., {Dobrzycka}, D., and {Hartmann}, L. (1994).
\newblock {A New Optical Extinction Law and Distance Estimate for the Taurus-Auriga Molecular Cloud}.
\newblock \emph{\aj} 108, 1872.
\newblock \doi{10.1086/117200}
\bibAnnoteFile{Kenyon1994}

\bibitem[{{Keyte} et~al.(2023){Keyte}, {Kama}, {Booth}, {Bergin}, {Cleeves}, {van Dishoeck} et~al.}]{Keyte2023}
{Keyte}, L., {Kama}, M., {Booth}, A.~S., {Bergin}, E.~A., {Cleeves}, L.~I., {van Dishoeck}, E.~F., et~al. (2023).
\newblock {Azimuthal C/O variations in a planet-forming disk}.
\newblock \emph{Nature Astronomy} 7, 684--693.
\newblock \doi{10.1038/s41550-023-01951-9}
\bibAnnoteFile{Keyte2023}

\bibitem[{{Lacy} et~al.(2020){Lacy}, {Kern}, and {Tobin}}]{Lacy2020}
{Lacy}, M., {Kern}, J., and {Tobin}, J.~J. (2020).
\newblock {The NRAO Science Ready Data Products Pilot Program}.
\newblock In \emph{Astronomical Data Analysis Software and Systems XXIX}, eds. R.~{Pizzo}, E.~R. {Deul}, J.~D. {Mol}, J.~{de Plaa}, and H.~{Verkouter}. vol. 527 of \emph{Astronomical Society of the Pacific Conference Series}, 519
\bibAnnoteFile{Lacy2020}

\bibitem[{{Law} et~al.(2026){Law}, {Le Gal}, {{\"O}berg}, {Zhang}, {Aikawa}, {Andrews} et~al.}]{Law2026}
{Law}, C.~J., {Le Gal}, R., {{\"O}berg}, K.~I., {Zhang}, K., {Aikawa}, Y., {Andrews}, S.~M., et~al. (2026).
\newblock {A Submillimeter Survey of CS Excitation in Protoplanetary Disks: Evidence of X-Ray-driven Sulfur Chemistry}.
\newblock \emph{\apj} 997, 91.
\newblock \doi{10.3847/1538-4357/ae1fdf}
\bibAnnoteFile{Law2026}

\bibitem[{{Le Gal} et~al.(2021){Le Gal}, {{\"O}berg}, {Teague}, {Loomis}, {Law}, {Walsh} et~al.}]{LeGal2021}
{Le Gal}, R., {{\"O}berg}, K.~I., {Teague}, R., {Loomis}, R.~A., {Law}, C.~J., {Walsh}, C., et~al. (2021).
\newblock {Molecules with ALMA at Planet-forming Scales (MAPS). XII. Inferring the C/O and S/H Ratios in Protoplanetary Disks with Sulfur Molecules}.
\newblock \emph{\apjs} 257, 12.
\newblock \doi{10.3847/1538-4365/ac2583}
\bibAnnoteFile{LeGal2021}

\bibitem[{{Lee} et~al.(2000){Lee}, {Mundy}, {Reipurth}, {Ostriker}, and {Stone}}]{Lee2000}
{Lee}, C.-F., {Mundy}, L.~G., {Reipurth}, B., {Ostriker}, E.~C., and {Stone}, J.~M. (2000).
\newblock {CO Outflows from Young Stars: Confronting the Jet and Wind Models}.
\newblock \emph{\apj} 542, 925--945.
\newblock \doi{10.1086/317056}
\bibAnnoteFile{Lee2000}

\bibitem[{{Lee} et~al.(2001){Lee}, {Stone}, {Ostriker}, and {Mundy}}]{Lee2001}
{Lee}, C.-F., {Stone}, J.~M., {Ostriker}, E.~C., and {Mundy}, L.~G. (2001).
\newblock {Hydrodynamic Simulations of Jet- and Wind-driven Protostellar Outflows}.
\newblock \emph{\apj} 557, 429--442.
\newblock \doi{10.1086/321648}
\bibAnnoteFile{Lee2001}

\bibitem[{{Li} and {Shu}(1996)}]{Li1996}
{Li}, Z.-Y. and {Shu}, F.~H. (1996).
\newblock {Interaction of Wide-Angle MHD Winds with Flared Disks}.
\newblock \emph{\apj} 468, 261.
\newblock \doi{10.1086/177688}
\bibAnnoteFile{Li1996}

\bibitem[{{Liu} et~al.(2025){Liu}, {van Dishoeck}, {Hogerheijde}, {van Gelder}, {Chen}, {Liu} et~al.}]{Liu2025}
{Liu}, X.-C., {van Dishoeck}, E.~F., {Hogerheijde}, M.~R., {van Gelder}, M.~L., {Chen}, Y., {Liu}, T., et~al. (2025).
\newblock {Sulfur oxides tracing streamers and shocks at low-mass protostellar disk─envelope interfaces}.
\newblock \emph{\aap} 701, A141.
\newblock \doi{10.1051/0004-6361/202554186}
\bibAnnoteFile{Liu2025}

\bibitem[{{Manigand} et~al.(2019){Manigand}, {Calcutt}, {J{\o}rgensen}, {Taquet}, {M{\"u}ller}, {Coutens} et~al.}]{Manigand2019}
{Manigand}, S., {Calcutt}, H., {J{\o}rgensen}, J.~K., {Taquet}, V., {M{\"u}ller}, H.~S.~P., {Coutens}, A., et~al. (2019).
\newblock {The ALMA-PILS survey: the first detection of doubly deuterated methyl formate (CHD$_{2}$OCHO) in the ISM}.
\newblock \emph{\aap} 623, A69.
\newblock \doi{10.1051/0004-6361/201832844}
\bibAnnoteFile{Manigand2019}

\bibitem[{{Massardi} et~al.(2021){Massardi}, {Stoehr}, {Bendo}, {Bonato}, {Brand}, {Galluzzi} et~al.}]{Massardi2021}
{Massardi}, M., {Stoehr}, F., {Bendo}, G.~J., {Bonato}, M., {Brand}, J., {Galluzzi}, V., et~al. (2021).
\newblock {The Additional Representative Images for Legacy (ARI-L) Project for the ALMA Science Archive}.
\newblock \emph{\pasp} 133, 085001.
\newblock \doi{10.1088/1538-3873/ac159c}
\bibAnnoteFile{Massardi2021}

\bibitem[{{McClure} et~al.(2016){McClure}, {Bergin}, {Cleeves}, {van Dishoeck}, {Blake}, {Evans} et~al.}]{McClure2016}
{McClure}, M.~K., {Bergin}, E.~A., {Cleeves}, L.~I., {van Dishoeck}, E.~F., {Blake}, G.~A., {Evans}, N.~J., II, et~al. (2016).
\newblock {Mass Measurements in Protoplanetary Disks from Hydrogen Deuteride}.
\newblock \emph{\apj} 831, 167.
\newblock \doi{10.3847/0004-637X/831/2/167}
\bibAnnoteFile{McClure2016}

\bibitem[{{Milam} et~al.(2005){Milam}, {Savage}, {Brewster}, {Ziurys}, and {Wyckoff}}]{Milam2005}
{Milam}, S.~N., {Savage}, C., {Brewster}, M.~A., {Ziurys}, L.~M., and {Wyckoff}, S. (2005).
\newblock {The $^{12}$C/$^{13}$C Isotope Gradient Derived from Millimeter Transitions of CN: The Case for Galactic Chemical Evolution}.
\newblock \emph{\apj} 634, 1126--1132.
\newblock \doi{10.1086/497123}
\bibAnnoteFile{Milam2005}

\bibitem[{{Miotello} et~al.(2019){Miotello}, {Facchini}, {van Dishoeck}, {Cazzoletti}, {Testi}, {Williams} et~al.}]{Miotello2019}
{Miotello}, A., {Facchini}, S., {van Dishoeck}, E.~F., {Cazzoletti}, P., {Testi}, L., {Williams}, J.~P., et~al. (2019).
\newblock {Bright C$_{2}$H emission in protoplanetary discs in Lupus: high volatile C/O > 1 ratios}.
\newblock \emph{\aap} 631, A69.
\newblock \doi{10.1051/0004-6361/201935441}
\bibAnnoteFile{Miotello2019}

\bibitem[{{Miotello} et~al.(2014){Miotello}, {Testi}, {Lodato}, {Ricci}, {Rosotti}, {Brooks} et~al.}]{Miotello2014}
{Miotello}, A., {Testi}, L., {Lodato}, G., {Ricci}, L., {Rosotti}, G., {Brooks}, K., et~al. (2014).
\newblock {Grain growth in the envelopes and disks of Class I protostars}.
\newblock \emph{\aap} 567, A32.
\newblock \doi{10.1051/0004-6361/201322945}
\bibAnnoteFile{Miotello2014}

\bibitem[{{M{\"u}ller} et~al.(2005){M{\"u}ller}, {Schl{\"o}der}, {Stutzki}, and {Winnewisser}}]{Muller2005}
{M{\"u}ller}, H. S.~P., {Schl{\"o}der}, F., {Stutzki}, J., and {Winnewisser}, G. (2005).
\newblock {The Cologne Database for Molecular Spectroscopy, CDMS: a useful tool for astronomers and spectroscopists}.
\newblock \emph{Journal of Molecular Structure} 742, 215--227.
\newblock \doi{10.1016/j.molstruc.2005.01.027}
\bibAnnoteFile{Muller2005}

\bibitem[{{M{\"u}ller} et~al.(2001){M{\"u}ller}, {Thorwirth}, {Roth}, and {Winnewisser}}]{Muller2001}
{M{\"u}ller}, H.~S.~P., {Thorwirth}, S., {Roth}, D.~A., and {Winnewisser}, G. (2001).
\newblock {The Cologne Database for Molecular Spectroscopy, CDMS}.
\newblock \emph{\aap} 370, L49--L52.
\newblock \doi{10.1051/0004-6361:20010367}
\bibAnnoteFile{Muller2001}

\bibitem[{{Murillo} et~al.(2018){Murillo}, {van Dishoeck}, {van der Wiel}, {J{\o}rgensen}, {Drozdovskaya}, {Calcutt} et~al.}]{Murillo2018}
{Murillo}, N.~M., {van Dishoeck}, E.~F., {van der Wiel}, M.~H.~D., {J{\o}rgensen}, J.~K., {Drozdovskaya}, M.~N., {Calcutt}, H., et~al. (2018).
\newblock {Tracing the cold and warm physico-chemical structure of deeply embedded protostars: IRAS 16293-2422 vs. VLA 1623-2417}.
\newblock \emph{\aap} 617, A120.
\newblock \doi{10.1051/0004-6361/201731724}
\bibAnnoteFile{Murillo2018}

\bibitem[{{Narang} et~al.(2026){Narang}, {Tyagi}, {Ohashi}, {Manoj}, {Megeath}, {Tobin} et~al.}]{Narang.Tyagi.ea2026}
{Narang}, M., {Tyagi}, H., {Ohashi}, N., {Manoj}, P., {Megeath}, S.~T., {Tobin}, J.~J., et~al. (2026).
\newblock {Investigating the Nested Structure of the Outflow from the Low Luminosity Protostar IRAS 16253-2429 using JWST and ALMA}.
\newblock \emph{arXiv e-prints} , arXiv:2602.09837\doi{10.48550/arXiv.2602.09837}
\bibAnnoteFile{Narang.Tyagi.ea2026}

\bibitem[{{{\"O}berg} et~al.(2023){{\"O}berg}, {Facchini}, and {Anderson}}]{Oberg2023}
{{\"O}berg}, K.~I., {Facchini}, S., and {Anderson}, D.~E. (2023).
\newblock {Protoplanetary Disk Chemistry}.
\newblock \emph{\araa} 61, 287--328.
\newblock \doi{10.1146/annurev-astro-022823-040820}
\bibAnnoteFile{Oberg2023}

\bibitem[{{{\"O}berg} et~al.(2021){{\"O}berg}, {Guzm{\'a}n}, {Walsh}, {Aikawa}, {Bergin}, {Law} et~al.}]{Oberg2021}
{{\"O}berg}, K.~I., {Guzm{\'a}n}, V.~V., {Walsh}, C., {Aikawa}, Y., {Bergin}, E.~A., {Law}, C.~J., et~al. (2021).
\newblock {Molecules with ALMA at Planet-forming Scales (MAPS). I. Program Overview and Highlights}.
\newblock \emph{\apjs} 257, 1.
\newblock \doi{10.3847/1538-4365/ac1432}
\bibAnnoteFile{Oberg2021}

\bibitem[{{Ohashi} et~al.(2014){Ohashi}, {Saigo}, {Aso}, {Aikawa}, {Koyamatsu}, {Machida} et~al.}]{Ohashi2014}
{Ohashi}, N., {Saigo}, K., {Aso}, Y., {Aikawa}, Y., {Koyamatsu}, S., {Machida}, M.~N., et~al. (2014).
\newblock {Formation of a Keplerian Disk in the Infalling Envelope around L1527 IRS: Transformation from Infalling Motions to Kepler Motions}.
\newblock \emph{\apj} 796, 131.
\newblock \doi{10.1088/0004-637X/796/2/131}
\bibAnnoteFile{Ohashi2014}

\bibitem[{{Ohashi} et~al.(2023){Ohashi}, {Tobin}, {J{\o}rgensen}, {Takakuwa}, {Sheehan}, {Aikawa} et~al.}]{Ohashi2023}
{Ohashi}, N., {Tobin}, J.~J., {J{\o}rgensen}, J.~K., {Takakuwa}, S., {Sheehan}, P., {Aikawa}, Y., et~al. (2023).
\newblock {Early Planet Formation in Embedded Disks (eDisk). I. Overview of the Program and First Results}.
\newblock \emph{\apj} 951, 8.
\newblock \doi{10.3847/1538-4357/acd384}
\bibAnnoteFile{Ohashi2023}

\bibitem[{{Oya} et~al.(2022){Oya}, {Kibukawa}, {Miyake}, and {Yamamoto}}]{Oya2022}
{Oya}, Y., {Kibukawa}, H., {Miyake}, S., and {Yamamoto}, S. (2022).
\newblock {FERIA: Flat Envelope Model with Rotation and Infall under Angular Momentum Conservation}.
\newblock \emph{\pasp} 134, 094301.
\newblock \doi{10.1088/1538-3873/ac8839}
\bibAnnoteFile{Oya2022}

\bibitem[{{Oya} et~al.(2015){Oya}, {Sakai}, {Lefloch}, {L{\'o}pez-Sepulcre}, {Watanabe}, {Ceccarelli} et~al.}]{Oya2015}
{Oya}, Y., {Sakai}, N., {Lefloch}, B., {L{\'o}pez-Sepulcre}, A., {Watanabe}, Y., {Ceccarelli}, C., et~al. (2015).
\newblock {Geometric and Kinematic Structure of the Outflow/Envelope System of L1527 Revealed by Subarcsecond-resolution Observation of CS}.
\newblock \emph{\apj} 812, 59.
\newblock \doi{10.1088/0004-637X/812/1/59}
\bibAnnoteFile{Oya2015}

\bibitem[{{Parise} et~al.(2006){Parise}, {Ceccarelli}, {Tielens}, {Castets}, {Caux}, {Lefloch} et~al.}]{Parise2006}
{Parise}, B., {Ceccarelli}, C., {Tielens}, A.~G.~G.~M., {Castets}, A., {Caux}, E., {Lefloch}, B., et~al. (2006).
\newblock {Testing grain surface chemistry: a survey of deuterated formaldehyde and methanol in low-mass class 0 protostars}.
\newblock \emph{\aap} 453, 949--958.
\newblock \doi{10.1051/0004-6361:20054476}
\bibAnnoteFile{Parise2006}

\bibitem[{{Perotti} et~al.(2023){Perotti}, {Christiaens}, {Henning}, {Tabone}, {Waters}, {Kamp} et~al.}]{Perotti2023}
{Perotti}, G., {Christiaens}, V., {Henning}, T., {Tabone}, B., {Waters}, L.~B.~F.~M., {Kamp}, I., et~al. (2023).
\newblock {Water in the terrestrial planet-forming zone of the PDS 70 disk}.
\newblock \emph{\nat} 620, 516--520.
\newblock \doi{10.1038/s41586-023-06317-9}
\bibAnnoteFile{Perotti2023}

\bibitem[{{Remijan} et~al.(2019){Remijan}, {Biggs}, {Cortes}, {Dent}, {Di Franceso}, {Fomalont} et~al.}]{ALMAtechnicalHandbook}
[Dataset] {Remijan}, A., {Biggs}, A., {Cortes}, P.~A., {Dent}, B., {Di Franceso}, J., {Fomalont}, E., et~al. (2019).
\newblock {ALMA Technical Handbook,ALMA Doc. 7.3, ver. 1.1}.
\newblock 2019, ALMA Technical Handbook,ALMA Doc. 7.3, ver. 1.1ISBN 978-3-923524-66-2.
\newblock \doi{10.5281/zenodo.4511522}
\bibAnnoteFile{ALMAtechnicalHandbook}

\bibitem[{{Rivi{\`e}re-Marichalar} et~al.(2026){Rivi{\`e}re-Marichalar}, {Fuente}, {le Gal}, {Neri}, {Esplugues}, {Semenov} et~al.}]{Riviere-Marichalar2026}
{Rivi{\`e}re-Marichalar}, P., {Fuente}, A., {le Gal}, R., {Neri}, R., {Esplugues}, G., {Semenov}, D., et~al. (2026).
\newblock {AB Aur, a Rosetta stone for studies of planet formation (IV): C/O estimates from CS and SO interferometric observations}.
\newblock \emph{arXiv e-prints} , arXiv:2602.06551\doi{10.48550/arXiv.2602.06551}
\bibAnnoteFile{Riviere-Marichalar2026}

\bibitem[{{Sai} et~al.(2023){Sai}, {Ohashi}, {Yen}, {Maury}, and {Maret}}]{Sai2023}
{Sai}, J. I.~C., {Ohashi}, N., {Yen}, H.-W., {Maury}, A.~J., and {Maret}, S. (2023).
\newblock {Probing Velocity Structures of Protostellar Envelopes: Infalling and Rotating Envelopes within Turbulent Dense Cores}.
\newblock \emph{\apj} 944, 222.
\newblock \doi{10.3847/1538-4357/acb3bd}
\bibAnnoteFile{Sai2023}

\bibitem[{{Sakai} et~al.(2017){Sakai}, {Oya}, {Higuchi}, {Aikawa}, {Hanawa}, {Ceccarelli} et~al.}]{Sakai2017}
{Sakai}, N., {Oya}, Y., {Higuchi}, A.~E., {Aikawa}, Y., {Hanawa}, T., {Ceccarelli}, C., et~al. (2017).
\newblock {Vertical structure of the transition zone from infalling rotating envelope to disc in the Class 0 protostar, IRAS 04368+2557}.
\newblock \emph{\mnras} 467, L76--L80.
\newblock \doi{10.1093/mnrasl/slx002}
\bibAnnoteFile{Sakai2017}

\bibitem[{{Sakai} et~al.(2014{\natexlab{a}}){Sakai}, {Oya}, {Sakai}, {Watanabe}, {Hirota}, {Ceccarelli} et~al.}]{Sakai2014}
{Sakai}, N., {Oya}, Y., {Sakai}, T., {Watanabe}, Y., {Hirota}, T., {Ceccarelli}, C., et~al. (2014{\natexlab{a}}).
\newblock {A Chemical View of Protostellar-disk Formation in L1527}.
\newblock \emph{\apjl} 791, L38.
\newblock \doi{10.1088/2041-8205/791/2/L38}
\bibAnnoteFile{Sakai2014}

\bibitem[{{Sakai} et~al.(2014{\natexlab{b}}){Sakai}, {Sakai}, {Hirota}, {Watanabe}, {Ceccarelli}, {Kahane} et~al.}]{Sakai2014_Nature}
{Sakai}, N., {Sakai}, T., {Hirota}, T., {Watanabe}, Y., {Ceccarelli}, C., {Kahane}, C., et~al. (2014{\natexlab{b}}).
\newblock {Change in the chemical composition of infalling gas forming a disk around a protostar}.
\newblock \emph{\nat} 507, 78--80.
\newblock \doi{10.1038/nature13000}
\bibAnnoteFile{Sakai2014_Nature}

\bibitem[{{Sakai} et~al.(2008){Sakai}, {Sakai}, {Hirota}, and {Yamamoto}}]{Sakai2008}
{Sakai}, N., {Sakai}, T., {Hirota}, T., and {Yamamoto}, S. (2008).
\newblock {Abundant Carbon-Chain Molecules toward the Low-Mass Protostar IRAS 04368+2557 in L1527}.
\newblock \emph{\apj} 672, 371--381.
\newblock \doi{10.1086/523635}
\bibAnnoteFile{Sakai2008}

\bibitem[{{Sakai} et~al.(2009){Sakai}, {Sakai}, {Hirota}, and {Yamamoto}}]{Sakai2009}
{Sakai}, N., {Sakai}, T., {Hirota}, T., and {Yamamoto}, S. (2009).
\newblock {Deuterated Molecules in Warm Carbon Chain Chemistry: The L1527 Case}.
\newblock \emph{\apj} 702, 1025--1035.
\newblock \doi{10.1088/0004-637X/702/2/1025}
\bibAnnoteFile{Sakai2009}

\bibitem[{{Sakai} et~al.(2010){Sakai}, {Sakai}, {Hirota}, and {Yamamoto}}]{Sakai2010}
{Sakai}, N., {Sakai}, T., {Hirota}, T., and {Yamamoto}, S. (2010).
\newblock {Distributions of Carbon-chain Molecules in L1527}.
\newblock \emph{\apj} 722, 1633--1643.
\newblock \doi{10.1088/0004-637X/722/2/1633}
\bibAnnoteFile{Sakai2010}

\bibitem[{{Sakai} and {Yamamoto}(2013)}]{Sakai2013}
{Sakai}, N. and {Yamamoto}, S. (2013).
\newblock {Warm Carbon-Chain Chemistry}.
\newblock \emph{Chemical Reviews} 113, 8981--9015.
\newblock \doi{10.1021/cr4001308}
\bibAnnoteFile{Sakai2013}

\bibitem[{{Semenov} et~al.(2018){Semenov}, {Favre}, {Fedele}, {Guilloteau}, {Teague}, {Henning} et~al.}]{Semenov2018}
{Semenov}, D., {Favre}, C., {Fedele}, D., {Guilloteau}, S., {Teague}, R., {Henning}, T., et~al. (2018).
\newblock {Chemistry in disks. XI. Sulfur-bearing species as tracers of protoplanetary disk physics and chemistry: the DM Tau case}.
\newblock \emph{\aap} 617, A28.
\newblock \doi{10.1051/0004-6361/201832980}
\bibAnnoteFile{Semenov2018}

\bibitem[{{Sharma} et~al.(2025){Sharma}, {J{\o}rgensen}, {van't Hoff}, {Lee}, {Aikawa}, {Gavino} et~al.}]{Sharma2025}
{Sharma}, R., {J{\o}rgensen}, J.~K., {van't Hoff}, M. L.~R., {Lee}, J.-E., {Aikawa}, Y., {Gavino}, S., et~al. (2025).
\newblock {Early Planet Formation in Embedded Disks (eDisk): XX. Constraining the chemical tracers of young protostellar sources}.
\newblock \emph{\aap} 704, A133.
\newblock \doi{10.1051/0004-6361/202553721}
\bibAnnoteFile{Sharma2025}

\bibitem[{{Sheehan} et~al.(2022){Sheehan}, {Tobin}, {Li}, {van't Hoff}, {J{\o}rgensen}, {Kwon} et~al.}]{Sheehan2022}
{Sheehan}, P.~D., {Tobin}, J.~J., {Li}, Z.-Y., {van't Hoff}, M. L.~R., {J{\o}rgensen}, J.~K., {Kwon}, W., et~al. (2022).
\newblock {A VLA View of the Flared, Asymmetric Disk around the Class 0 Protostar L1527 IRS}.
\newblock \emph{\apj} 934, 95.
\newblock \doi{10.3847/1538-4357/ac7a3b}
\bibAnnoteFile{Sheehan2022}

\bibitem[{{Shu} et~al.(1991){Shu}, {Ruden}, {Lada}, and {Lizano}}]{Shu1991}
{Shu}, F.~H., {Ruden}, S.~P., {Lada}, C.~J., and {Lizano}, S. (1991).
\newblock {Star Formation and the Nature of Bipolar Outflows}.
\newblock \emph{\apjl} 370, L31.
\newblock \doi{10.1086/185970}
\bibAnnoteFile{Shu1991}

\bibitem[{{Slavicinska} et~al.(2025){Slavicinska}, {Tychoniec}, {Navarro}, {van Dishoeck}, {Tobin}, {van Gelder} et~al.}]{Slavicinska2025}
{Slavicinska}, K., {Tychoniec}, {\L}., {Navarro}, M.~G., {van Dishoeck}, E.~F., {Tobin}, J.~J., {van Gelder}, M.~L., et~al. (2025).
\newblock {HDO Ice Detected toward an Isolated Low-mass Protostar with JWST}.
\newblock \emph{\apjl} 986, L19.
\newblock \doi{10.3847/2041-8213/addb45}
\bibAnnoteFile{Slavicinska2025}

\bibitem[{{Spezzano} et~al.(2016){Spezzano}, {Bizzocchi}, {Caselli}, {Harju}, and {Br{\"u}nken}}]{Spezzano2016}
{Spezzano}, S., {Bizzocchi}, L., {Caselli}, P., {Harju}, J., and {Br{\"u}nken}, S. (2016).
\newblock {Chemical differentiation in a prestellar core traces non-uniform illumination}.
\newblock \emph{\aap} 592, L11.
\newblock \doi{10.1051/0004-6361/201628652}
\bibAnnoteFile{Spezzano2016}

\bibitem[{{Spezzano} et~al.(2020){Spezzano}, {Caselli}, {Pineda}, {Bizzocchi}, {Prudenzano}, and {Nagy}}]{Spezzano2020}
{Spezzano}, S., {Caselli}, P., {Pineda}, J.~E., {Bizzocchi}, L., {Prudenzano}, D., and {Nagy}, Z. (2020).
\newblock {Distribution of methanol and cyclopropenylidene around starless cores★}.
\newblock \emph{\aap} 643, A60.
\newblock \doi{10.1051/0004-6361/201936598}
\bibAnnoteFile{Spezzano2020}

\bibitem[{{Tabone} et~al.(2023){Tabone}, {Bettoni}, {van Dishoeck}, {Arabhavi}, {Grant}, {Gasman} et~al.}]{Tabone2023}
{Tabone}, B., {Bettoni}, G., {van Dishoeck}, E.~F., {Arabhavi}, A.~M., {Grant}, S., {Gasman}, D., et~al. (2023).
\newblock {A rich hydrocarbon chemistry and high C to O ratio in the inner disk around a very low-mass star}.
\newblock \emph{Nature Astronomy} 7, 805--814.
\newblock \doi{10.1038/s41550-023-01965-3}
\bibAnnoteFile{Tabone2023}

\bibitem[{{Terebey} et~al.(2025){Terebey}, {Sandoval Ascencio}, {Flores-Rivera}, {Turner}, and {Barajas}}]{Terebey2025}
{Terebey}, S., {Sandoval Ascencio}, L., {Flores-Rivera}, L., {Turner}, N.~J., and {Barajas}, A. (2025).
\newblock {The Dynamics of Infall and Accretion Shocks in the Outer Disk}.
\newblock \emph{\apj} 990, 53.
\newblock \doi{10.3847/1538-4357/ade3c6}
\bibAnnoteFile{Terebey2025}

\bibitem[{{Tobin} et~al.(2013){Tobin}, {Bergin}, {Hartmann}, {Lee}, {Maret}, {Myers} et~al.}]{Tobin_N2H+}
{Tobin}, J.~J., {Bergin}, E.~A., {Hartmann}, L., {Lee}, J.-E., {Maret}, S., {Myers}, P.~C., et~al. (2013).
\newblock {Resolved Depletion Zones and Spatial Differentiation of N$_{2}$H$^{+}$ and N$_{2}$D$^{+}$}.
\newblock \emph{\apj} 765, 18.
\newblock \doi{10.1088/0004-637X/765/1/18}
\bibAnnoteFile{Tobin_N2H+}

\bibitem[{{Tobin} et~al.(2012){Tobin}, {Hartmann}, {Chiang}, {Wilner}, {Looney}, {Loinard} et~al.}]{Tobin2012}
{Tobin}, J.~J., {Hartmann}, L., {Chiang}, H.-F., {Wilner}, D.~J., {Looney}, L.~W., {Loinard}, L., et~al. (2012).
\newblock {A \raisebox{-0.5ex}\textasciitilde0.2-solar-mass protostar with a Keplerian disk in the very young L1527 IRS system}.
\newblock \emph{\nat} 492, 83--85.
\newblock \doi{10.1038/nature11610}
\bibAnnoteFile{Tobin2012}

\bibitem[{{Tuan-Anh} et~al.(2016){Tuan-Anh}, {Nhung}, {Hoai}, {Diep}, {Phuong}, {Thao} et~al.}]{Tuan-Anh2016}
{Tuan-Anh}, P., {Nhung}, P.~T., {Hoai}, D.~T., {Diep}, P.~N., {Phuong}, N.~T., {Thao}, N.~T., et~al. (2016).
\newblock {Morphology and kinematics of the gas envelope of protostar L1527 as obtained from ALMA observations of the C$^{18}$O(2-1) line emission}.
\newblock \emph{\mnras} 463, 3563--3572.
\newblock \doi{10.1093/mnras/stw2206}
\bibAnnoteFile{Tuan-Anh2016}

\bibitem[{{Tychoniec} et~al.(2020){Tychoniec}, {Manara}, {Rosotti}, {van Dishoeck}, {Cridland}, {Hsieh} et~al.}]{Tychoniec2020}
{Tychoniec}, {\L}., {Manara}, C.~F., {Rosotti}, G.~P., {van Dishoeck}, E.~F., {Cridland}, A.~J., {Hsieh}, T.-H., et~al. (2020).
\newblock {Dust masses of young disks: constraining the initial solid reservoir for planet formation}.
\newblock \emph{\aap} 640, A19.
\newblock \doi{10.1051/0004-6361/202037851}
\bibAnnoteFile{Tychoniec2020}

\bibitem[{{Tychoniec} et~al.(2021){Tychoniec}, {van Dishoeck}, {van't Hoff}, {van Gelder}, {Tabone}, {Chen} et~al.}]{Tychoniec2021}
{Tychoniec}, {\L}., {van Dishoeck}, E.~F., {van't Hoff}, M. L.~R., {van Gelder}, M.~L., {Tabone}, B., {Chen}, Y., et~al. (2021).
\newblock {Which molecule traces what: Chemical diagnostics of protostellar sources}.
\newblock \emph{\aap} 655, A65.
\newblock \doi{10.1051/0004-6361/202140692}
\bibAnnoteFile{Tychoniec2021}

\bibitem[{{Ulrich}(1976)}]{Ulrich1976}
{Ulrich}, R.~K. (1976).
\newblock {An infall model for the T Tauri phenomenon.}
\newblock \emph{\apj} 210, 377--391.
\newblock \doi{10.1086/154840}
\bibAnnoteFile{Ulrich1976}

\bibitem[{{van Dishoeck} et~al.(1995){van Dishoeck}, {Blake}, {Jansen}, and {Groesbeck}}]{vanDishoeck1995}
{van Dishoeck}, E.~F., {Blake}, G.~A., {Jansen}, D.~J., and {Groesbeck}, T.~D. (1995).
\newblock {Molecular Abundances and Low-Mass Star Formation. II. Organic and Deuterated Species toward IRAS 16293-2422}.
\newblock \emph{\apj} 447, 760.
\newblock \doi{10.1086/175915}
\bibAnnoteFile{vanDishoeck1995}

\bibitem[{{van 't Hoff} et~al.(2020){van 't Hoff}, {Harsono}, {Tobin}, {Bosman}, {van Dishoeck}, {J{\o}rgensen} et~al.}]{vantHoff2020}
{van 't Hoff}, M. L.~R., {Harsono}, D., {Tobin}, J.~J., {Bosman}, A.~D., {van Dishoeck}, E.~F., {J{\o}rgensen}, J.~K., et~al. (2020).
\newblock {Temperature Structures of Embedded Disks: Young Disks in Taurus Are Warm}.
\newblock \emph{\apj} 901, 166.
\newblock \doi{10.3847/1538-4357/abb1a2}
\bibAnnoteFile{vantHoff2020}

\bibitem[{{van 't Hoff} et~al.(2022){van 't Hoff}, {Leemker}, {Tobin}, {Harsono}, {J{\o}rgensen}, and {Bergin}}]{vantHoff2022}
{van 't Hoff}, M. L.~R., {Leemker}, M., {Tobin}, J.~J., {Harsono}, D., {J{\o}rgensen}, J.~K., and {Bergin}, E.~A. (2022).
\newblock {The Young Embedded Disk L1527 IRS: Constraints on the Water Snowline and Cosmic-Ray Ionization Rate from HCO+ Observations}.
\newblock \emph{\apj} 932, 6.
\newblock \doi{10.3847/1538-4357/ac63b4}
\bibAnnoteFile{vantHoff2022}

\bibitem[{{van 't Hoff} et~al.(2018){van 't Hoff}, {Tobin}, {Harsono}, and {van Dishoeck}}]{vantHoff2018}
{van 't Hoff}, M. L.~R., {Tobin}, J.~J., {Harsono}, D., and {van Dishoeck}, E.~F. (2018).
\newblock {Unveiling the physical conditions of the youngest disks. A warm embedded disk in L1527}.
\newblock \emph{\aap} 615, A83.
\newblock \doi{10.1051/0004-6361/201732313}
\bibAnnoteFile{vantHoff2018}

\bibitem[{{van 't Hoff} et~al.(2023){van 't Hoff}, {Tobin}, {Li}, {Ohashi}, {J{\o}rgensen}, {Lin} et~al.}]{vantHoff_eDisk}
{van 't Hoff}, M. L.~R., {Tobin}, J.~J., {Li}, Z.-Y., {Ohashi}, N., {J{\o}rgensen}, J.~K., {Lin}, Z.-Y.~D., et~al. (2023).
\newblock {Early Planet Formation in Embedded Disks (eDisk). III. A First High-resolution View of Submillimeter Continuum and Molecular Line Emission toward the Class 0 Protostar L1527 IRS}.
\newblock \emph{\apj} 951, 10.
\newblock \doi{10.3847/1538-4357/accf87}
\bibAnnoteFile{vantHoff_eDisk}

\bibitem[{{Wilson} and {Rood}(1994)}]{Wilson1994}
{Wilson}, T.~L. and {Rood}, R. (1994).
\newblock {Abundances in the Interstellar Medium}.
\newblock \emph{\araa} 32, 191--226.
\newblock \doi{10.1146/annurev.aa.32.090194.001203}
\bibAnnoteFile{Wilson1994}

\bibitem[{{Yoshida} et~al.(2019){Yoshida}, {Sakai}, {Nishimura}, {Tokudome}, {Watanabe}, {Sakai} et~al.}]{Yoshida2019}
{Yoshida}, K., {Sakai}, N., {Nishimura}, Y., {Tokudome}, T., {Watanabe}, Y., {Sakai}, T., et~al. (2019).
\newblock {An unbiased spectral line survey observation toward the low-mass star-forming region L1527}.
\newblock \emph{\pasj} 71, S18.
\newblock \doi{10.1093/pasj/psy136}
\bibAnnoteFile{Yoshida2019}

\bibitem[{{Youdin} and {Zhu}(2025)}]{Youdin2025}
{Youdin}, A.~N. and {Zhu}, Z. (2025).
\newblock {Formation of Giant Planets}.
\newblock \emph{arXiv e-prints} , arXiv:2501.13214\doi{10.48550/arXiv.2501.13214}
\bibAnnoteFile{Youdin2025}

\bibitem[{{Zhang} et~al.(2019){Zhang}, {Bergin}, {Schwarz}, {Krijt}, and {Ciesla}}]{Zhang2019}
{Zhang}, K., {Bergin}, E.~A., {Schwarz}, K., {Krijt}, S., and {Ciesla}, F. (2019).
\newblock {Systematic Variations of CO Gas Abundance with Radius in Gas-rich Protoplanetary Disks}.
\newblock \emph{\apj} 883, 98.
\newblock \doi{10.3847/1538-4357/ab38b9}
\bibAnnoteFile{Zhang2019}

\bibitem[{{Zhang} et~al.(2021){Zhang}, {Booth}, {Law}, {Bosman}, {Schwarz}, {Bergin} et~al.}]{Zhang2021}
{Zhang}, K., {Booth}, A.~S., {Law}, C.~J., {Bosman}, A.~D., {Schwarz}, K.~R., {Bergin}, E.~A., et~al. (2021).
\newblock {Molecules with ALMA at Planet-forming Scales (MAPS). V. CO Gas Distributions}.
\newblock \emph{\apjs} 257, 5.
\newblock \doi{10.3847/1538-4365/ac1580}
\bibAnnoteFile{Zhang2021}

\bibitem[{{Zhang} et~al.(2020){Zhang}, {Schwarz}, and {Bergin}}]{Zhang2020}
{Zhang}, K., {Schwarz}, K.~R., and {Bergin}, E.~A. (2020).
\newblock {Rapid Evolution of Volatile CO from the Protostellar Disk Stage to the Protoplanetary Disk Stage}.
\newblock \emph{\apjl} 891, L17.
\newblock \doi{10.3847/2041-8213/ab7823}
\bibAnnoteFile{Zhang2020}

\bibitem[{{Zhang} et~al.(2018){Zhang}, {Higuchi}, {Sakai}, {Oya}, {L{\'o}pez-Sepulcre}, {Imai} et~al.}]{Zhang2018}
{Zhang}, Y., {Higuchi}, A.~E., {Sakai}, N., {Oya}, Y., {L{\'o}pez-Sepulcre}, A., {Imai}, M., et~al. (2018).
\newblock {Rotation in the NGC 1333 IRAS 4C Outflow}.
\newblock \emph{\apj} 864, 76.
\newblock \doi{10.3847/1538-4357/aad7ba}
\bibAnnoteFile{Zhang2018}

\bibitem[{{Zhang} et~al.(2024){Zhang}, {Sakai}, {Ohashi}, {Murillo}, {Chandler}, {Svoboda} et~al.}]{Zhang2024}
{Zhang}, Z.~E., {Sakai}, N., {Ohashi}, S., {Murillo}, N.~M., {Chandler}, C.~J., {Svoboda}, B., et~al. (2024).
\newblock {FAUST. XIV. Probing the Flared Disk in L1527 with Sulfur-bearing Molecules}.
\newblock \emph{\apj} 966, 207.
\newblock \doi{10.3847/1538-4357/ad3921}
\bibAnnoteFile{Zhang2024}

\bibitem[{{Zucker} et~al.(2019){Zucker}, {Speagle}, {Schlafly}, {Green}, {Finkbeiner}, {Goodman} et~al.}]{Zucker2019}
{Zucker}, C., {Speagle}, J.~S., {Schlafly}, E.~F., {Green}, G.~M., {Finkbeiner}, D.~P., {Goodman}, A.~A., et~al. (2019).
\newblock {A Large Catalog of Accurate Distances to Local Molecular Clouds: The Gaia DR2 Edition}.
\newblock \emph{\apj} 879, 125.
\newblock \doi{10.3847/1538-4357/ab2388}
\bibAnnoteFile{Zucker2019}

\end{thebibliography}
